\documentclass[preprint,12pt]{elsarticle}
\usepackage[left=1 in,right=1 in, top=1 in, bottom=1 in]{geometry}

\usepackage{hyperref}       

\usepackage{graphicx,color} 

\usepackage{amsmath,amsfonts,amssymb,theorem}
\usepackage{url,subfig,anysize,wrapfig}
\usepackage{multirow,tabularx,afterpage,fancyhdr}
\usepackage{cases}
\usepackage{epstopdf}

\usepackage[linewidth=1pt]{mdframed}
\usepackage{lipsum}

\usepackage{lineno}
\usepackage{placeins}

\usepackage{tikz}
\usetikzlibrary{calc}

\usepackage[titletoc]{appendix}

\begin{document}

\begin{frontmatter}



\title{Geometry of escape and transition dynamics in the presence of dissipative and gyroscopic forces in two degree of freedom systems}

\date{\today} 


\author[1]{Jun Zhong\corref{cor1}}

\author[2]{Shane D. Ross}

\address[1]{Engineering Mechanics Program, Virginia Tech, Blacksburg, VA 24061, USA}

\address[2]{Department of Aerospace and Ocean Engineering, Virginia Tech, Blacksburg, VA 24061, USA}

\cortext[cor1]{Corresponding author: junzhong@vt.edu (J. Zhong)}

\begin{abstract}
Escape from a potential well can occur in different physical systems, such as capsize of ships, resonance transitions in celestial mechanics, and dynamic snap-through of arches and shells, as well as molecular reconfigurations in chemical reactions. 
The criteria and routes of escape in one-degree of freedom systems has been well studied theoretically with reasonable agreement with experiment. 
The trajectory can only transit from the hilltop of the one-dimensional potential energy surface. 
The situation becomes more complicated when the system has higher degrees of freedom since it has multiple routes to escape through an equilibrium of saddle-type, specifically, an index-1 saddle. 
This paper  summarizes the geometry of escape across a saddle in some widely known physical systems with two degrees of freedom and establishes the criteria of escape providing both  a methodology and results under the conceptual framework known as tube dynamics. 
These problems are classified into two categories based on whether the saddle projection and focus projection in the symplectic eigenspace  are coupled or not when damping and/or gyroscopic effects are considered. 
For simplicity, only the linearized system around the saddle points is analyzed, but the results generalize to the nonlinear system.  
We define a transition region, $\mathcal{T}_h$, as the region of initial conditions of a given initial energy $h$ which transit from one side of a saddle to the other.  
We find that in conservative systems, the boundary of the transition region, 
$\partial \mathcal{T}_h$, is a cylinder, while in dissipative systems, $\partial \mathcal{T}_h$ is an ellipsoid. 	
\end{abstract}


\end{frontmatter}

\newpage

\tableofcontents

\newpage
\
\section{Introduction}

Transition events are very common in both the natural world, daily life and even industrial applications. 
Examples of transition are the snap-through of plant leaves and engineering structures  responding to stimuli \cite{virgin2017geometric,zhong2018tube}, the flipping over of umbrellas on a windy day, reaction rates in chemical reaction dynamics \cite{wiggins2001impenetrable}, the escape and recapture of comets and spacecraft in celestial mechanics \cite{jaffe2002statistical,KoLoMaRo2000,KoLoMaRo2011},  and the capsize of ships \cite{soliman1991transient,NaRo2017}. Better understanding and prediction of transitions, or escape, have significance in both utilization and evasion of such events, such as how to transfer spacecraft in specific space missions from one prescribed initial orbit to a desired final orbit with lower energy,  or in structural mechanics, how to avoid collapse of structures. From the perspective of mechanics, such behavior can be interpreted as the escape from one local minimum of potential energy (i.e., a potential well) to another, which has been widely been studied as `escape dynamics' \cite{waalkens2005efficient,contopoulos2013order,zotos2014escapes,zotos2017overview,barrio2009bifurcations}. 
Escape in a one degree of freedom system, like a double-well oscillator, is unambiguous, as the phase space is two dimensional and the hilltop equilibrium becomes a saddle point in phase space.  
The only way the system state can escape from the potential energy minimum is by passing over the hilltop to another local minimum. 
Therefore, all trajectories which have an energy above that of the hilltop, as evaluated as they pass through the location of the hilltop, transit from one side to the other.
This situation has been studied by both experiments and theory with good agreement between the two \cite{gottwald1995routes,virgin2000introduction,mann2009energy}.

Higher degree of freedom systems, however, are  more complicated since there are multiple paths to transition through an index-1 saddle equilibrium point, as the phase space is now four dimensions or more. 
For such systems, it is of importance to establish systematic methods and criteria to predict the escape from a potential well. 
In this paper, we focus on two degrees of freedom systems, an intermediate situation, to simplify the analysis procedure, and consider the effect of damping and gyroscopic forces both in isolation and in combination. 
We take a Hamiltonian point of view and use canonical Hamiltonian variables, even when dissipation is included.

Generally, escape can occur only when the system has energy higher than the escape energy which is the critical energy that allows escape, the energy of the saddle point \cite{zhong2018tube,NaRo2017,zotos2014escapes,zotos2017overview}. If the energy is lower than the escape energy, the zero velocity curve (or surface)---
which is the boundary of the projection of the energy manifold onto position space---is closed, allowing no open neck region around the saddle point. 
In this case, all of the trajectories are bounded to only evolve within their potential wells of origin and no trajectory can escape from the well. 
For initial conditions with energy higher than the escape energy, the equipotential surfaces open around the saddle point in a neck region, 
and trajectories have a chance to escape the potential well  to another or even to infinity. 
However, the energy criterion alone is not sufficient to guarantee  escape. 
The dynamic boundary between transition and non-transition of a system with energy higher than critical energy can be thoroughly understood under the conceptual framework of transition dynamics or sometimes known as tube dynamics \cite{zhong2018tube,NaRo2017,GaKoMaRoYa2006,ross2004cylindrical,OnYoRo2017}. In conservative two degrees of freedom systems with energy higher than the critical energy, there is an unstable periodic orbit in the bottleneck region. Emanating from the periodic orbit are its stable and unstable manifolds which have cylindrical or ``tube'' geometry within the conserved energy manifold. The tube manifold, sometimes called a transition tube in tube dynamics, consists of pieces of asymptotic orbits.  As stated in \cite{contopoulos2013order}, the best systematic way to study the escape from such a system is by calculating the asymptotic orbits of the periodic orbit. The reason is that the transition tube, acting like a separatrix, separates two distinct types of orbits: transit orbits and non-transit orbits. Transit orbits are those inside of the tube which can escape from one potential well to another, while non-transit, those outside of the tube, cannot pass through the bottleneck region, and thus return to their region of origin.

Although we have made it clear that the phase space structure, known as a transition tube, governs the escape in conservative systems of two degrees of freedom, it is just an ideal case since energy fluctuations and dissipation cannot be avoided in the real world. Thus,
it is natural to consider how the situation will change
if  dissipative forces are considered. 
Ref.\ \cite{zhong2018tube} has addressed this, in part, for the example of  dynamic snap-through of a shallow arch. By using the bisection method, transition boundaries for both the nonlinear conservative system and dissipative system were obtained. The transition `tube' for the dissipative system was found to be different from that for the conservative system. The transition tube of the conservative system not only gives all the initial conditions for transit orbits in phase space, but also gives the boundary of their evolution, while the transition `tube' for the dissipative system merely gives the boundary of the initial conditions of a specific initial energy for transit orbits on a specific Poincar\'e section and the evolution of the transit orbits with those initial conditions is not along an invariant energy manifold any longer. As for the global structure of the phase space in the dissipative system that governs the initial conditions of transit orbits, this was not addressed in  \cite{zhong2018tube}. In the current study, we  continue this study and answer in more detail the concern of how the situation changes when dissipation is present, finding that the transition tube in the conservative system becomes a transition ellipsoid in the dissipative system. 

On the other hand, when the system is rotating or magnetic forces are present, gyroscopic forces 
must be considered. 
Gyroscopic forces, widely found in rotating systems \cite{greenwood2006advanced,bloch2004gyroscopically,krechetnikov2007dissipation,bottema1976stability,kirillov2011brouwers} as well as electromagnetic systems, are non-dissipative and the gyroscopic coefficients enter the equations of motion in a skew-symmetric manner \cite{greenwood2006advanced}. Some researchers have studied escape in conservative gyroscopic systems (e.g., \cite{gabern2005theory,KoLoMaRo2011}). 
There exist transition tubes controlling the escape which are topologically the same as in an inertial system \cite{zhong2018tube,NaRo2017,Ross2018experimental}. 
However, to the best knowledge of the authors, no study has been carried out to study the escape in systems with {\it both} dissipative and gyroscopic forces present. In fact, gyroscopic systems are 
interesting due to some unexpected phenomena which have some uncommon features. In conservative gyroscopic systems, motion near an unstable point of the potential energy surface, such as an index-1 saddle point, can be stabilized via gyroscopic forces, e.g., rotation with large enough angular velocity \cite{bottema1976stability,kirillov2011brouwers,MuDe1999,Szebehely1967}. But small dissipation can make the system lose the stability which is called dissipation-induced instability \cite{krechetnikov2007dissipation}, different from the common notion that dissipation makes a system more stable. 
Considering this difference in dynamical behavior, another concern is whether 
  the dynamical behavior of the dissipative system will be the same if the gyroscopic forces are included. This study will also partially answer this concern.

In this paper, we will establish criteria and present methods to estimate the transition in different physical problems with two degrees of freedom. The systems are: an idealized rolling ball on both stationary and rotating saddle surfaces, the pitch and roll dynamics of a ship near the capsize state with equal and unequal damping, the snap-through of a shallow arch, and potential well transitions in the planar circular restricted three-body problem (PCR3BP). 
The focus of this analysis  is the local behavior near the neck region around the saddle point, obtained via the linearized dynamics.  
The corresponding global behavior are left for future work. 
In such linearized systems, the equilibrium point is of type saddle $\times$ center in the conservative system (i.e., an index-1 saddle) which becomes a saddle $\times$ focus when dissipation is considered.
In other words, the equilibrium point changes from one with a one-dimensional stable, one-dimensional unstable, and two-dimensional center manifold, to  one with a three-dimensional stable and one-dimensional unstable manifold.
To compare the similarities and differences between the conservative and dissipative system in each setting, we  introduce the same change of variables that uses the generalized eigenvectors of the corresponding conservative system, which we refer to as the {\it symplectic eigenspace}. 

In the symplectic eigenspace, the dynamics in the saddle and focus projections are coupled for some dissipative systems, while for others, they remain  uncoupled. 
Thus, this paper classifies different systems into two categories depending on the resulting linear coupling between the saddle and focus variables of the transformed dissipative system. 
The example problems considered share the same dynamic behavior so that we only need to give the full analysis for just one as an exemplar representative. 
Among the problems we will discuss, the idealized ball rolling on a saddle surface is of special interest since it can be either an inertial system or gyroscopic system,
depending on whether the surface is stationary or rotating so that it can have the properties of both types of problems. 
Thus, we will focus on analyzing the idealized ball rolling on a surface, where the rotation is about the saddle point itself.  
The other examples will be shown to be equivalent to a standard form derived for the idealized ball rolling on a surface.
The PCR3BP from celestial mechanics is a final special case as it involves rotation, but not about the saddle point.  
When a certain kind of dissipation  is included, the saddle point changes location compared with the conservative system and special care needs to be taken for this case, using an effective quadratic Hamiltonian about the saddle point.

\section{Transition region for the conservative case}
\label{general phase space}
A linear two degrees of freedom conservative system with a saddle-center type equilibrium point (i.e., index-1 or rank-1 saddle) \cite{zhong2018tube,wiggins2001impenetrable,jaffe2002statistical,KoLoMaRo2000,NaRo2017} can be transformed via a canonical transformation to normal form coordinates $(q_1,q_2,p_1,p_2)$ such that the  quadratic Hamiltonian, $\mathcal{H}_2$, can be written in the normal form,
\begin{equation}
\mathcal{H}_2= \lambda q_1 p_2 + \tfrac{1}{2}\omega_p \left(q_2^2 + p_2^2 \right),
\label{phase-space Hamiltonian}
\end{equation}
where $q_i$ and $p_i$ are the generalized coordinates and corresponding momenta. The Hamiltonian equations are defined as
\begin{equation}
\dot q_i = \frac{\partial \mathcal{H}_2}{\partial p_i}, \hspace{0.2in} \dot p_i = - \frac{\partial \mathcal{H}_2}{\partial q_i},
\end{equation}
which yields the following equations of motion,
\begin{equation}
\begin{aligned}
\dot q_1 & = \lambda q_1, \hspace{0.2 in} && \dot p_1 = - \lambda p_1,\\
\dot q_2 & = \omega_p p_2, \hspace{0.2in} && \dot p_2 = -\omega_p q_2,
\label{phase-space Hamiltonian Equations}
\end{aligned}
\end{equation}
where the dot over the variable denotes the derivative with respect to time. In the above equations, $\lambda$ is the real eigenvalue corresponding to the saddle coordinates spanned by $(q_1,p_1)$ and $\omega_p$ is the frequency associated with the center coordinates $(q_2,p_2)$. The solutions can be written as,
\begin{equation}
\begin{aligned}
& q_1 =q_1^0 e^{\lambda t}, \hspace{0.2in} p_1= p_1^0 e^{-\lambda t},\\
& q_2 + i p_2 = \left(q_2^0 + i p_2^0 \right) e^{- i \omega_p t}.
\label{phase-space solutions}
\end{aligned}
\end{equation}
Note that,
\begin{equation}
f_1=q_1 p_1, \hspace{0.2in} f_2 = q_2^2 + p_2^2
\end{equation}
are two independent constants of motion under the Hamiltonian system \eqref{phase-space Hamiltonian} with $\mathcal{H}_2$ itself trivially a constant of motion.

\subsection{Boundary of transit and non-transit orbits}
\label{R: sec:separatrix}

\paragraph{The linearized phase space} 
For positive $h$ and $c$,
the equilibrium or bottleneck region $\mathcal{R}$ (sometimes just called the neck region), which is determined by,
\[
\mathcal{H}_2=h, \quad  \mbox{and} \quad |p_1-q_1|\leq c,
\]
where $c>0$, is homeomorphic to the product of a 2-sphere and an interval $I \in \mathbb{R}$,
$S^2\times I$;
namely, for each fixed value of $p_1 -q_1 $ in the interval $I=[-c,c]$,
we see that  the equation $\mathcal{H}_2=h$ determines a 2-sphere,
\begin{equation}\label{2-sphere}
\tfrac{\lambda }{4}(q_1 +p_1 )^2
+ \tfrac{1}{2}\omega_p (q_2^2+p_2^2)
=h+\tfrac{\lambda }{4}(p_1 -q_1 )^2.
\end{equation}
Suppose $a \in I$, then \eqref{2-sphere} can be re-written as,
\begin{equation}\label{2-sphere2}
x_1^2 + q_2^2+p_2^2
= r^2,
\end{equation}
where $x_1 = \sqrt{\tfrac{1 }{2}\tfrac{\lambda}{\omega_p}}(q_1 +p_1 )$ and 
$r^2=\tfrac{2}{\omega_p}(h+\tfrac{\lambda }{4}a^2)$, which defines a 2-sphere of radius $r$ in the three variables $x_1$, $q_2$, and $p_2$.

\begin{figure}[!t]
	\begin{center}
		\includegraphics[width=\textwidth]{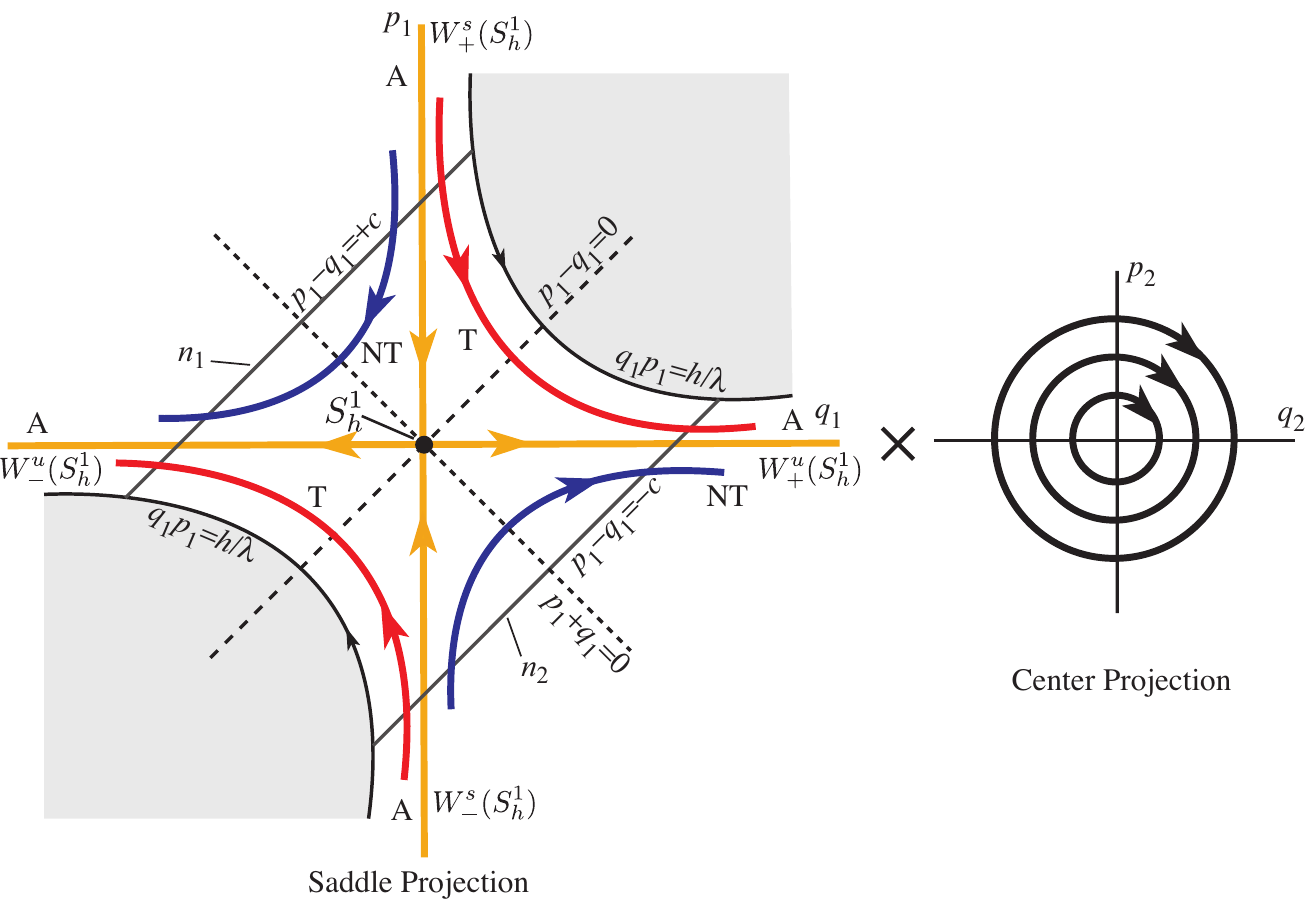}
	\end{center}
	\caption{{\footnotesize 
			The flow in the equilibrium region for the conservative system has the form
			saddle $\times$ center.
			On the left is shown a schematic of the projection onto the $(q_1,p_1)$-plane, the saddle projection.
			For the conservative dynamics, the Hamiltonian function $\mathcal{H}_2$ remains constant at $h>0$. Shown are the periodic orbit
			(black dot at the center), the asymptotic orbits (labeled A), two
			transit orbits (T) and two non-transit orbits (NT).
	}}
	\label{conservative eigen flow}
\end{figure}

The bounding 2-sphere of $\mathcal{R}$ for which $p_1 -q_1 = c$ will
be called $n_1$ (the ``left'' bounding 2-sphere), and where $p_1 -q_1 = -c$,
$n_2$ (the ``right'' bounding 2-sphere). 
Therefore, $\partial \mathcal{R} =\{ n_1, n_2 \}$.
See Figure \ref{conservative eigen flow}.  

We call the set of points on each
bounding 2-sphere where $q_1 + p_1 = 0$ the equator, and the sets where
$q_1 + p_1 > 0$ or $q_1 + p_1 < 0$ will be called the northern and 
southern hemispheres, respectively.

\paragraph{The linear flow in $\mathcal{R}$} 
To analyze the flow in
$\mathcal{R}$,   consider
the projections on the ($q_1, p_1$)-plane and the $(q_2,p_2)$-plane, respectively.
In the first case we see the standard
picture of a saddle point in two dimensions,
and in the second, of a center consisting of 
harmonic oscillator motion.
Figure \ref{conservative eigen flow} schematically illustrates the flow.
With regard to the first projection we
see that $\mathcal{R}$ itself projects
to a set bounded on two sides by
the hyperbolas
$q_1p_1 = h/\lambda $
(corresponding to $q_2^2+p_2^2=0$, see \eqref{phase-space Hamiltonian}) and on two
other sides by the line segments
$p_1-q_1= \pm c$, which correspond to the bounding 2-spheres, $n_1$ and $n_2$, respectively.

Since $q_1p_1$ is an integral
of the equations in $\mathcal{R}$,
the projections of
orbits in the $(q_1,p_1)$-plane
move on the branches of the corresponding
hyperbolas $q_1p_1 =$ constant,
except in the case $q_1p_1=0$, where $q_1 =0$ or $p_1 =0$.
If $q_1p_1 >0$, the branches connect
the bounding line segments $p_1 -q_1 =\pm c$ and if $q_1p_1 <0$, they
have both end points on the same segment.  A check of equation
\eqref{phase-space solutions} shows that the orbits move
as indicated by the arrows in Figure \ref{conservative eigen flow}.

To interpret Figure \ref{conservative eigen flow} as a flow in $\mathcal{R}$,  notice
that each point in the $(q_1,p_1)$-plane projection 
corresponds to a 1-sphere, $S^1$, or circle, in
$\mathcal{R}$ given by, 
\[
q_2^2+p_2^2
=\tfrac{2 }{\omega_p}(h-\lambda q_1p_1) .
\]
Of course, for points on the bounding  hyperbolic
segments ($q_1p_1 =h/\lambda $), the
1-sphere collapses to a point. Thus, the segments of the lines 
$p_1-q_1 =\pm c$  in the projection correspond to the 2-spheres
bounding $\mathcal{R}$.  This is because each corresponds to a
1-sphere crossed with an interval where the two end 1-spheres are
pinched to a point.

We distinguish nine classes of orbits grouped into the following four
categories:
\begin{enumerate}
	\item The point $q_1 =p_1 =0$ corresponds to an invariant
	1-sphere $S^1_h$, an unstable {\bf periodic orbit} in
	$\mathcal{R}$ of energy $\mathcal{H}_2=h$.  This 1-sphere is given by,
	\begin{equation}\label{3-sphere}
	q_2^2+p_2^2=\tfrac{2 }{\omega_p}h, 
	\hspace{0.3in} q_1 =p_1 =0.
	\end{equation}
	It is an example of a 
	normally hyperbolic invariant manifold (NHIM) (see \cite{Wiggins1994}).
	Roughly, this means that the stretching and contraction rates under
	the linearized dynamics transverse to the 1-sphere dominate those tangent
	to the 1-sphere.  This is clear for this example since the dynamics normal
	to the 1-sphere are described by the exponential contraction and expansion
	of the saddle point dynamics.  Here the 1-sphere acts as a ``big
	saddle point''.  
	See the black dot at the center of the $(q_1,p_1)$-plane on the left side 
	of Figure
	\ref{conservative eigen flow}.
	
	\item The four half open segments on the axes, $q_1p_1 =0$, 
	correspond to four 
	cylinder surfaces of orbits asymptotic to this invariant 1-sphere 
	$S^1_h$ either as time
	increases ($q_1 =0$) or as time decreases ($p_1 =0$).  These are called {\bf
		asymptotic} orbits and they are the stable and the unstable manifolds of
	$S^1_h$.  The stable manifolds, $W^s_{\pm}(S^1_h)$, are given by,
	\begin{equation}\label{stable_manifold}
	q_2^2+p_2^2=\tfrac{2 }{\omega_p}h, 
	\hspace{0.3in} q_1 =0,
	\hspace{0.3in} p_1 ~{\rm arbitrary}.
	\end{equation}
	$W^s_+(S^1_h)$ (with $p_1>0$) is the branch entering from $n_1$ and
	$W^s_-(S^1_h)$ (with $p_1<0$) is the branch entering from $n_2$.
	The unstable manifolds, $W^u_{\pm}(S^1_h)$, 
	are given by,
	\begin{equation}\label{unstable_manifold}
	q_2^2+p_2^2=\tfrac{2 }{\omega_p}h, 
	\hspace{0.3in} p_1 =0,
	\hspace{0.3in} q_1 ~{\rm arbitrary}
	\end{equation}
	$W^u_+(S^1_h)$ (with $q_1>0$) is the branch exiting from $n_2$ and
	$W^u_-(S^1_h)$ (with $q_1<0$) is the branch exiting from $n_1$.
	See the four orbits labeled A of Figure \ref{conservative eigen flow}.
	
	\item The hyperbolic segments determined by
	$q_1p_1 ={\rm constant}>0$ correspond
	to two solid cylinders of orbits 
	which cross $\mathcal{R}$ from one bounding 2-sphere to the
	other, meeting both in the same hemisphere; the northern hemisphere
	if they go from
	$p_1-q_1 =+c$ to $p_1-q_1 =-c$, and the southern hemisphere
	in the other case. Since
	these orbits transit from one realm to another, we call  them {\bf transit}
	orbits.  See the two orbits labeled T of Figure \ref{conservative eigen flow}.
	
	\item Finally the hyperbolic segments determined by $q_1p_1 = {\rm
		constant}<0$ correspond to two cylinders of orbits in
	$\mathcal{R}$ each of which runs from one hemisphere to the other hemisphere
	on the same bounding 2-sphere.  Thus if $q_1 >0$, the 2-sphere is $n_2$ ($p_1
	-q_1 =-c$) and orbits run from the southern hemisphere
	($q_1 +p_1 <0$) to the northern hemisphere ($q_1
	+p_1 >0$) while the converse holds if $q_1 <0$, where the 
	2-sphere is
	$n_1$. Since these orbits return to the same realm, we call them {\bf
		non-transit} orbits.  See the two orbits labeled NT of Figure \ref{conservative eigen flow}.
\end{enumerate}

We define the transition region, $\mathcal{T}_h$, as the region of initial conditions of a given initial energy $h$ which transit from one side of the neck region to the other.  
This is the set of all transit orbits, which has the geometry of a solid cylinder.   
The transition region, $\mathcal{T}_h$, is made up of one half which goes to the right 
(from $n_1$ to $n_2$),
$\mathcal{T}_{h+}$, defined by $q_1p_1 ={\rm constant}>0$ with both $q_1>0$ and $p_1>0$, and the other half which goes to the left
(from $n_2$ to $n_1$), 
$\mathcal{T}_{h-}$, 
defined by $q_1p_1 ={\rm constant}>0$ with both $q_1<0$ and $p_1<0$.
The boundaries are $\partial \mathcal{T}_{h+}$ and $\partial \mathcal{T}_{h-}$, respectively.
The closure of $\partial \mathcal{T}_{h}$, $\overline{\partial \mathcal{T}_{h}}$, is equal to the boundaries 
$\partial \mathcal{T}_{h+}$ and $\partial \mathcal{T}_{h-}$, along with the periodic orbit $S^1_h$, i.e., 
$\partial \mathcal{T}_{h-} \cup \partial \mathcal{T}_{h+} \cup S^1_h$.

In summary, for the conservative case, the boundary of the transition region, 
$\partial \mathcal{T}_h$, has the topology of a cylinder.  The topology of $\partial \mathcal{T}_h$ will be different for the dissipative case, as will be shown in later sections.
For convenience, we may refer to $\partial \mathcal{T}_{h}$ and $\overline{\partial \mathcal{T}_{h}}$ interchangeably.

\subsection{McGehee representation of the equilibrium region}
\label{Conservative McGehee representaion}
\begin{figure}
	\begin{center}
		\includegraphics[width=\textwidth]{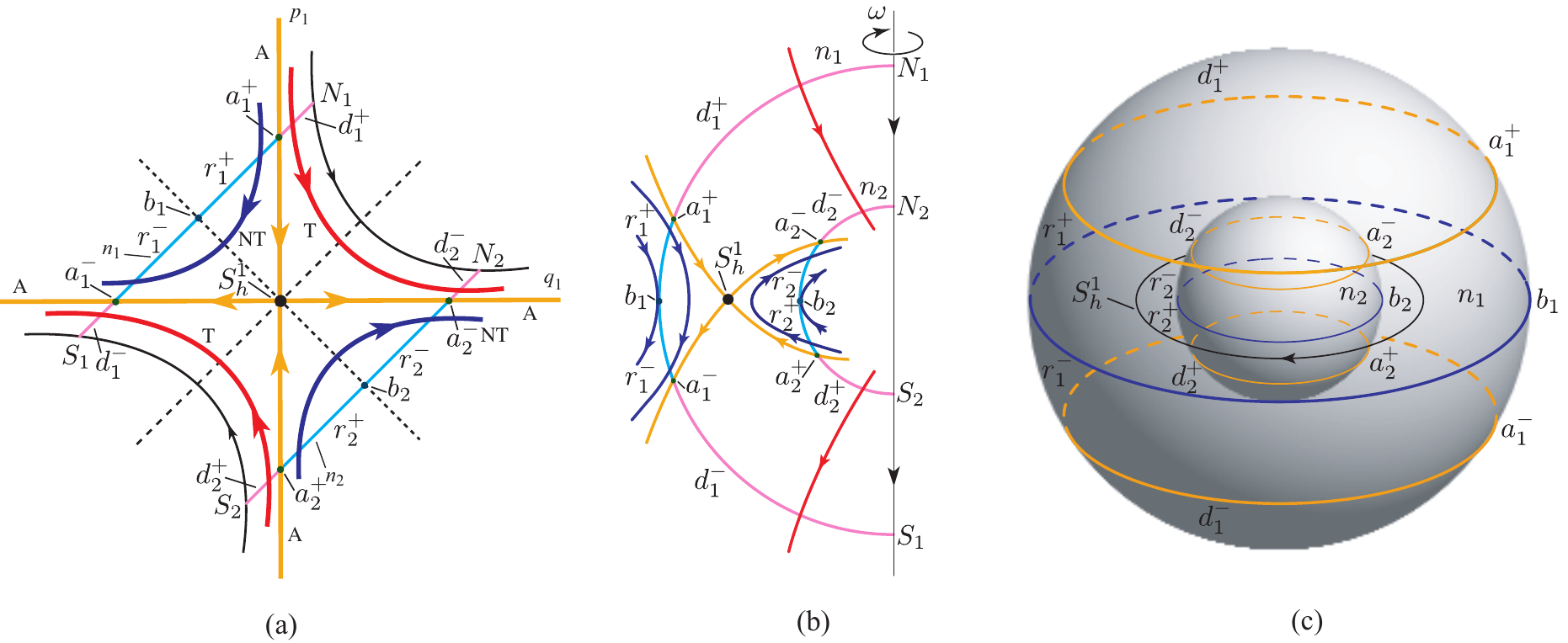}
	\end{center}
	\caption{{\footnotesize 
			(a) The projection onto the $(q_1,p_1)$-plane, the saddle projection, with labels consistent with the text and (b) and (c).
			(b) The cross-section of the flow in the
			$\mathcal{R}$ region of the energy surface. The north and south poles of bounding sphere $n_i$ are labeled as $N_i$ and $S_i$, respectively.
			(c) The McGehee representation of the flow on the boundaries of the $\mathcal{R}$ region, highlighting the features on the bounding spheres $n_1$ and $n_2$ for $h>0$.
	}}
	\label{conservative McGehee}
\end{figure}

McGehee \cite{McGehee1969}, building on the work of Conley  \cite{Conley1968}, proposed a
representation which makes it easier to visualize the region $\mathcal{R}$.
Recall that $\mathcal{R}$ is a 3-dimensional manifold that is homeomorphic to $S^2\times I$. In \cite{McGehee1969},
it is represented by a spherical annulus  bounded by the 
two 2-spheres $n_1, n_2$, as shown in Figure \ref{conservative McGehee}(c).

Figure \ref{conservative McGehee}(a) is a cross-section of $\mathcal{R}$. Notice that this
cross-section is qualitatively the same as the saddle projection illustration in Figure
\ref{conservative eigen flow}.  
The full picture (Figure \ref{conservative McGehee}(c)) is obtained by rotating
this cross section, Figure \ref{conservative McGehee}(b), about the indicated axis, where the azimuthal angle $\omega$ roughly describes the angle in the center projection in Figure \ref{conservative eigen flow}.
The following
classifications of orbits correspond to the previous four categories:
\begin{enumerate}
	\item  There is an invariant
	1-sphere $S^1_h$, a {\it periodic orbit} 
	in the region $\mathcal{R}$ corresponding to the black dot in the
	middle of Figure \ref{conservative McGehee}(a).
	Notice that this 1-sphere is the equator of the
	central 2-sphere given by $p_1 -q_1 =0$. 
	
	\item Again let $n_1,n_2$ be the bounding 2-spheres
	of region $\mathcal{R}$, and let $n$ denote either $n_1$ or $n_2$.
	We can divide $n$ into two hemispheres: $n^+$, where the flow enters
	$\mathcal{R}$,
	and $n^-$, where the flow leaves $\mathcal{R}$.  
	There are four cylinders of orbits 
	asymptotic to the invariant 1-sphere $S^1_h$.
	They form the stable and unstable manifolds which are {\it asymptotic} to the invariant 1-sphere
	$S^1_h$.  
	Topologically, both invariant manifolds 
	look like 2-dimensional cylinders or ``tubes'' ($S^1\times {\mathbb R}$) inside
	a 3-dimensional energy manifold.
	The interior of the stable manifolds  
	$W^s_{\pm}(S^1_h)$ and unstable manifolds
	$W^u_{\pm}(S^1_h)$ can be given as follows 
	\begin{equation}\label{interior}
	\begin{split}
	{\rm int}(W^s_+(S^1_h))
	&=
	\{(q_1,p_1,q_2,p_2)\in \mathcal{R}\mid 
	\hspace{.1in} p_1>q_1>0\},  \\
	{\rm int}(W^s_-(S^1_h))
	&=
	\{(q_1,p_1,q_2,p_2)\in \mathcal{R} \mid 
	\hspace{.1in} p_1<q_1<0\},  \\
	{\rm int}(W^u_+(S^1_h))
	&=
	\{(q_1,p_1,q_2,p_2)\in \mathcal{R}\mid 
	\hspace{.1in} q_1>p_1>0\},  \\
	{\rm int}(W^u_-(S^1_h))
	&=
	\{(q_1,p_1,q_2,p_2)\in \mathcal{R}\mid 
	\hspace{.1in} q_1<p_1<0\}.
	\end{split}
	\end{equation}
	The exterior of these invariant manifolds can be given similarly from
	studying Figure \ref{conservative McGehee}(a) and (b).
	
	\item
	Let $a^+$ and $a^-$
	(where $q_1 =0$ and $p_1 =0$ respectively) be the
	intersections of the stable and unstable manifolds with 
	the bounding sphere $n$.
	Then $a^+$ appears as a 1-sphere in $n^+$, and $a^-$ appears as a 
	1-sphere in $n^-$.
	Consider the two spherical caps on each bounding 
	2-sphere given by
	\begin{align*}
	d_1^+&=\{(q_1,p_1,q_2,p_2)\in \mathcal{R}
	\mid \hspace{.1in}p_1-q_1 =+c, 
	\hspace{.1in}p_1>q_1 >0\},\\
	d_1^-&=\{(q_1,p_1,q_2,p_2)\in \mathcal{R}
	\mid \hspace{.1in}p_1 -q_1 =+c,
	\hspace{.1in}q_1<p_1<0\},\\
	d_2^+ &= \{(q_1,p_1,q_2,p_2)\in \mathcal{R}\mid 
	\hspace{.1in}p_1 -q_1 =-c, \hspace{.1in}
	p_1<q_1 <0\}, \\
	d_2^-&=\{(q_1,p_1,q_2,p_2)\in \mathcal{R}
	\mid \hspace{.1in}p_1 -q_1 =-c,
	\hspace{.1in} q_1 >p_1>0\}.
	\end{align*}
	Since $d_1^+$ is  the spherical cap 
	in $n_1^+$ bounded by $a_1^+$, then
	the {\it transit} orbits
	entering $\mathcal{R}$ on $d_1^+$ exit on $d_2^-$ of the other bounding sphere.
	Similarly, since 
	$d_1^-$ is  the spherical cap in $n_1^-$ bounded by $a_1^-$, the transit
	orbits leaving on $d_1^-$ have
	come from $d_2^+$ on the other bounding sphere.
	Note that all spherical caps where the transit orbits pass through are
	in the interior of stable and unstable manifold tubes. 
	
	\item Let $b$ be the intersection
	of $n^+$ and $n^-$ (where $q_1 +p_1 =0$).  Then, $b$
	is a 1-sphere of tangency
	points.  Orbits tangent at this 1-sphere ``bounce off,'' i.e., do not
	enter
	$\mathcal{R}$ locally.  Moreover, if we let $r^+$ be a spherical zone which
	is bounded by $a^+$ and $b$, then {\it non-transit} orbits
	entering $\mathcal{R}$ on $r^+$ 
	exit on the same bounding 2-sphere through $r^-$ 
	which is bounded by
	$a^-$ and $b$.  It is easy to show that all the spherical zones where  
	non-transit orbits bounce off are in the exterior of stable and unstable
	manifold tubes.
\end{enumerate}

The McGehee representation provides an additional, perhaps clearer, visualization of the dynamics in the equilibrium region. In particular, the features on the two spheres, $n_1$ and $n_2$, which form $\partial \mathcal{R}$ for a constant $h>0$, can be considered in the dissipative case as well, and compared with the situation in the conservative case, as shown for some examples below.  The spheres $n_1$ and $n_2$ can be considered as spherical Poincar\'e sections parametrized by their distance from the saddle point, $c$, which reveal the topology of the transition region boundary, $\partial \mathcal{T}_h$, particularly through how the geometry of $a_i^+$ and $a_i^-$ (for $i=1,2$) change as $c$ changes.

\section{Uncoupled systems in the dissipative case}
\label{uncoupling systems}
As pointed out in the introduction, when applying the symplectic change of variables consisting of the generalized eigenvectors of the conservative system to the dissipative system, the saddle projection and focus projection are coupled in some systems, while in others systems they are not. According to the coupling conditions, the systems are classified into two categories: uncoupled systems and coupled systems. In this section, we will discuss the uncoupled systems first.

\subsection{Ball rolling on a stationary surface}
\label{Ball rolling}
Among the examples of escape from potential wells, a small ball or particle moving in an idealized fashion on a surface is an easy one from the perspective of both theory and experiment. The tracking of the moving object is easily executed by using a high-speed digital camera which is much easier than measurements of structural snap-through or ship motion, not to mention the motion of spacecraft in space. It can be either an inertial system or a gyroscopic system depending on whether the surface is stationary or rotating, due to a turntable, for instance \cite{lewis1995variational}. The easy switch between non-gyroscopic system and gyroscopic system makes it easy to compare their similarities and differences in escape from potential wells. The mathematical model of a rolling ball on a stationary surface was established in \cite{virgin2010nonlinear}.  Experiments \cite{Ross2018experimental,Xu2019}  regarding escape from the potential wells on similar surfaces were shown to validate the theory of the phase space conduits   predicted by the  mathematical model, which mediate the transitions between wells in the system. The dissipation of energy cannot be avoided in any physical experiment, but over small enough time-scales of interest, \cite{Ross2018experimental} justified that dissipation could be ignored. The good agreement between the theory and experiment to within $1\%$ indicates the robustness of the transition tube in the conservative systems. However, it is still not clear how dissipation affects the transition of a rolling ball on a surface and what the phase space structure controlling the transition in the corresponding dissipative system is. In the current example, we will present the answers.

\subsubsection{Governing equations}
\begin{figure}
	\begin{center}
		\includegraphics[width=4in]{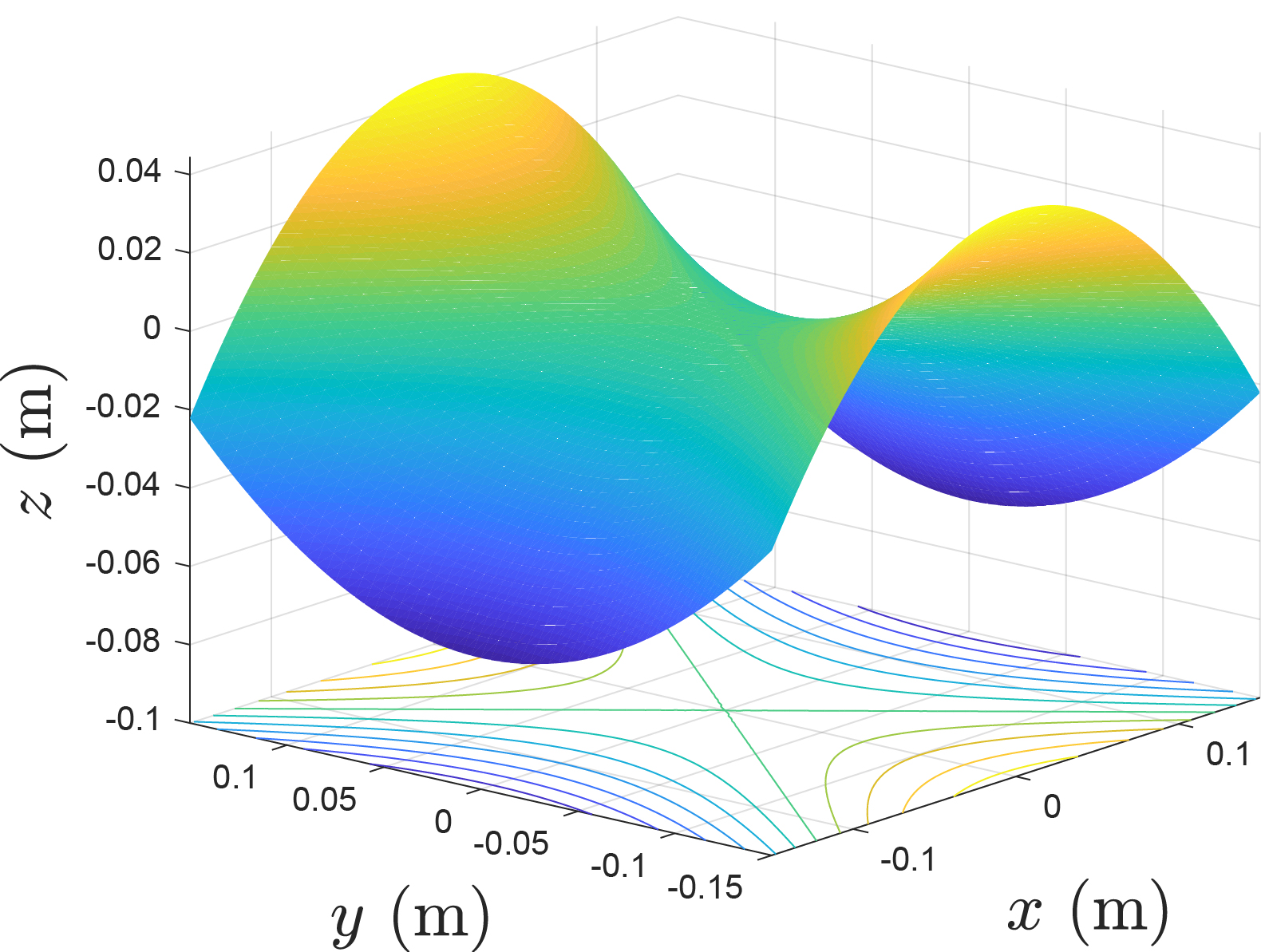}
	\end{center}
	\caption{{\footnotesize 
			The graph of the example saddle surface considered, based on \eqref{surface equation}. The contours of the surface are projected on the bottom plane.  The $z$ direction is shown scaled by a factor of  2 compared to $x$ and $y$ in order to highlight the saddle nature of the surface.
	}}
	\label{B_surface}
\end{figure}

Here we consider a ball with unit mass rolling on a surface without slipping. Before analyzing the dynamical behavior of the rolling ball, a Cartesian coordinate system $o\mbox{-}xyz$ with $z$ oriented upward is established. Thus, the equations of the surface can be determined by $z=H(x,y)$. In the current study, a saddle surface of the following form is selected,
\begin{equation}
H(x,y)=\tfrac{1}{2} \left(k_1 x^2 + k_2 y^2 \right), \hspace{0.3 in}k_1=-5.91~\text{m}^{-1}, k_2=3.94~\text{m}^{-1},
\label{surface equation}
\end{equation}
which is shown in Figure \ref{B_surface}.

Before analyzing the dynamical behavior of the system, one needs to obtain the equations of motion. To do so, one can use either the Lagrangian approach or Hamiltonian approach \cite{greenwood2006advanced}. In the Lagrangian approach, the kinetic energy and potential energy are needed to get the Lagrangian function which will yield the Euler-Lagrangian equations. In the Hamiltonian approach, the generalized momenta should be defined by introducing a Legendre transformation from the Lagrangian and then the Hamiltonian function can be given which will generate the Hamilton's equations. In Section \ref{rotating surface}, we will consider a more complicated system where the surface is not stationary, but it is rotating with a constant angular velocity $\omega$ where the gyroscopic force is included. Since the stationary surface is just a special case  of the 
 rotating surface where one takes angular velocity as zero, we do not separately derive the governing equations for the stationary surface and rotating surface. The derivation of the equations of motion will be briefly described for the current problem and readers can refer to Section \ref{rotating surface} for more details.

From the analysis in Section \ref{rotating surface}, one can set the angular velocity of the rotating surface as zero to obtain the kinetic energy (the translational plus rotational without slipping), $\mathcal{K}= \frac{1}{2} I \left(\dot x^2 + \dot y^2 + \dot z^2\right)$, and potential energy, $\mathcal{U}=gz$, where $g=9.81 \mathrm{m/s^2}$ is the gravitational acceleration and $z$ and $\dot z$ are written in terms of $x$, $y$, $\dot x$ and $\dot y$ via the relationship $z=H(x,y)$. The factor $I=7/5$ is introduced by including rotational kinetic energy for a ball rolling without slipping. See details in the supplemental material in \cite{Ross2018experimental}. If we consider a particle sliding on the surface, we have $I=1$.
The kinetic energy $\mathcal{K}$ and potential energy $\mathcal{U}$ are,
\begin{equation}
\begin{split}
\mathcal{K}(x,y)&= \tfrac{1}{2} I \left[ \dot x ^2 + \dot y ^2 + \left(H_{,x} \dot x + H_{,y} \dot y \right) ^2 \right],\\
\mathcal{U} (x,y)&= g H(x,y).
\label{B:energy}
\end{split}
\end{equation}
where $H_{,x} = \partial H/\partial x$ and $H_{,y} = \partial H/\partial y$.
Thus, one can define the Lagrangian function by,
\begin{equation}
\mathcal{L} \left(x,y \right)= \mathcal{K} \left(x,y,\dot x,\dot y \right) - \mathcal{U} (x,y)
\label{Lagrangian},
\end{equation}
which generates the Euler-Lagrange equations,
\begin{equation}
\frac{\mathrm{d}}{\mathrm{d}t} \left(\frac{\partial \mathcal{L}}{\partial \dot q_i} \right) - \frac{\partial \mathcal{L}}{\partial q_i}= Q_i
\label{general Lagrange equations},
\end{equation}
where $q_i$ are the generalized coordinates $(x,y)$ and $Q_i$ are the non-conservative forces.
In the current problem, a small linear viscous damping, proportional to the magnitude of the inertial velocity, is considered, with the form given via a Rayleigh dissipation function as,
\begin{equation}
\begin{split}
Q_x &= -c_d \left[ \left(1 + H_{,x}^2 \right) \dot x + H_{,x} H_{,y} \dot y \right],\\
Q_y &= -c_d \left[ \left(1 + H_{,y}^2 \right) \dot y + H_{,x} H_{,y} \dot x \right],
\end{split}
\end{equation}
where $c_d$ is the coefficient of damping.
The equations of motion for the current problem are,
\begin{equation}
\begin{aligned}
& I \left(1 + k_1^2 x^2 \right) \ddot x + I k_1 k_2 x y \ddot y + I k_1^2 x \dot x^2 + I k_1 k_2 x \dot y^2 + g k_1 x + c_d \left[ \left(1+k_1^2 x^2 \right) \dot x + k_1 k_2 x y \dot y\right]=0,\\
& I k_1 k_2 x y \ddot x + I \left(1 + k_2^2 y^2 \right) \ddot y+ I k_1 k_2 y \dot x^2 + I k_2^2 y \dot y^2 + g k_2 y + c_d \left[ \left(1+k_2^2 y^2 \right) \dot y + k_1 k_2 x y \dot x\right]=0.
\end{aligned}
\end{equation}

Once the Lagrangian system is established, one can transform it to a Hamiltonian system by use of the Legendre transformation,
\begin{equation}
p_i=\frac{\partial \mathcal{L}}{\partial \dot q_i}, \hspace{0.5in} \mathcal{H} \left(q_i,p_i \right)= \sum_{i=1}^n p_i \dot q_i - \mathcal{L} \left( q_i ,p_i \right),
\label{general Legendre stransformation}
\end{equation}
where $p_i$ are called the generalized momenta conjugate to the generalized coordinates $q_i$ and $\mathcal{H}$ the Hamiltonian function. In the current case, the Legendre transformation is given by,
\begin{equation}
\begin{split}
p_x &= \frac{\partial \mathcal{L}}{\partial \dot x}= \dot x - y \omega + H_{,x}^2 \dot x + H_{,x} H_{,y} \dot y, \\
p_y &=\frac{\partial \mathcal{L}}{\partial \dot y}= \dot y + x \omega + H_{,x} H_{,y} \dot x + H_{,y}^2 \dot y.
\end{split}
\end{equation}
Therefore, one obtains the Hamiltonian function,
\begin{equation}
\begin{split}
\mathcal{H} = \frac{\left[ p_x^2 \left(1+ H_{,y}^2\right) - 2 p_x p_y H_{,x} H_{,y} +  p_y^2 \left(1+ H_{,x}^2 \right) \right]}{2 I \left(1+ H_{,x}^2 + H_{,y}^2 \right)}  + g H,
\end{split}
\end{equation}
where $p_x$ and $p_y$ are the momenta conjugate to $x$ and $y$, respectively. The comma in the subscript means the partial derivative with respect to the following coordinate. The general form of the Hamilton's equations with damping \cite{greenwood2006advanced} are given by,
\begin{equation}
\dot q_i=\frac{\partial \mathcal{H}}{\partial p_i}, \hspace{0.5in} \dot p_i = - \frac{\partial \mathcal{H}}{\partial q_i} + Q_i.
\label{general Hamilton equations}
\end{equation}
where $Q_i$ is the same non-conservative generalized force written in terms of $(q,p)$ variables. For simplicity, the specific form of Hamilton's equations for the current problem are not listed here.

For the surface adopted in \eqref{surface equation}, it has a saddle type equilibrium point at the origin $\left(0,0\right)$. To study the transition from one side of the bottleneck to the other, the local dynamical behavior near the equilibrium point plays a critical role. Thus, we will obtain the linearized Hamiltonian equations around the equilibrium point to study the local properties. A short computation for \eqref{general Hamilton equations} gives the linearized equations of motion in Hamiltonian form as,
\begin{equation}
\begin{split}
\dot x &=  p_x/I,\\
\dot y &=  p_y/I,\\
\dot p_x & = -g k_1 x - c_d p_x/I ,\\
\dot p_y & = -g k_2 y - c_d p_y/I.
\end{split}
\end{equation}

We introduce the following re-scaled parameters,
\begin{equation}
\begin{split}
\left(\bar q_1, \bar q_2 \right)= \left(x,y \right),  \left(\bar p_1, \bar p_2 \right)= \left(p_x, p_y \right)/I, \left(c_x, c_y \right)=- g\left( k_1, k_2 \right)/I, c_h= c_d/I,
\end{split}
\label{B: nondimensional parameters}
\end{equation}
and the equations of motion can be rewritten in the simpler re-scaled form,
\begin{equation}
\begin{split}
\dot{\bar{q}}_1  &=        \bar{p}_1,\\
\dot{\bar{q}}_2 & =        \bar{p}_2,\\
\dot{\bar{p}}_1 & = c_x \bar{q}_1  - c_h \bar{p}_1,\\
\dot{\bar{p}}_2 & = c_y \bar{q}_2  - c_h \bar{p}_2.
\label{B:linearized nondimen eqns}
\end{split}
\end{equation}
Written in matrix form, with column vector $\bar z = \left( \bar q_1, \bar q_2, \bar p_1, \bar p_2 \right)^T$, we have $\dot{\bar{z}}=A \bar z$, where $A = M + D$, i.e.,
\begin{equation}
\dot{\bar{z}}= M \bar z + D \bar z,
\end{equation}
where,
\begin{equation}
M=
\begin{pmatrix}
0 & 0 & 1 & 0\\
0 & 0 & 0 & 1\\
c_x & 0 & 0 & 0\\
0 & c_y & 0 & 0
\end{pmatrix}
,\hspace{0.1in}
D= c_h \begin{pmatrix}
0 & 0 & 0 & 0\\
0 & 0 & 0 & 0\\
0 & 0 & -1 & 0\\
0 & 0 & 0 & -1
\end{pmatrix}.
\end{equation}
The corresponding quadratic  Hamiltonian for the linearized system is,
\begin{equation}
\mathcal{H}_2(\bar q_1, \bar q_2, \bar p_1, \bar p_2)=\tfrac{1}{2} \left( \bar p_1^2 + \bar p_2^2 \right) - \tfrac{1}{2}  \left(c_x \bar q_1^2 + c_y \bar q_2^2 \right).
\label{B:rescale Hamiltonian}
\end{equation}

\subsubsection{Analysis in the conservative system}
First we analyze the behavior in the conservative system which can be obtained by taking zero damping, $c_h=0$. 
It is straightforward to obtain the eigenvalues of the conservative system which are of the form $\pm \lambda$ and $\pm i \omega_p$ as expected, since the linearization matrix $A=M$ is an infinitesimal symplectic matrix (also known as a Hamiltonian matrix) \cite{marsden2013introduction,meiss2007differential} where $\lambda$ and $\omega_p$ are positive constants given by $\lambda=\sqrt{c_x}$ and $\omega_p=\sqrt{- c_y}$. The corresponding eigenvectors are defined as $u_{\pm \lambda} $ and $ u_{\omega_p} \pm i v_{\omega_p}$, where $u_{\pm \lambda}$, $u_{\omega_p}$, and $v_{\omega_p}$ are real vectors with the following form,
\begin{equation}
\begin{split}
u_{+ \lambda} &= \left(\lambda^2 - c_y, 0, \lambda^3 - \lambda c_y, 0 \right),\\
u_{- \lambda} &= \left(-\lambda^2 + c_y , 0, \lambda^3 - \lambda c_y, 0 \right),\\
u_{\omega_p} &= \left(0, \omega_p^2 + c_x, 0, 0 \right),\\
v_{\omega_p} &= \left(0, 0, 0, \omega_p^3 + \omega_p c_x \right).
\end{split}
\end{equation}

Considering the change of variables defined by,
\begin{equation}
\bar z=C z,
\label{change of variables}
\end{equation}
where $\bar z = \left(\bar q_1, \bar q_2, \bar p_1, \bar p_2 \right)^T$ and $z=\left(q_1, q_2, p_1, p_2\right)^T$,
with $C= \left(u_{\lambda}, u_{\omega_p}, u_{-\lambda}, v_{\omega_p} \right)$ where $u_{\lambda}$, etc, are understood as column vectors, one can find,
\begin{equation*}
C^T J C =
\begin{pmatrix}
0 & \bar D \\- \bar D & 0
\end{pmatrix}
, \hspace{0.2in}
\bar{D}=\begin{pmatrix}
d_{\lambda} & 0\\
0 & d_{\omega_p}
\end{pmatrix},
\end{equation*}
where,
\begin{equation*}
\begin{split}
d_{\lambda} &=2 \lambda \left[(c_x - c_y) \lambda^2 - c_x c_y + c_y^2 \right],\\
d_{\omega_p} &=\omega_p \left[(c_x - c_y) \omega_p^2 + c_x^2 - c_x c_y  \right],
\end{split}
\end{equation*}
and $J$ is the $4 \times 4$ canonical symplectic matrix,
\begin{equation}
J=\begin{pmatrix}
0 & I_2\\
-I_2 & 0
\end{pmatrix},
\end{equation}
where $I_2$ is the $2 \times 2$ identity matrix.

We can introduce two factors $s_1=\sqrt{d_{\lambda}}$ and $s_2=\sqrt{d_{\omega_p}}$ to the columns in $C$ which makes it a symplectic matrix, i.e., satisfying $C^T J C=J$. The final form of the symplectic matrix is,
\begin{equation}
C=\begin{pmatrix}
\tfrac{\lambda^2 - c_y }{s_1} & 0 & \tfrac{-\lambda^2 + c_y }{s_1} & 0\\
0 & \tfrac{\omega_p^2 + c_x }{s_2} & 0 & 0\\
\tfrac{\lambda^3 - \lambda c_y}{s_1} & 0 & \tfrac{  \lambda^3 - \lambda c_y}{s_1} & 0\\
0 & 0 & 0 &  \tfrac{\omega_p^3 + \omega_p c_x}{s_2}
\end{pmatrix}.
\label{B: symp tranform}
\end{equation}
The equations of motion in the {\bf symplectic eigenspace} (i.e., the $z$ variables) can be obtained as,
\begin{equation}
\dot z = \Lambda z,
\label{B:EOM with no damping in phase space}
\end{equation}
where $\Lambda=C^{-1}MC$ is the conservative part of the dynamics,
\begin{equation}
\Lambda = 
\begin{pmatrix}
\lambda & 0 & 0 & 0\\
0 & 0 & 0 & \omega_p\\
0 & 0 & -\lambda & 0\\
0 & -\omega_p & 0 & 0
\end{pmatrix}. \label{Lambda_standard}
\end{equation}
Thus, via the transformation \eqref{change of variables}, the equations of motion in the conservative system can be rewritten in a normal form given in \eqref{phase-space Hamiltonian Equations} with Hamiltonian \eqref{phase-space Hamiltonian} whose solutions are given by \eqref{phase-space solutions}.

\paragraph{Behavior in the position space} 
Recalling the solutions in \eqref{phase-space solutions} and the symplectic matrix $C$ in \eqref{B: symp tranform}, we  obtain the general (real) solutions of the conservative system in phase space in the form,
\begin{equation}
\begin{aligned}
\bar z(t) &= \left(\bar q_1, \bar q_2, \bar p_1, \bar p_2 \right)^T\\
&= q_1^0 e^{\lambda t} u_{+\lambda} + p_1^0 e^{-\lambda t} u _{-\lambda} + \mathrm{Re} \left[\beta_0 e^{-i \omega_p t} \left(u_{\omega_p} - i v_{\omega_p} \right) \right],
\label{conser_general_sol}
\end{aligned}
\end{equation}
where $q_1^0$, $p_1^0$, $q_2^0$, $p_2^0$ are real and determined by initial conditions, where $\beta_0=q_2^0+i p_2^0$. In particular, we have,
\begin{equation}
\begin{aligned}
\bar q_1(t)&= \frac{\lambda^2 - c_y }{s_1} q_1^0 e^{\lambda t} - \frac{\lambda^2 - c_y }{s_1} p_1^0 e^{-\lambda t},\\
\bar q_2(t)&= \frac{\omega_p^2 + c_x }{s_2} \left(q_2^0 \cos \omega_p t + p_2^0 \sin \omega_p t \right).
\end{aligned}
\end{equation}
Notice that all trajectories in the configuration space in $\mathcal{R}$ must evolve within the energy manifold which is bounded by the \textbf{zero velocity curve} (corresponding to $\bar p_1=\bar p_2=0$) \cite{zhong2018tube,KoLoMaRo2000,zotos2014escapes,zotos2017overview} given by solving \eqref{B:rescale Hamiltonian} as,
\begin{equation}
\bar q_2=\pm \sqrt{\frac{-2 h-c_x \bar q_1^2}{c_y}}.
\end{equation}

\begin{figure}[!t]
	\begin{center}
		\includegraphics[width=4in]{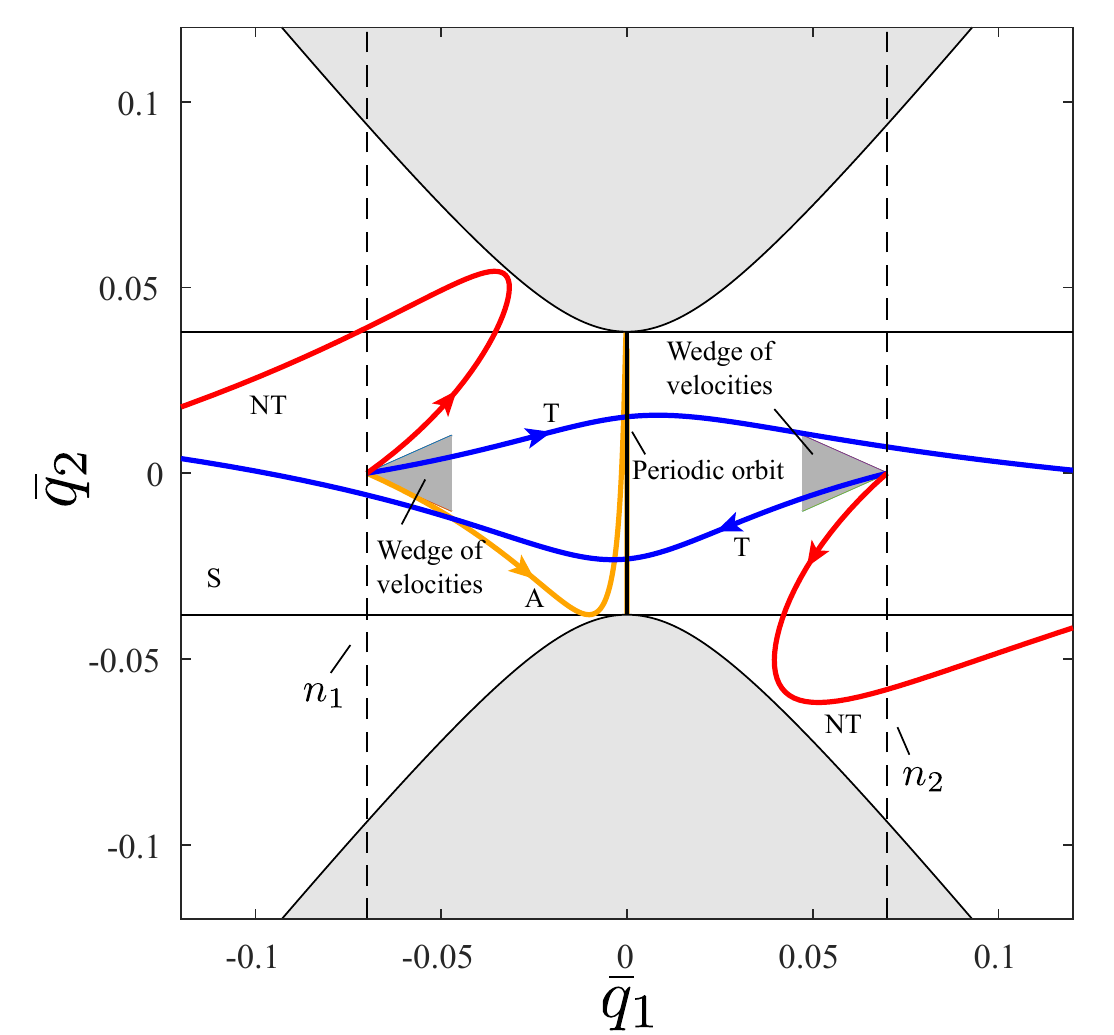}
	\end{center}
	\caption{{\footnotesize 
			The flow in the equilibrium region $\mathcal{R}$ projected onto position space $(\bar q_{1},\bar q_2)$ in the conservative system with fixed positive energy, $\mathcal{H}_2=h>0$, for a ball rolling on a stationary surface. Shown are the unstable periodic orbit (vertical segment in the center), a typical asymptotic orbit winding onto the the periodic orbit; two transit orbits (blue); and two non-transit orbits (red). At each point on the bounding lines $n_1$ or $n_2$ (dashed) inside the strip $S$, there is a wedge of velocity dividing different types of orbits, inside of which are transit orbits, and outside of which are non-transit orbits; specifically, the trajectories with initial conditions on the boundary are the orbits asymptotic to the unstable periodic orbit. See the text for the explanation of the details.
	}}
	\label{B: conserfative confguration flow}
\end{figure}

By examining the general solution, we can see the solutions on the energy surface fall into different classes depending upon the limiting behavior of $\bar q_1$ as $t$ goes to plus or minus infinity according to the fact that $\bar q_1(t)$ is dominated by the $q_1^0$ and $p_1^0$ terms when $t\rightarrow + \infty$ and $t\rightarrow - \infty$, respectively. Thus, the nine classes of orbits determined by varying the signs of $q_1^0$ and $p_1^0$ are classified into four categories.
\begin{enumerate}
	\item If $q_1^0=p_1^0=0$, we obtain a periodic solution with energy $h$. The periodic orbit, $S^1_h$, projects onto the $\left(\bar q_1, \bar q_2 \right)$ plane as a segment with length $\sqrt{-2h/c_y}$.
	
	\item Orbits with $q_1^0 p_1^0=0$ are asymptotic orbits. They are asymptotic to the periodic orbit, which is the origin, labeled $S_h^1$ in Figure \ref{conservative eigen flow}. Asymptotic orbits with either $q_1^0=0$ or $p_1^0=0$ project into a strip $S$, as shown in Figure \ref{B: conserfative confguration flow}, bounded by lines,
	\begin{equation}
	\bar q_2 = \pm \frac{\omega_p^2 + c_x}{s_2} \sqrt{\frac{2h}{\omega_p}}.
	\end{equation}
	
	\item Orbits with $q_1^0 p_1^0 >0$ are transit orbits because they cross the equilibrium region $\mathcal{R}$ from $- \infty$ (the left-hand side) to $+ \infty$ (the right-hand side) or vice versa.
	
	\item Orbits with $q_1^0 p_1^0 <0$ are non-transit orbits. 
\end{enumerate}

Figure \ref{B: conserfative confguration flow} gives the four categories of orbits mentioned above. In the figure, $S$ is the strip confining the asymptotic orbits. Outside of the strip, the situation is simple and only non-transit orbits exist which means the signs of $q_1^0$ and $p_1^0$ are independent of the direction of the velocity and we always have $q_1^0 p_1^0<0$. The signs in each component of the equilibrium region $\mathcal{R}$ complementary to the strip can be determined by limiting behavior of $\bar q_1$ for positive and negative infinite time. For example, in the left two components the non-transit orbits stay on the left side for $t\rightarrow \pm \infty$ which indicates $q_1^0<0$ and $p_1^0>0$. Similarly, in the right two components are $q_1^0>0$ and $p_1^0<0$. As one can determine  from the discussions in the phase space of the equilibrium region, the asymptotic orbits are the stable and unstable manifolds of a periodic orbit, which acts as a separatrix, the boundary of transition orbits and non-transit orbits. Denoting $\left(\bar q_{10}, \bar q_{20}, \bar p_{10}, \bar p_{20} \right)$ as the initial conditions in phase space, the Hamiltonian function for asymptotic orbits in the phase space for the conservative system can be rewritten using the initial conditions as,
\begin{equation}
\frac{\bar q_{20}^2}{b_e^c} + \frac{\bar p_{20}^2}{c_e^2}=0,
\label{B: transition tube}
\end{equation}
where $b_e$ and $c_e$ can be found in \eqref{B:axis of ellipse}. The form of \eqref{B: transition tube} is a cylinder or tube which will be discussed later.

Inside the strip, the situation is more complicated because the signs of $q_1^0$ $p_1^0$ are no longer independent of the direction of velocity. At each position inside the strip, there is a {\bf wedge of velocity},  as proved in \cite{zhong2018tube,KoLoMaRo2000,ross2004cylindrical,KoLoMaRo2011,Conley1968}, separating the transit orbits and non-transit orbits whose two boundaries are given by the angles $\theta_\pm=\arctan \left(\bar p_{20\pm}/ \bar p_{10} \right)$ with respect to the $\bar q_1$-axis, where,
\begin{equation}
\bar p_{10}=- \bar q_{10} \sqrt{c_x}, \hspace{0.2in} \bar p_{20\pm}= \pm \sqrt{2 h + c_y \bar q_{20}^2},
\end{equation}
See the shaded wedges in Figure \ref{B: conserfative confguration flow}. Here, the derivations are ignored for simplicity (they can be found in the analysis for the dissipative system in \cite{zhong2018tube}). As a visualization and example,  wedges on the two vertical bounding line segments are given. For example, consider the intersection of strip $S$ with the left-most vertical line, $n_1$. On this subsegment,
  there exists  a non-empty wedge of velocity at each position. Orbits with their velocity inside the wedge are transit orbits $\left(q_1^0 p_1^0>0 \right)$, while orbits with velocity outside of the wedge are non-transit $\left(q_1^0 p_1^0<0 \right)$. Orbits with their velocity on the boundary of the wedge are asymptotic $\left(q_1^0 p_1^0=0 \right)$. The situation on the right-hand side subsegment is similar. Notice that the magnitude of the wedge depends on the initial positions $\left(\bar q_{10}, \bar q_{20} \right)$. On the boundary of the strip, only one result of $\bar p_{20\pm}$ exists which indicates the wedge becomes a line along the boundary.

\subsubsection{Analysis in the dissipative system}
\label{Dissipative system in eigenspace}
For the dissipative system, we still use the symplectic matrix $C$ in \eqref{B: symp tranform} to perform a transformation, via \eqref{change of variables}, to the symplectic eigenspace, even though this is no longer the true eigenspace of the dissipative linearization matrix $A=M+D$. The equations of motion in the symplectic eigenspace  are, 
\begin{equation}
\dot z = \Lambda z + \Delta z,
\label{B:EOM with damping in phase space}
\end{equation}
where $\Lambda=C^{-1}MC$ is the conservative part of the dynamics, as before, and the transformed damping matrix is,
\begin{equation}
\Delta= C^{-1}DC= -c_h
\begin{pmatrix}
\tfrac{1}{2} & 0 & \tfrac{1}{2} & 0\\
0 & 0 & 0 & 0\\
\tfrac{1}{2} & 0 & \tfrac{1}{2} & 0\\
0 & 0 & 0 & 1
\end{pmatrix}.\label{standard_damping_matrix}
\end{equation}

To analyze the behavior in the dissipative eigenspace (as opposed to the symplectic eigenspace), the eigenvalues and eigenvectors, $\beta_i$ and $u_{\beta_i}$, respectively, $(i=1,...,4)$, are,
\begin{equation}
\begin{aligned}
\beta_{1,2} &=-\delta \mp \tfrac{1}{2}\sqrt{c_h^2 + 4 \lambda^2}, 
\hspace{0.2in} && u_{\beta_{1,2}}=\left(\delta,0, \lambda \pm \tfrac{1}{2} \sqrt{c_h^2 + 4 \lambda^2},0 \right),\\
\beta_{3,4}& =-\delta \pm i \omega_d, \hspace{0.2in} &&u_{\beta_{3,4}}=\left(0, \omega_p, 0, - \delta \pm i \omega_d \right),
\end{aligned}
\end{equation}
where $\delta=\tfrac{1}{2}c_h$, $\omega_d=\omega_p \sqrt{1-\xi_d^2}$ and $\xi_d=\delta/\omega_p$.
Thus, the general (real) solutions are,
\begin{equation}
\begin{aligned}
& q_1(t) =k_1 e^{\beta_1 t} + k_2 e^{\beta_2 t}, 
\hspace{0.2in} 
p_1(t)=k_3 e^{\beta_1 t} + k_4 e^{\beta_2 t},\\
& q_2(t) = k_5  e^{- \delta t} \cos{\omega_d t} + k_6 e^{- \delta t} \sin{\omega_d t},\\
& p_2(t) = \frac{k_5 }{\omega_p} e^{- \delta t} \left(-\delta \cos{\omega_d t} - \omega_d \sin{\omega_d t} \right) +\frac{k_6 }{\omega_p}  e^{- \delta t} \left(\omega_d \cos{\omega_d t - \delta \sin{\omega_d t}} \right),\\
\end{aligned}
\end{equation}
where,
\begin{equation*}
\begin{aligned}
k_1 &= \frac{q_1^0 \left(2 \lambda + \sqrt{c_1^2 + 4 \lambda^2} \right)-c_1 p_1^0 }{2\sqrt{c_1^2 + 4 \lambda^2}}, \hspace{0.2in}&& k_2 =\frac{q_1^0 \left(-2 \lambda + \sqrt{c_1^2 + 4 \lambda^2} \right)+c_1 p_1^0 }{2\sqrt{c_1^2 + 4 \lambda^2}},\\
k_3 &= \frac{p_1^0 \left(-2 \lambda + \sqrt{c_1^2 + 4 \lambda^2} \right)-c_1 q_1^0 }{2\sqrt{c_1^2 + 4 \lambda^2}}, && k_4 = \frac{p_1^0 \left(2 \lambda + \sqrt{c_1^2 + 4 \lambda^2} \right)+c_1 q_1^0 }{2\sqrt{c_1^2 + 4 \lambda^2}},\\
k_5&=q^0_2 , \hspace{0.5in} k_6=\frac{p^0_2 \omega_p + q^0_2 \delta}{\omega_d}.
\end{aligned}
\end{equation*}
Taking the total derivative of the Hamiltonian with respective to time along trajectories and using \eqref{B:EOM with damping in phase space}, we have,
\begin{equation*}
\frac{\mathrm{d} \mathcal{H}_2}{\mathrm{d} t}= - \tfrac{1}{2} c_h \lambda \left(q_1 + p_1 \right)^2 - c_h \omega_p p_2^2 \le 0,
\end{equation*}
which means the Hamiltonian is generally decreasing (more precisely, non-increasing) due to damping. 

\paragraph{The linear flow in $\mathcal{R}$} 
Similar to the discussions in the conservative system, we still choose the same equilibrium region $\mathcal{R}$ to consider the projections on the $\left(q_1, p_1\right)$-plane and $\left(q_2, p_2\right)$-plane, respectively. Different from the saddle $\times$ center projections in the conservative system, here we see saddle $\times$ focus projections in the dissipative system. The stable focus is a damped oscillator with frequency of $\omega_d=\omega_p \sqrt{1-\xi_d^2}$. Different classes of orbits can also be grouped into the following four categories:
\begin{enumerate}
	\item The point $q_1=p_1=0$ corresponds to a  {\bf focus-type asymptotic} orbit with motion purely in the $(q_2,p_2)$-plane (see black dot at the origin of
	the $(q_1,p_1)$-plane in Figure \ref{B: dissipative eigen flow}).  
	\begin{figure}[!t]
		\begin{center}
			\includegraphics[width=\textwidth]{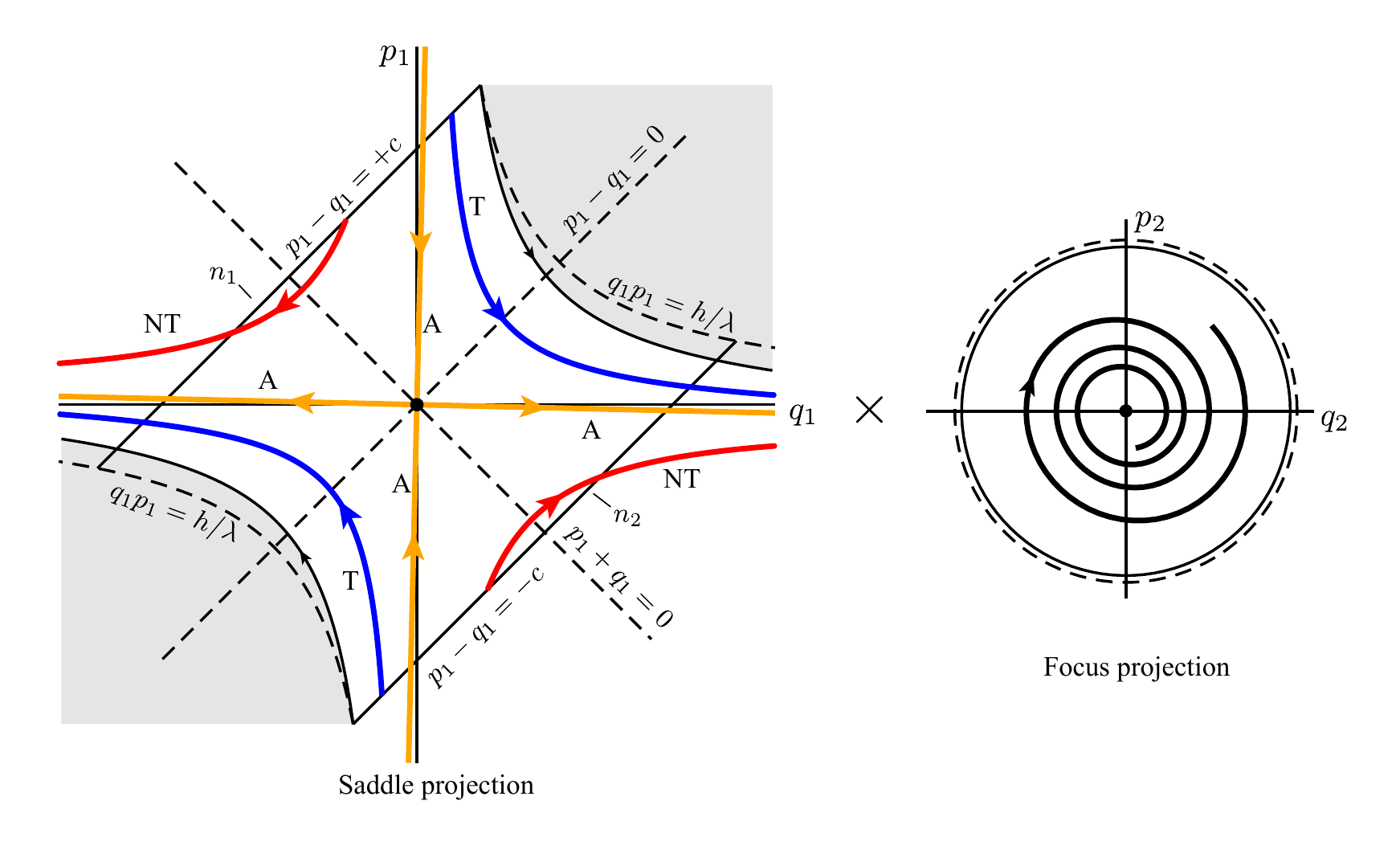}
		\end{center}
		\caption{{\footnotesize 
				The flow in the equilibrium region for the dissipative system has the form saddle $\times$ focus. On the left is shown the saddle projection onto the $(q_1,p_1)$-plane. The black dot at the origin represents focus-type asymptotic orbits with only a focus projection, thus oscillatory dynamics decaying towards the equilibrium point. The asymptotic orbits (labeled A) are the saddle-type asymptotic orbits which are tilted clockwise compared to the conservative system. They still form the separatrix between transit orbits (T) and non-transit orbits (NT). The hyperbolas, $q_1 p_1=h/\lambda$, are no longer the boundary of trajectories with initial conditions on the bounding sphere ($n_1$ or $n_2$) due to the dissipation of the energy. The boundary of the shaded region are still the fastest trajectories with initial conditions on the bounding sphere, but are not strictly hyperbolas. Note that the saddle projection and focus projection are uncoupled in this dissipative system.
		}}
		\label{B: dissipative eigen flow}
	\end{figure}
	Such orbits are asymptotic to the equilibrium point itself, rather than a periodic orbit of energy $h$ as in the conservative case.
	Due to the effect of damping, the periodic orbits on each energy manifold of energy $h$ do not exist.
	The 1-sphere $S_h^1$ still exists, but is no longer invariant. Instead, it corresponds to all the initial conditions of initial energy $h$ which are focus-type asymptotic orbits.  
	The projection of $S_h^1$ to the configuration space in the dissipative system is the same as the projection of the periodic orbit in the conservative system.
	
	\item The four half open segments on the lines governed by $q_1=  c_h p_1/(2 \lambda \pm \sqrt{c_1^2 + 4 \lambda^2}) $ correspond to {\bf saddle-type asymptotic} orbits.
	See the four orbits labeled A in Figure \ref{B: dissipative eigen flow}. 
	
	\item The segments which cross $\mathcal{R}$ from one boundary to the other, i.e., from $p_1 - q_1=+c$ to $p_1 - q_1=-c$ in the northern hemisphere, and vice versa in the southern hemisphere, correspond to {\it transit} orbits. See the two orbits labeled $T$ of Figure \ref{B: dissipative eigen flow}.
	
	\item  Finally the segments which run from one hemisphere to the other hemisphere on the same boundary, namely which start from $p_1 - q_1 = \pm c$ and return to the same boundary, correspond to {\it non-transit} orbits. See the two orbits labeled NT of Figure \ref{B: dissipative eigen flow}.
\end{enumerate}

As done in Section \ref{R: sec:separatrix}, we define the transition region, $\mathcal{T}_h$, as the region of initial conditions of a given initial energy $h$ which transit from one side of the neck region to the other.  
As before, the transition region, $\mathcal{T}_h$, is made up of one half which goes to the right,
$\mathcal{T}_{h+}$, 
and the other half which goes to the left, 
$\mathcal{T}_{h-}$. 
The boundaries are $\partial \mathcal{T}_{h+}$ and $\partial \mathcal{T}_{h-}$, respectively.
The closure of $\partial \mathcal{T}_{h}$, $\overline{\partial \mathcal{T}_{h}}$, is equal to the boundaries 
$\partial \mathcal{T}_{h+}$ and $\partial \mathcal{T}_{h-}$, along with the focus-type asymptotic initial conditions $S^1_h$, i.e., as before,
$\partial \mathcal{T}_{h-} \cup \partial \mathcal{T}_{h+} \cup S^1_h$.

As shown below, for the dissipative case, the closure of the boundary of the transition region, 
$\partial \mathcal{T}_h$, has the topology of an ellipsoid, rather than a cylinder as in the conservative case.  
As before, for convenience, we may refer to $\partial \mathcal{T}_{h}$ and $\overline{\partial \mathcal{T}_{h}}$ interchangeably.

\paragraph{McGehee representation}
Similar to the McGehee representation for the conservative system given in Section \ref{Conservative McGehee representaion} to visualize the region $\mathcal{R}$, here we utilize the McGehee representation again to illustrate the behavior in same region for the dissipative system. 
All labels are consistent throughout the paper.

\begin{figure}
	\begin{center}
		\includegraphics[width=\textwidth]{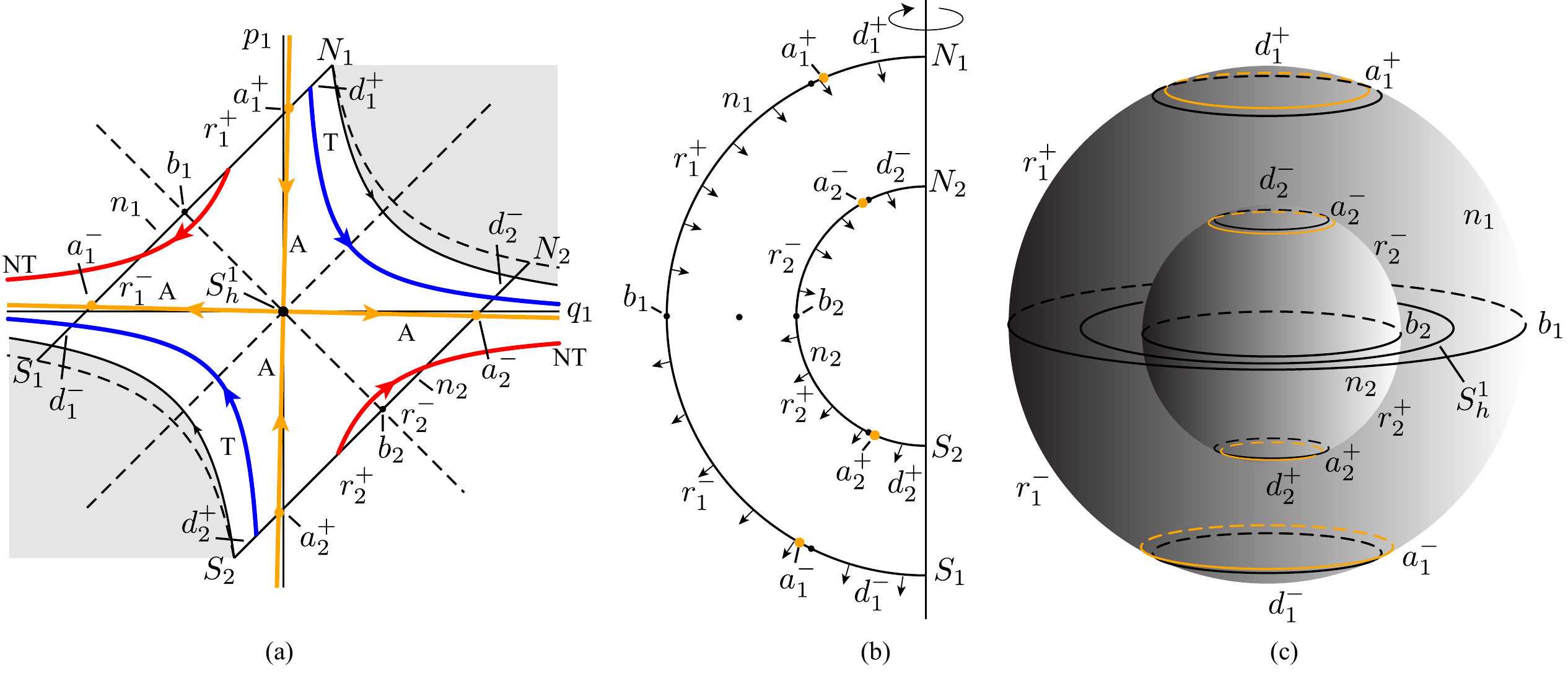}
	\end{center}
	\caption{{\footnotesize 
			(a) The projection onto the $(q_1,p_1)$-plane, the saddle projection, with labels consistent with the text and (b) and (c).
			(b) The cross-section of the flow in the
			$\mathcal{R}$ region of the energy surface. The north and south poles of bounding sphere $n_i$ are labeled as $N_i$ and $S_i$, respectively.
			(c) The McGehee representation of the
			flow in the region $\mathcal{R}$.
	}}
	\label{B: McGehee representation}
\end{figure}

Note that since the McGehee representation uses spheres with the same energy to show the dynamical behavior in phase space, while the energy of any particular trajectory in the dissipative system decreases gradually during evolution, Figures \ref{B: McGehee representation}(b) and \ref{B: McGehee representation}(c) show only the initial conditions at a given initial energy. Therefore, in the present McGehee representation, only the initial conditions on the two bounding spheres are shown and discussed in the next part. In addition, the black dot near the orange dots $a_i^{\pm}$ and $b_i^{\pm}$ ($i=1,2$) in Figure \ref{B: McGehee representation}(b) are the corresponding dots in the conservative system which are used to show how damping affects the transition. 

The following classifications of orbits correspond to the previous four categories:
\begin{enumerate}
	\item 1-sphere $S_h^1$ exists in the region $\mathcal{R}$ corresponding to the black dot in the middle of Figure \ref{B: McGehee representation}(b) and the equator of the central 2-sphere given by $p_1-q_1=0$ in \ref{B: McGehee representation}(c). The 1-sphere gives the initial conditions of the initial energy $h$ for all focus-type asymptotic orbits. The same 1-sphere in the conservative system is invariant under the flow, that is, a periodic orbit of constant energy $h$. However, the corresponding $S_h^1$ is not invariant in the dissipative system, since the energy is decreasing during evolution due to the damping.

	\item There are four 1-spheres in the region $\mathcal{R}$ starting in the bounding 2-spheres $n_1$ and $n_2$ which give the initial conditions for orbits asymptotic to the equilibrium point. Two of them in $n^+$, labeled by $a^+$, are stable saddle-type asymptotic orbits and the other two in $n^-$, labeled by $a^-$, are unstable  asymptotic orbits, where $a^+$ and $a^-$ are given by,
	\begin{equation}
	\begin{aligned}
	&a_1^+=\left\{(q_1,p_1,q_2,p_2)\in \mathcal{R} | \hspace{0.1in} (q_1,p_1)=(k_p,1)c/(1-k_p)\right\},\\
	&a_1^-=\left\{(q_1,p_1,q_2,p_2)\in \mathcal{R} | \hspace{0.1in} (q_1,p_1)=(-1,k_p)c/(1+k_p)\right\},\\
	&a_2^+=\left\{(q_1,p_1,q_2,p_2)\in \mathcal{R} | \hspace{0.1in} (q_1,p_1)=(k_p,1)c/(k_p-1)\right\},\\
	&a_2^-=\left\{(q_1,p_1,q_2,p_2)\in \mathcal{R} | \hspace{0.1in} (q_1,p_1)=(1,-k_p)c/(1+k_p)\right\},\\
	\end{aligned}
	\end{equation}
	where $k_p = c_h /(2 \lambda + \sqrt{c_h^2 + 4 \lambda^2} )$. As shown in Figure \ref{B: McGehee representation}(c), $a^+$ appears as an orange circle in $n^+$, and $a^-$ appears as an orange circle in $n^-$. The corresponding curves for the same energy in the conservative system are shown as black curves.
	
	\item Consider the two spherical caps on each bounding 2-sphere, $n_1$ and $n_2$, given by,
	\begin{equation}
	\begin{aligned}
	& d_1^+=\left\{(q_1,p_1,q_2,p_2) \in \mathcal{R} \mid  \hspace{0.1in}p_1-q_1=c,  \hspace{0.1in} q_1>ck_p/(1-k_p)\right\},\\
	& d_1^-=\left\{(q_1,p_1,q_2,p_2) \in \mathcal{R} \mid  \hspace{0.1in}p_1-q_1=c,  \hspace{0.1in} q_1<-c/(1+k_q)\right\},\\
	& d_2^+=\left\{(q_1,p_1,q_2,p_2) \in \mathcal{R} \mid  \hspace{0.1in}p_1-q_1=-c,  \hspace{0.1in} q_1<ck_p/(k_p-1)\right\},\\
	& d_2^-=\left\{(q_1,p_1,q_2,p_2) \in \mathcal{R} \mid  \hspace{0.1in}p_1-q_1=-c,  \hspace{0.1in} q_1>c/(1+k_p)\right\}.
	\end{aligned}
	\end{equation}
	The spherical cap $d_1^+$, bounded by the $a_1^+$ on $n_1^+$, gives all initial conditions of initial energy $h$ for the transit orbits starting from the bounding sphere $n_1^+$ and entering $\mathcal{R}$. Similarly, the spherical cap $b_1^-$ in $n_1^-$, bounded by $a_1^-$, determines all initial conditions of initial energy $h$ for transit orbits starting on the bounding sphere $n_1^-$ and leaving $\mathcal{R}$. The spherical caps $d_2^+$ and $d_2^-$ on $n_2$ have similar dynamical behavior. Note that in the conservative system the transit orbits entering $\mathcal{R}$ on $d^+$ will leave on $d^-$ in the same 2-sphere. However, those transit orbits with the same initial conditions in the dissipative system will not leave on the corresponding 2-sphere, but leave on another sphere with lower energy. Moreover, the spherical caps $d^+$ shrink and $d^-$ expand compared to that of the conservative system. Since the area of the caps $d^+$ and $b^-$ determines the amount of transit orbits and non-transit orbits respectively, the shrinkage of the caps $d^+$ and expansion of the caps $d^-$ means the damping reduces the probability of transition and increase the probability of non-transition, respectively. 
	
	\item Let $b$ be the intersection of $n^+$ and $n^-$ (where $q_1+p_1=0$). Then, $b$ is 1-sphere of tangency points. Orbits tangent at this 1-sphere ``bounce off'', i.e., do not enter $\mathcal{R}$ locally. 
	The spherical zones $r_1$ and $r_2$, bounded by $a^+_i$ and $a^-_i$, give the initial conditions for non-transit orbits zone. $r^+$, bounded by $a^+_i$ and $b_i$, are the initial conditions of initial energy $h$ for non-transit orbits entering $\mathcal{R}$ and $r^-_i$ are the initial conditions of initial energy $h$ for non-transit orbits leaving $\mathcal{R}$. Note that unlike the shift of the spherical caps in the dissipative system compared to that of the conservative system, the tangent spheres $b_1$ and $b_2$ do not move when damping is taken into account. Moreover, in the conservative system, non-transit orbits enter $\mathcal{R}$ on $r^+$ and then exit on the same energy bounding 2-sphere through $r^-$, but the non-transit orbits in the dissipative system exit on a different 2-sphere with different energy determined by the damping and the initial conditions.
\end{enumerate}

\paragraph{Trajectories in the equilibrium region}
From the analysis in the eigenspace, we  obtain the general solution for the dissipative system in the original coordinates, that is,
\begin{equation}
\begin{split}
\bar q_1(t) &= \frac{\lambda^2 - c_y }{s_1}  \left(\bar k_1e^{\beta_1 t} - \bar k_2e^{\beta_2 t} \right),\\
\bar q_2(t) &= \frac{\omega_p^2 + c_x}{s_2} e^{- \delta t} \left( k_5  \cos{\omega_d t} + k_6 \sin{\omega_d t} \right),
\label{diss sol}
\end{split}
\end{equation}
where $\bar k_1 = k_1 - k_3$ and $\bar k_2 = k_4 - k_2$.

Analogous to the situation in the conservative system, we can still classify the orbits into different classes depending on the limiting behavior of $\bar q_1$ as $t$ tends to plus or minus infinity. Four different categories of orbits can be obtained:

\begin{enumerate}
	\item Orbits with $\bar k_1= \bar k_2 = 0$ are 	{\it focus-type asymptotic} orbits.
	\item Orbits with $ \bar k_1\bar k_2= 0$ are 	{\it saddle-type asymptotic} orbits.
	\item Orbits with $\bar k_1\bar k_2>0$ are 	{\it transit} orbits.
	\item Orbits with $\bar k_1\bar k_2<0$ are 	{\it non-transit} orbits.
\end{enumerate}

\paragraph{Wedge of velocity and ellipse of transition}
As discussed in Section \ref{Dissipative system in eigenspace}, the initial conditions of stable asymptotic orbits in the saddle projection of the phase space should be governed by,
\begin{equation}
q_1 = k_p p_1,
\label{B: stable asymptotic line}
\end{equation}
which governs the stable asymptotic orbits which is the boundary of the transit orbits. For the initial conditions in the position space and symplectic eigenspace, denoted by $\left(\bar q_{10}, \bar q_{20}, \bar p_{10}, \bar p_{20} \right) $ and $\left( q_{10}, q_{20},  p_{10}, p_{20}  \right)$, respectively, they can be connected by the symplectic matrix \eqref{B: symp tranform}. By using \eqref{B: stable asymptotic line} and the change of variables \eqref{change of variables}, the Hamiltonian function for asymptotic orbits in the symplectic eigenspace can be rewritten by eliminating $q_{10}, q_{20},  p_{10}, p_{20}$ and $\bar p_{10}$, as,
\begin{equation}
\frac{\bar q_{10}^2}{a_e^2} + \frac{\bar q_{20}^2}{b_e^2} + \frac{\bar p_{20}^2}{c_e^2}=1,
\label{B: transition ellipsoid}
\end{equation}
where,
\begin{equation}
a_e=\sqrt{\frac{h \left( k_p - 1 \right)^2 \left(\lambda^2 - c_y \right)^2}{k_p s_1^2 \lambda} }, \hspace{0.2in} b_e = \sqrt{\frac{2 h \left(\omega_p^2 +c_x \right)^2}{s_2^2 \omega_p} }, \hspace{0.2 in} c_e=\sqrt{\frac{2h \omega_p \left(\omega_p^2 + c_x \right)^2}{s_2^2}},
\label{B:axis of ellipse}
\end{equation}
which is geometrically an ellipsoid (topologically a 2-sphere).  
As \eqref{B: transition ellipsoid} is the boundary between transit and non-transit orbits starting at an
initial energy $h$, we therefore refer to the object described by \eqref{B: transition ellipsoid}
as the \textbf{transition ellipsoid} of energy $h$.
The critical condition for the existence of real solutions for $\bar p_{20}$ requires zero discriminant for \eqref{B: transition ellipsoid}, that is,
\begin{equation}
\frac{\bar q_{10}^2}{a_e^2} + \frac{\bar q_{20}^2}{b_e^2} =1, \hspace{0.2in} \bar p_{20}=0,
\label{B: Ellipse of transition}
\end{equation}
which is an ellipse in the configuration space called the \textbf{ellipse of transition}, and is merely the configuration space projection of the transition ellipsoid \eqref{B: transition ellipsoid},  first found in \cite{zhong2018tube}. The ellipse of the transition confines the existence of transit orbits of a given initial energy which means the transit orbits can just exist inside the ellipse.
For a specific position $\left(\bar q_{10}, \bar q_{20} \right)$ inside the ellipse, $(\bar q_{10}/a_e)^2+(\bar q_{20}/ b_e)^2<1$, the solutions of $\left(\bar p_{10}, \bar p_{20} \right)$ are written as,
\begin{equation}
\begin{aligned}
\bar p_{20} = \pm c_e \sqrt{1 - \frac{\bar q_{10}^2}{a_e^2} - \frac{\bar q_{20}^2}{b_e^2}}, \hspace{0.1 in} \bar p_{10} = \frac{k_p + 1}{k_p - 1} \lambda \bar q_{10}.
\end{aligned}
\end{equation}
Each pair of $(\bar p_{10} , \bar p_{20})$ determines an angle: $\theta=\arctan(\bar p_{20}/\bar p_{10})$, which together defines the wedge of velocity. The boundary of the wedge gives the two asymptotic orbits at that position.

\begin{figure}[!ht]
	\begin{center}
		\includegraphics[width=4in]{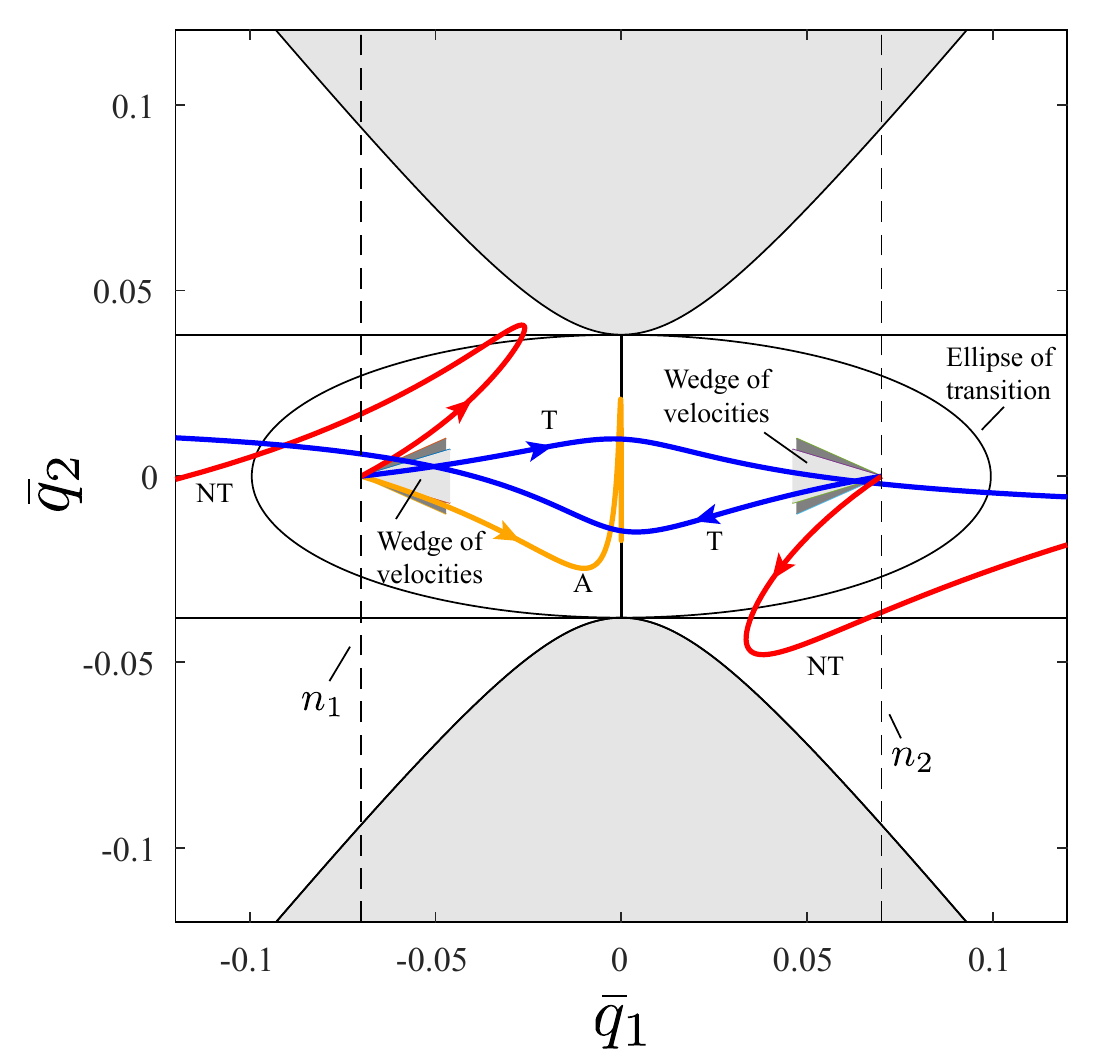}
	\end{center}
	\caption{{\footnotesize 
			The flow in the equilibrium region $\mathcal{R}$ projected onto position space $(\bar q_1, \bar q_2)$ in the dissipative system with fixed positive energy, $\mathcal{H}_2=h>0$, for a ball rolling on a stationary surface. Shown are different types of orbits as discussed in the text. Notice that due to the dissipation of energy, the periodic orbit in the conservative system does not exist, but is replaced by the initial conditions of initial energy $h$ of the focus-type asymptotic orbits. Moreover, the strip for the conservative system---which is the position space projection of the tubes of transition at initial energy $h$---is replaced by the ellipse of transition. It means that the existence of transit orbits are constrained by the ellipse so that the wedge of velocity, determining the permissible direction of the transit orbits, only exist inside the ellipse. For a given fixed energy $h$, the wedge of velocity for the dissipative system is a subset of the wedge for the conservative system which is shown as a darker wedge.
	}}
	\label{B: dissipative comfiguration flow}
\end{figure}

Figure \ref{B: dissipative comfiguration flow} gives the projection on the position space in the equilibrium region. 
The strip projected onto configuration space in the conservative system which is the boundary of the asymptotic orbits is replaced by the ellipse of transition, which restricts the existence of transition for initial conditions of initial energy $h$ to a locally bounded region. Outside the ellipse, the situation is simple: only non-transit orbits exist. Inside the ellipse, the situation is more complicated since there is a wedge of velocity restricting the direction of transit orbits. The orbits with velocity interior to the wedge are transit orbits, while orbits with velocity outside the wedge are non-transit orbits. The boundary of the wedge gives velocity for the asymptotic orbits. Note that for different point in the position space, the size of the wedge of velocity varies. The closer the wedge is to the boundary of the ellipse of transition, the smaller it is. Clearly, on the ellipse the wedge becomes a line which means only one asymptotic orbit exists there. Note that in the figure, the light grey shaded wedges are the wedges for the dissipative system, while the dark grey shaded wedges partially covered by the light grey ones are for the conservative system of the same initial energy $h$. The significant shrinking of the wedges from the conservative system to the dissipative system is caused by damping. It means an increase in damping decreases the size of the ellipse of transition and wedges on a specific point, which confirms our expectation.

\subsubsection{Transition tube and transition ellipsoid}
\label{B: tube and ellipsoid}

In the position space, we discussed how damping affects the transition. In fact, the strip in the conservative system and ellipse in the dissipative system associated with respective wedges of velocity can predict the transition and non-transition in the corresponding system for a given energy in the position space. 

\begin{figure}[!t]
	\begin{center}
		\includegraphics[width=\textwidth]{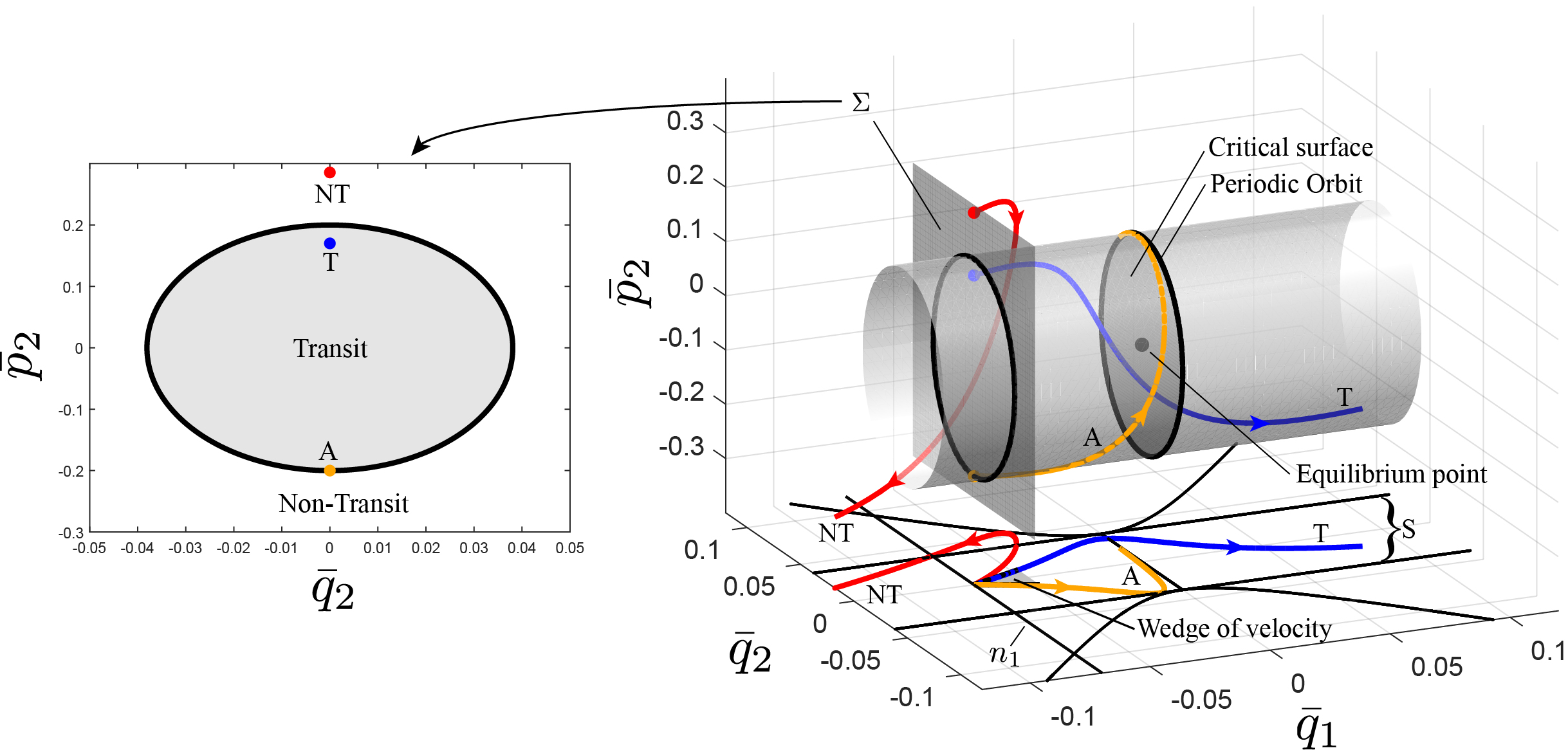}
	\end{center}
	\caption{{\footnotesize 
			Transition region boundary $\partial \mathcal{T}_h$ which is a tube (cylinder) for the conservative system of an idealized  ball rolling on a stationary surface with initial energy $h$. The left figure shows tube boundary (the ellipse) separating the transit and non-transit orbits on the Poincar\'e section $\Sigma$, where the dots are the initial conditions for the corresponding trajectories. The right figure shows the transition tube for a given energy. The critical surface divides the transition tubes into two parts whose left part gives the initial conditions for orbits transitioning to the right, and right part gives the initial conditions for orbits transitioning to the left. Some trajectories are given to show how the transition tube controls the transition whose initial conditions are shown as dots on the left Poincar\'e section with same color.
	}}
	\label{B: tube figure}
\end{figure}

\begin{figure}[!t]
	\begin{center}
		\includegraphics[width=\textwidth]{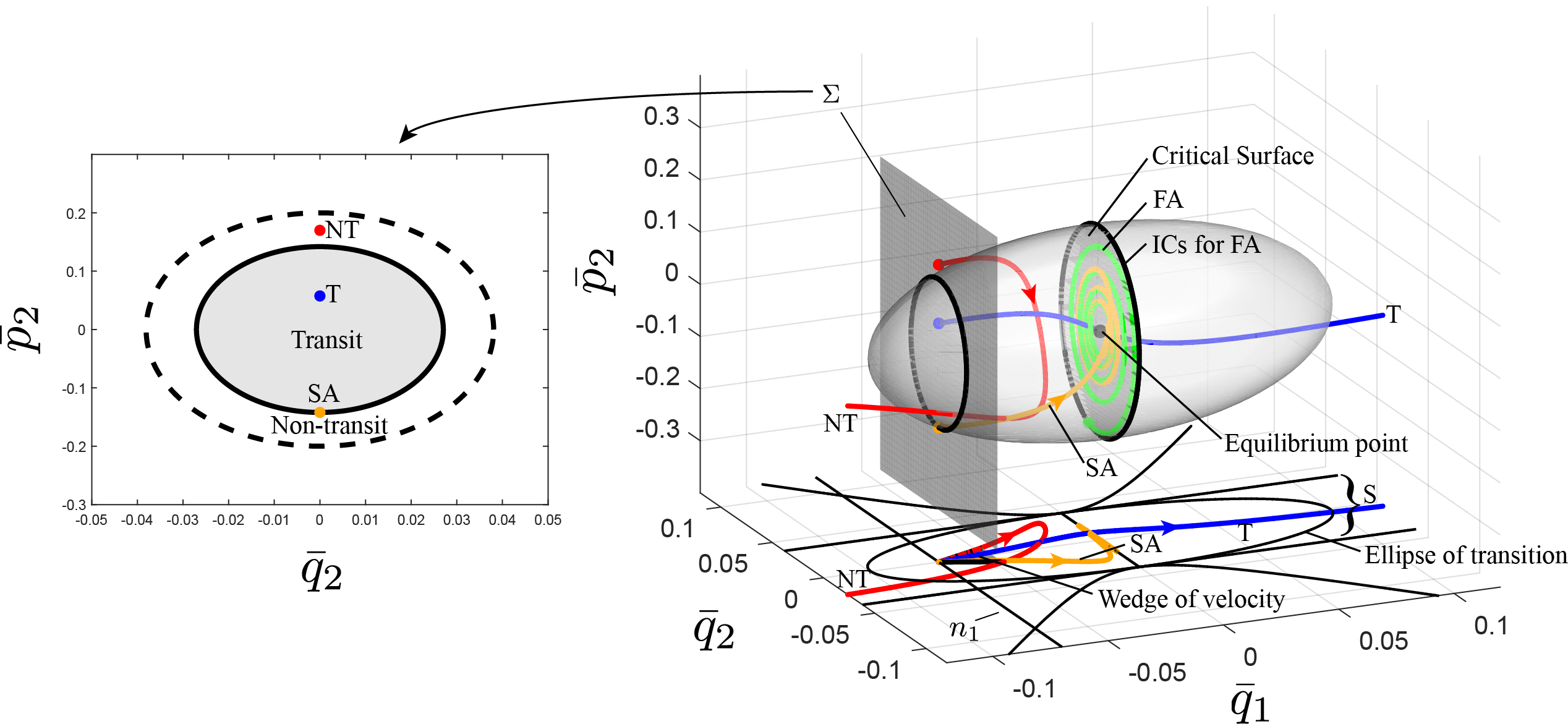}
	\end{center}
	\caption{{\footnotesize 
			Transition ellipsoid for the dissipative system of a rolling ball on a stationary surface. The left figure shows the Poincar\'e section $\Sigma$, where the dots are the initial conditions for the corresponding trajectories with the same color in the right figure and the solid ellipse is the set of initial conditions for saddle-type asymptotic orbits. For comparison, the dashed ellipse of the tube boundary for the conservative system with the same energy $h$ is also given. On the right is the ellipsoid giving the initial conditions for all transit orbits. The critical surface divides the ellipsoid into two parts. Each side of the ellipsoid gives the initial conditions of transit orbits passing through the critical surface to the other side. In this figure, SA and FA denote the saddle-type and focus-type asymptotic orbits, respectively.
	}}
	\label{B: ellipsoid figure}
\end{figure}

To obtain the initial conditions for asymptotic orbits, the Hamiltonian function for asymptotic orbits has been rewritten in the form of a tube in \eqref{B: transition tube} for the conservative system and the form of an ellipsoid in \eqref{B: transition ellipsoid} for the dissipative system, respectively. Here we refer to them as the {\it transition tube} and {\it transition ellipsoid}, respectively. 
Compactly, both are $\partial \mathcal{T}_h$. 
See the tube and ellipsoid in Figure \ref{B: tube figure} and Figure \ref{B: ellipsoid figure}, respectively. 
In the figures, the tube and the ellipsoid give the boundaries of the initial conditions for transit orbits starting with a given initial energy $h$ in the conservative and the dissipative systems, respectively; all transit orbits must have initial conditions inside the transition tube or transition ellipsoid, respectively; non-transit orbits have initial conditions outside the boundary and asymptotic orbits have initial conditions on the boundary; of course, the periodic orbit not only has initial conditions on the boundary of the transition tube, but also evolves on the boundary. Note that there is a \textbf{critical surface boundary}, given by $S^1_h$, dividing the tube and ellipsoid into two parts. The left side part is composed of transit orbits \textit{going to the right} and the right part for transit orbits \textit{going to the left}. 



The orbits with initial conditions on the critical surface $S^1_h$ are periodic orbits if in the conservative system or focus-type asymptotic orbits if in the dissipative system. The periodic orbit keeps evolving on the critical surface, while the focus-type asymptotic orbit gradually approaches the equilibrium point and finally stops there. The critical surface also plays another important role separating the motion of transit orbits and non-transit orbits. Transit orbits can cross the surface, while non-transit orbits will bounce back before reaching it. Of course, the asymptotic orbits moves asymptotically towards the surface.


\paragraph{Illustration of effectiveness} 
To illustrate the effectiveness of the transition tube and transition ellipsoid, we choose a specific Poincar\'e section $\Sigma$ revealing the transit region and initial conditions (see dots) of the trajectories shown in the insets of the conservative and dissipative case, respectively. For both the conservative and dissipative systems, the trajectories with initial conditions inside the boundary of the transition can transit from left to right, while trajectories with initial conditions outside of the boundary bounce back to the region where they start; the trajectories with initial conditions on the boundary are asymptotic to a periodic orbit or equilibrium point, for a conservative or dissipative system, respectively. This proves the transition tube and transition ellipsoid can effectively estimate the transition initial conditions in the conservative system and dissipative system, respectively.

It should be noted from the Poincar\'e section in the dissipative system that the transit region for the dissipative system (see the area encompassed by the solid closed curve) is smaller than the transit region for the conservative system (see the area encompassed by the dashed closed curve) for the same initial energy $h$. The decrease in the area for the transition is caused by the dissipation of the energy.  In fact the transit orbit in the conservative system and the non-transit orbit in the dissipative system plotted in the figure have the same initial conditions which means the dissipation of energy can make a transit orbit in the conservative system become a non-transit orbit if dissipation is added.

Up to now, we give the geometry governing the transition in both the position space and phase space. In the position space the strip in the conservative system and the ellipse in the dissipative system are the projections of the outline of the transition tube and transition ellipsoid, respectively. The wedge of velocity on a specific position $(\bar q_1, \bar q_2)$ has two boundaries. The boundaries are the projections of the upper and lower bounds on the corresponding Poincar\'e section at $\bar q_2$. 

\subsection{Snap-through buckling of a shallow arch}
Curved structures, like arches/buckled beams \cite{WiVi2016,moghaddasie2013equilibria}, shells \cite{yiming2010damage} and domes \cite{plaut2018snap,guan2018structural}, have many engineering applications. This type of structures can withstand larger transverse loading mainly through membrane stresses compared to flat structures mainly through bending moments. The arch, as an example in this paper, can be at rest in a local minimum of underlying potential energy in unloaded state or under small loading. If subjected to large input of energy or external forces, it may suddenly jump (snap-through) dynamically to another remote local minimum or stable equilibrium. The transition of a buckled conservative  nanobeam \cite{collins2012isomerization} and macroscopic arch \cite{zhong2018tube} have been studied under the frame of tube dynamics. This section will review the results in \cite{zhong2018tube} where dissipative forces were considered.

\paragraph{Governing equations} 
In this analysis a slender arch with thickness $d$, width $b$ and length $L$ is considered. A Cartesian coordinate system $xyz$ is established on the mid-plane of the beam in which 
$x,y$ are the directions along the length and width directions and $z$ the downward direction normal to the mid-plane. Let $u$ and $w$ be the axial and transverse displacements of an arbitrary point on the mid-plane of the beam, respectively, and $w_0$ the initial deflection. Based on Euler-Bernoulli beam theory \cite{zhong2016analysis,WiVi2016}, the nonlinear integro-differential governing equation \cite{WiVi2016,zhong2018tube,moghaddasie2013equilibria} of the beam with in-plane immovable ends is given by,
\begin{equation}
\begin{split}
\rho A \frac{\partial^2 w}{\partial t^2} + c_d \frac{\partial w}{\partial t} + & EI \left(\frac{\partial^4 w}{\partial x^4} - \frac{\partial^4 w_0}{\partial x^4}\right) \\
& + \left[N_T - \frac{EA}{2L}  \int_0^L \left( \left(\frac{\partial w}{\partial x} \right)^2 - \left(\frac{\partial w_0}{\partial x} \right)^2 \right) \mathrm{d} x \right] \frac{\partial^2 w}{\partial x^2} =0,
\label{S:PDE}
\end{split}
\end{equation}
where the boundary conditions of the in-plane immovable ends, $u(0)=u(L)=0$, are applied. See the details of the derivation in \cite{zhong2018tube}. In the equation of motion $\rho$ and $E$ are the mass density and Young's modulus, respectively; $c_d$ is the coefficient of linear viscous damping. $A$ and $I$ are the area and the moment of inertia of the cross-section, respectively, so that $EA$ and $EI$ are the extensional stiffness and bending stiffness. Finally, $N_T$ is the axial thermal loading as a convenient way of controlling the initial deflection which replaces the external axial force due to the impossibility of applying such force to the beam with immovable ends. For different types of end constraints, the boundary conditions can be written as,
\begin{equation}
\begin{aligned}
&w=0, \quad \frac{\partial^2 w}{\partial x^2}=0, \hspace{0.5in} &&\text{for simply-simply supported},\\
&w=0, \quad \frac{\partial w}{\partial x}=0, \hspace{0.5in} &&\text{for clamped-clamped supported}.
\end{aligned}
\end{equation}

To capture the symmetric and asymmetric snap-through behavior of the arch, the first two mode shapes, $\phi_1(x)$ and $\phi_2(x)$, will be used. Refs. \cite{zhong2018tube,virgin2000introduction,WiVi2016} list the specific forms of $\phi_i$ satisfying the boundary conditions of simply-simply supports and clamped-clamped supports which will not be given here for simplification. Assume the deflection and initial imperfection have the following forms,
\begin{equation}
\begin{split}
w(x,t)&= X(t) \phi_1(x) +Y(t) \phi_2(x),\\
w_0(x)&= \gamma_1 \phi_1(x) + \gamma_2 \phi_2(x),
\label{First_two_modes}
\end{split}
\end{equation}
where $X(t)$ and $Y(t)$ are the amplitudes corresponding to the first two mode shapes of the deflection and $\gamma_i$ are the imperfection coefficients. Applying the Galerkin method, one can obtain the following equations of motion for the amplitudes,
\begin{equation}
\begin{split}
& M_1 \ddot X + C_1 \dot X + K_1 \left(X - \gamma_1 \right) - N_T G_1 X - \frac{EA}{2L}G_1^2 \left(\gamma_1^2 X -X^3 \right) - \frac{EA}{2L} G_1 G_2 \left(\gamma_2^2 X -X Y^2 \right)=0,\\
& M_2 \ddot Y + C_2 \dot Y + K_2 \left(Y - \gamma_2 \right) - N_T G_2 Y - \frac{EA}{2L}G_2^2 \left(\gamma_2^2 Y -Y^3 \right) - \frac{EA}{2L} G_1 G_2 \left(\gamma_1^2 Y -X^2 Y \right)=0,
\label{odes}
\end{split}
\end{equation}
where the coefficients are defined by,
\begin{equation}
\begin{split}
\left(M_i, C_i \right) = \left(\rho A, c_d \right) \int_0^L \phi_i^2 \mathrm{d}x, \ \ K_i = EI \int_0^L \left(\frac{\partial^2 \phi_i}{\partial x^2} \right)^2 \mathrm{d}x, \ \ G_i = \int_0^L \left(\frac{\partial \phi_i}{\partial x} \right)^2 \mathrm{d}x.
\label{Galerkin-coefficient}
\end{split}
\end{equation}

The equations of motion can be re-cast in Hamiltonian form, using the Hamiltonian function $\mathcal{H}=\mathcal{K} + \mathcal{U}$, with
$\mathcal{K}$ and $\mathcal{U}$ as in Appendix \ref{app_arch}.

\begin{figure}
	\begin{center}
		\includegraphics[width=\textwidth]{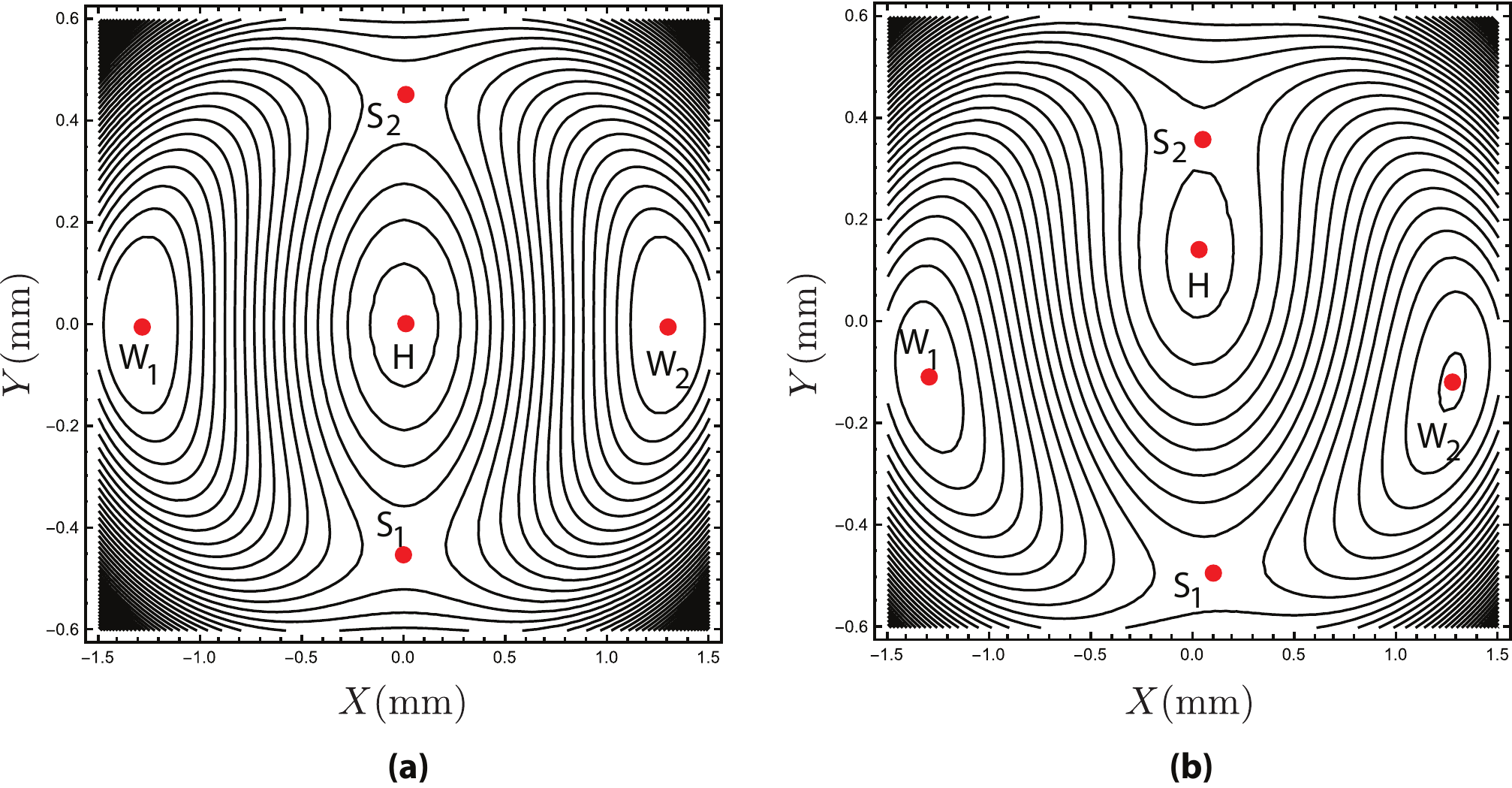}
	\end{center}
	\caption{{\footnotesize 
			Contours of potential energy, $\mathcal{U}(X,Y)$, of dynamic arch model: (a) the symmetric system, $\gamma_1= \gamma_2=0$; (b) with small initial imperfections in both modes, i.e., $\gamma_1$ and $\gamma_2$ are nonzero.
	}}
	\label{contours_of_arch}
\end{figure}

For the parameters selected in \cite{zhong2018tube}, we have a fixed two-dimensional potential energy landscape, 
$\mathcal{U}(X,Y)$, as illustrated in Figure \ref{contours_of_arch}. For the arch without initial imperfections, Figure \ref{contours_of_arch}(a) shows a symmetric potential energy surface about both $X$ and $Y$. In this system, there are five equilibrium points shown as dots, some of which are stable and some of which are unstable. Of the five points, W$_1$ and W$_2$ are the stable equilibrium points, each within its own
potential well; 
S$_1$ and S$_2$ are (unstable) saddle points; and H is the unstable hilltop. If the system starts at rest at W$_1$, it will remain there. If a large impulse with the right size and direction is applied to the arch, it may snap-through, or as understood in phase space, jump to the remote stable equilibrium at W$_2$,  passing close to S$_1$ or S$_2$ along the way, and generally avoiding H. 
If an initial imperfection in both modes is considered, the symmetry of the energy surface about $X$ and $Y$ is  broken, as in 
Figure \ref{contours_of_arch}(b). 
Since we are most interested in the behavior near the saddle points, we linearize the equations about a saddle, either S$_1$ or S$_2$, with position $(X_e,Y_e)$, which gives the following linearized equations,
\begin{equation}
\begin{split}
\dot x&= \frac{p_x}{M_1},\\
\dot y&= \frac{p_y}{M_2},\\
\dot p_x&= A_{31} x + A_{32} y - C_H p_x,\\
\dot p_y&= A_{32} x + A_{42} y - C_H p_y,
\label{linearization}
\end{split}
\end{equation}
where $(x,y,p_x,p_y)= (X,Y,p_X,p_Y) - (X_e,Y_e,0,0)$ is the displacement from the saddle point in configuration-momentum phase space and the parameters $(A_{31}, A_{32}, A_{42}, C_H)$ are given in Appendix \ref{app_arch}.

Since Ref.\ \cite{zhong2018tube} carried out a detailed study on the transition of shallow arch from both a global view, and a local view near the saddle, for both the conservative  and dissipative system, the corresponding discussions about the transition are not repeated here . The reader should refer to Ref.\ \cite{zhong2018tube} for more details. We merely point out that 
\eqref{linearization} can be non-dimensionalized and
transformed into the standard form of \eqref{B:EOM with damping in phase space} in the symplectic eigenspace via a symplectic transformation.  The details are given in Appendix \ref{app_arch}.

\subsection{Ship motion with equal damping}
\label{Ship with equal damping}
The stability of ship motion plays an important role in delivery, fishing, transport, and military applications. 
The phenomenon of capsize has attracted a great amount of attention, due to the ensuing catastrophic losses of life and property. 
While some studies have only considered single degree-of-freedom dynamics in the roll motion, several 
studies have concluded that pitch-roll coupling is a much better approximation \cite{nayfeh1973nonlinear,nayfeh1974perturbation,mccue2005probabilistic,thompson1996suppression}.
However, the consideration of now two coupled degrees of freedom 
makes the analysis challenging. 
Under the framework of tube dynamics, Ref.\ \cite{NaRo2017} studied nonlinear ship motion and the transition tube for capsize. 
In addition, the effect of stochastic forcing was taken into account and the skeleton formed by the tube dynamics was shown to persist. However, damping was not taken into consideration in \cite{NaRo2017}. 
In this section, we derive the equations of motion with the influence of equal damping along the roll and pitch directions. 

\paragraph{Governing equations}
Based on \cite{NaRo2017,nayfeh1973nonlinear,nayfeh1974perturbation}, we consider the coupled roll and pitch equations for the ship motion of the form,
\begin{equation}
\begin{split}
\ddot \phi&=-\omega_{\phi}^2 \phi + 2K_1 \phi \theta + m_{\phi} (t),\\
\ddot \theta&=-\omega_{\theta}^2 \theta + K_1 \frac{I_{xx}}{I_{yy}}\phi^2 + m_{\theta}(t),
\label{S: dimensional ODE}
\end{split}
\end{equation}
where $\phi$ and $\theta$ are roll and pitch angles measured in radians. The coefficients are defined as,
\begin{equation*}
\omega_{\phi}=\sqrt{\frac{K_{\phi}}{I_{xx}}}, \hspace{0.1in} \omega_{\theta}=\sqrt{\frac{K_{\theta}}{I_{yy}}}, \hspace{0.1in} K_1=-\frac{K_{\phi \theta}}{2I_{xx}}, \hspace{0.2in} m_{\phi} (t)=\frac{\tau_{\phi}(t)}{I_{xx}}, \hspace{0.1in} m_{\theta}(t)=\frac{\tau_{\theta}(t)}{I_{yy}},
\end{equation*}
where $I_{xx}$ and $I_{yy}$ are the sums of the second moments of inertia and hydrostatic inertia; $K_{\phi}$ and $K_{\theta}$ are the linear rotational stiffness related to the square of the corresponding natural frequency; $K_{\phi \theta}$ is the nonlinear coupling coefficient; $\tau_{\phi} \left(t \right)$ and $\tau_{\theta} \left(t \right)$ are generalized possibly time-dependent torques in the roll and pitch directions, respectively, and $\omega_{\phi}$ and $\omega_{\theta}$ are called the natural roll and natural pitch frequencies, respectively.

For the conservative system, i.e., $m_{\phi}=m_{\theta}=0$, the system has two saddle points at $\left(\pm \phi_e, \theta_e \right)$, with,
\begin{equation}
\phi_e=\frac{\omega_{\phi} \omega_{\theta}}{\sqrt{2} K_1} \sqrt{\frac{I_{yy}}{I_{xx}}}, \hspace{0.2in} \theta_e=\frac{\omega_{\phi}^2}{2 K_1}.
\end{equation}
Here $\phi_e$ is called the roll angle of vanishing stability and $\theta_e$ is the corresponding pitch angle. 

The equations of motion \eqref{S: dimensional ODE} can be re-cast in a non-dimensional Hamiltonian form, using a Hamiltonian function $\mathcal{H}$ as given in Appendix \ref{app_ship}.  
The linearized equations about the saddle points can be written
written in matrix form, 
\begin{equation}
\dot{\bar z}= M \bar z + D \bar z, \label{eq_ship_matrix}
\end{equation}
where $\bar z= \left(\bar q_1, \bar q_2, \bar p_1, \bar p_2 \right)^T$ is the displacement from the saddle point in the phase space, and where,
\begin{equation}
M=\begin{pmatrix}
0 & 0 & 1 & 0\\
0 & 0 & 0 & 1\\
0 & 1 & 0 & 0\\
1 & c_y & 0 & 0
\end{pmatrix},
\hspace{0.5in}
D=\begin{pmatrix}
0 & 0 & 0 & 0\\
0 & 0 & 0 & 0\\
0 & 0 & -c_{h_1} & 0\\
0 & 0 & 0 & -c_{h_2}\\
\end{pmatrix}.
\end{equation}
The corresponding quadratic Hamiltonian function 
is given by,
\begin{equation}
\mathcal{H}_2=\tfrac{1}{2} \bar p_1^2 + \tfrac{1}{2} \bar p_2^2- \bar q_1 \bar q_2 - \tfrac{1}{2} c_y \bar q_2^2 .
\end{equation}

\paragraph{Conservative system} 
For the conservative system, i.e. $c_{h_1}=c_{h_2}=0$, one can introduce a change of variables \eqref{change of variables} with the  symplectic matrix $C$ given by \eqref{ship sym matrix}
which casts  the equations of motion in a simple form in the symplectic eigenspace \eqref{phase-space Hamiltonian Equations} with Hamiltonian function \eqref{phase-space Hamiltonian} and solutions \eqref{phase-space solutions}. 
The dynamical behavior 
near the saddle point in both position space and eigenspace are similar to the rolling ball on a stationary surface. 
Readers can also consult 
\cite{NaRo2017} for more details. Note that here further nondimensional parameters were introduced, while Ref.\
\cite{NaRo2017} kept them unchanged.

\paragraph{Dissipative system with equal damping}
If the coefficients of the viscous damping along both the pitch and roll directions happen to be proportional to the second moments of inertia and hydrostatic inertia, $c_{h_1}$ and $c_{h_2}$ are exactly the same, denoted by $c_h$. Thus, using the same symplectic matrix in \eqref{ship sym matrix} one gets the same equations of motion as the standard uncoupled form, 
\eqref{B:EOM with damping in phase space}, in the symplectic eigenspace. 
For the general case of unequal damping, $c_{h_1}\neq c_{h_2}$, one will get {\it coupled dynamics} on the saddle and focus planes, as shown below in Section \ref{Ship motion with unequal damping}.

\section{Coupled systems in the dissipative case}
In Section \ref{uncoupling systems}, we investigated the geometry of escape/transition in uncoupled systems (in the symplectic eigenspace) which are generally inertial systems with equal damping in each degree of freedom. Due to the uncoupled property, it is easy to obtain the analytical solutions and the dynamical behavior. We have found the transition tube and transition ellipsoid governing the escape in the conservative and dissipative systems, respectively. Another category of system is one in which the saddle and focus are coupled with each other when the system is transformed to the corresponding eigenspace. The situation is more complicated but important and interesting. The first kind is an inertial system with unequal damping, like the ship motion discussed in Section \ref{Ship motion with unequal damping}. Another one is a system with both gyroscopic and dissipative forces present. Such systems can display non-intuitive phenomena, like dissipation-induced instabilities \cite{krechetnikov2007dissipation} as discussed in the introduction. In this section, we establish the mathematical models for some physical problems and reveal the geometry of escape/transition in such systems.

\subsection{Ball rolling on a rotating surface}
\label{rotating surface}
In Section \ref{Ball rolling}, the rolling ball on a stationary surface was studied and the effect of dissipative forces was considered. We established it as a standard example to investigate the escape from a potential well in inertial systems with equal damping and revealing the escape mechanism in such systems. Here we  further expand the framework regarding escape to a more complicated situation where the surface is rotating such that  gyroscopic forces exist. 
Several researchers have investigated a ball or particle moving on a rotating surface \cite{krechetnikov2007dissipation,bottema1976stability,kirillov2011brouwers,thompson2002rotating,brouwer1918motion}, mainly due to the unexpected dissipation-induced instabilities. The combination of the dissipative and gyroscopic forces enriches the behavior in escape dynamics.

\subsubsection{Governing equations}

\begin{figure}
	\begin{center}
		\includegraphics[width=5in]{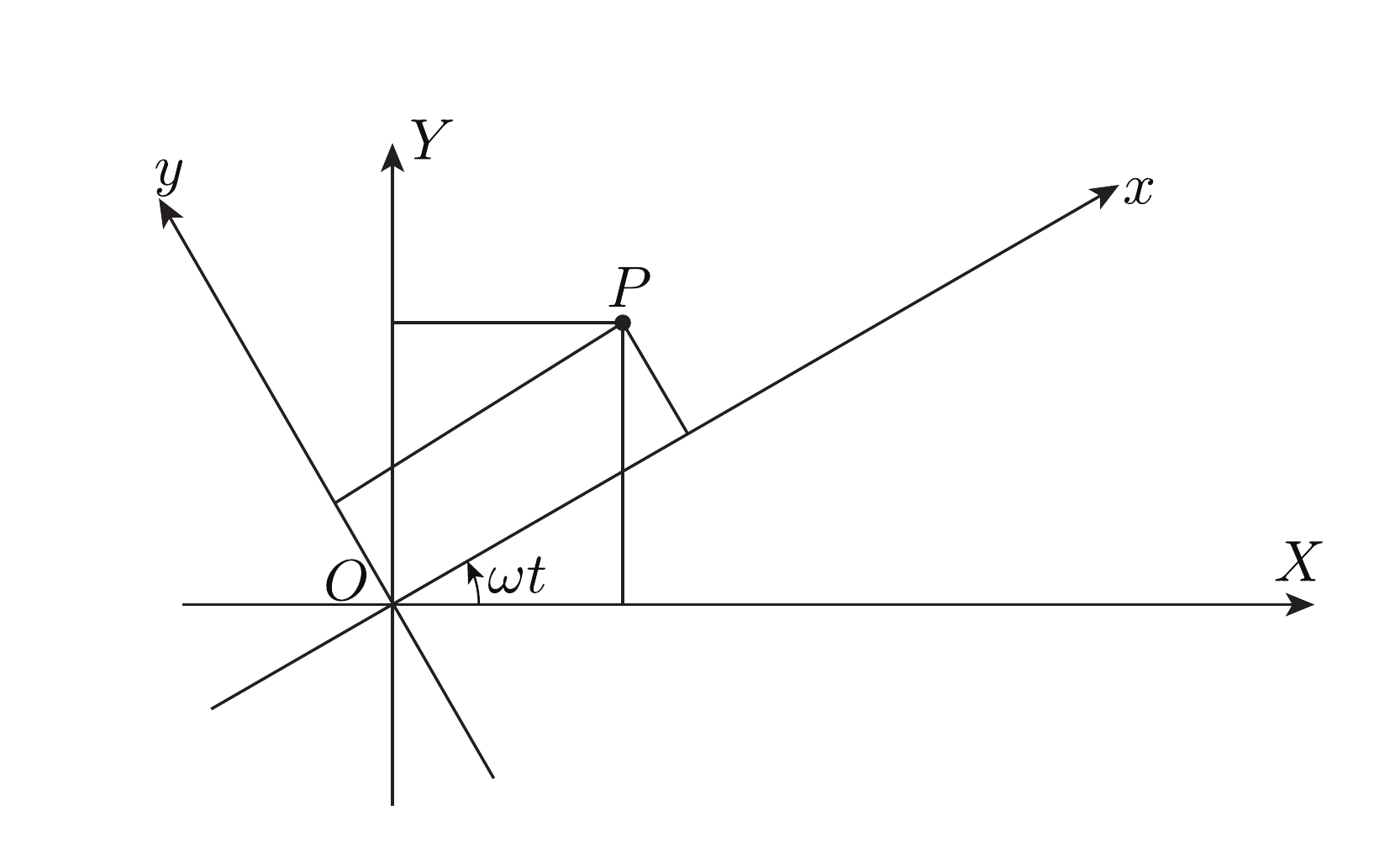}
	\end{center}
	\caption{\label{Rotating surface coordinate}{\footnotesize 
			Inetial and rotating frames. The rotating coordinate system of $x$ and $y$ axes moves counterclockwise with constant angular velocity $\omega$ relative to the inertial frame with $X$ and $Y$ axes. The $z$ axis coincides with the $Z$ axis which is pointing out of the plane and is not shown here. We denote the unit vectors along $x$, $y$, $z$ by $\mathbf{e}_1$, $\mathbf{e}_2$ and $\mathbf{e}_3$, respectively.
	}}
\end{figure}

Consider a rotating surface with counterclockwise angular velocity $\omega$ as shown in Figure \ref{Rotating surface coordinate}. 
Let $X \mbox{-} Y \mbox{-}Z$ be an inertial frame, denoted as the $N$ frame, with origin $O$, where $X \mbox{-} Y$ plane is horizontal and $Z$ is vertical to the plane. Establish another rotating frame $x \mbox{-} y \mbox{-} z$, denoted as the $R$ frame, with the same origin $O$ fixed on the rotating surface, where $Oz$  coincides with $OZ$. In this study, the geometrical parameters of the rotating surface are the same as before given in \eqref{surface equation}.

The angular velocity vector of the $R$ frame relative to the $N$ frame is, 
\begin{equation}
\text{\boldmath$\omega$}^{R/N}= \omega \mathbf{e}_3.
\end{equation}

A particle (or ball), denoted by $P$, with unit mass, moves on the rotating surface, with a position vector described in the $R$ frame as,
\begin{equation}
\mathbf{P}(x,y,z,t)= x(t) \mathbf{e}_1 + y(t) \mathbf{e}_2 + z(t) \mathbf{e}_3,
\end{equation}
where $(x,y,z)$ is the position of the mass in the $R$ frame. 
The inertial velocity of the mass can be written in the $R$ frame as,
\begin{equation}
\begin{split}
^N \mathbf{v}^{P} &= \dot x \mathbf{e}_1+ \dot y \mathbf{e}_2 + \dot z \mathbf{e}_3 +\text{\boldmath$\omega$}^{R/N} \times \mathbf{P}\\
&= \left(\dot x - y \omega \right) \mathbf{e}_1 + \left(y + x \omega \right) \mathbf{e}_2 + \dot z \mathbf{e}_3.
\end{split}
\end{equation}

Considering the motion is constrained on the rotating surface, here $z$ is not an independent variable, but depends on $x$ and $y$ via $z=H(x,y)$. Thus, the kinetic energy $\mathcal{K}$ and potential energy $\mathcal{U}$ are,
\begin{equation}
\begin{split}
\mathcal{K}(x,y)&=\tfrac{1}{2} I |^N \mathbf{v}^{P}|^2= \tfrac{1}{2} I \left[\left( \dot x - y \omega\right)^2 + \left(\dot y + x \omega \right)^2 + \left(H_{,x} \dot x + H_{,y} \dot y \right) ^2 \right],\\
\mathcal{U} (x,y)&= g H(x,y).
\label{R:energy}
\end{split}
\end{equation}

After obtaining the Lagrangian function, $\mathcal{L}=\mathcal{K}-\mathcal{U}$, we can derive the Euler-Lagrange equations given in \eqref{general Lagrange equations}. As discussed in \cite{bottema1976stability}, two types of damping can be considered in the rotating surface system, i.e., \textbf{internal damping} and \textbf{external damping}. Internal damping is proportional to the relative velocity measured in the rotating frame, while external damping is proportional to the inertial velocity.
Thus, the mathematical form of two types of the generalized damping forces are, 
\begin{equation}
\begin{split}
Q_x^{\rm int} &= -c_d\left[  \left(1+ H_{,x}^2 \right) \dot x +  H_{,x} H_{,y}\dot y \right], \\
Q_y^{\rm int} &= -c_d\left[  \left(1+ H_{,y}^2 \right) \dot y +  H_{,x} H_{,y}\dot x \right],
\end{split}
\hspace{0.1in} \text{ for internal damping}
\end{equation}
and, 
\begin{equation}
\begin{split}
Q_x^{\rm ext} &= -c_d\left[  \left(1+ H_{,x}^2 \right) \dot x + H_{,x} H_{,y}\dot y -  \omega y  \right], \\
Q_y^{\rm ext} &= -c_d\left[  \left(1+ H_{,y}^2 \right) \dot y + H_{,x} H_{,y}\dot x + \omega x  \right],
\end{split}
\hspace{0.1in} \text{ for external damping}
\label{inter-exter damping}
\end{equation}
where $c_d$ is the coefficient of damping.
In the current problem, we only consider internal damping, $(Q_x,Q_y)=(Q_x^{\rm int},Q_y^{\rm int})$,  due to the friction between the mass and the moving surface, as the most physically relevant. 

The equations of motion can be written in non-dimensional Hamiltonian form, using a Hamiltonian function $\mathcal{H}$ as given in Appendix \ref{app_ball}.  
Following the same procedure as for the ball rolling on a stationary surface, we linearize the equations of motion around the saddle point at the origin which gives the linearized non-dimensional Hamilton's equation
in matrix form, 
\begin{equation}
\dot{\bar{z}}= M \bar z + D \bar z, \label{linear_ball}
\end{equation}
where $\bar z = \left( \bar q_1, \bar q_2, \bar p_1, \bar p_2 \right)^T$ is the displacement from the saddle point,
and
where,
\begin{equation}
M=
\begin{pmatrix}
0 & \omega & 1 & 0\\
- \omega & 0 & 0 & 1\\
c_x & 0 & 0 & \omega\\
0 & c_y & -\omega & 0
\end{pmatrix}
,\hspace{0.1in}
D= c_h \begin{pmatrix}
0 & 0 & 0 & 0\\
0 & 0 & 0 & 0\\
0 & - \omega & -1 & 0\\
\omega & 0 & 0 & -1
\end{pmatrix}.
\label{conservative and damping matrix}
\end{equation}
The 
quadratic Hamiltonian function corresponding to matrix $M$ is,
\begin{equation}
\mathcal{H}_2(\bar q_1, \bar q_2, \bar p_1, \bar p_2)=\tfrac{1}{2} \left( \bar p_1^2 + \bar p_2^2 \right)+ \omega \bar p_1 \bar q_2 - \omega \bar p_2 \bar q_1 - \tfrac{1}{2}  \left(c_x \bar q_1^2 + c_y \bar q_2^2 \right).
\label{R:rescale Hamiltonian}
\end{equation}

\subsubsection{Analysis in the conservative system}
In this section, the dynamic behavior in the conservative system will be analyzed. Here the damping $c_h$ is set to zero which gives,
\begin{equation}
\dot{\bar z} = M \bar z.
\label{R: conservative ODEs}
\end{equation} 
Curiously, we are able to use the eigenvectors of $M$ in \eqref{conservative and damping matrix} and use them to construct a symplectic linear change of variables which changes \eqref{R: conservative ODEs} into the simple normal  form \eqref{phase-space Hamiltonian Equations}, with the simple Hamiltonian function  \eqref{phase-space Hamiltonian}  and with solutions as given in \eqref{phase-space solutions}. The details are in Appendix \ref{app_ball}.


\paragraph{Trajectories in the equilibrium region}
\label{conservative wedge and strips}
The flow in the equilibrium region $\mathcal{R}$ in the symplectic eigenspace was performed for the normal form in Section \ref{general phase space} and will not be repeated here. 
However, it is instructive to study the appearance of the orbits in the position space for this particular problem, i.e.,  the $\left(\bar q_1, \bar q_2\right)$ plane. Note that the evolution of all trajectories must be restricted by the given energy $h$ which forms the zero velocity curves \cite{KoLoMaRo2011} (corresponding to $\dot v_x = \dot v_y =0$) which bound the motion in the position space projection and are  determined by the following function,
\begin{equation}
\bar q_2 \left(\bar q_1 \right)=\pm \sqrt{\frac{-2h - \left(  c_x + \omega^2 \right) \bar q_1^2}{c_y + \omega^2}},
\end{equation}
which is obtained from \eqref{R:rescale Hamiltonian}.

From the solutions in the symplectic eigenspace \eqref{phase-space solutions}, we can obtain the general (real) solutions in the position space by using the transformation matrix $C$ in \eqref{R: symp tranform} which yields the general (real) solutions with the form  \eqref{conser_general_sol}. 
Thus, we can obtain the solutions for $\bar q_1$ and $\bar q_2$, given the initial conditions in the eigenspace, $(q_1^0,q_2^0,p_1^0,p_2^0)$,
\begin{equation}
\begin{split}
\bar q_1(t)&=\frac{\lambda^2 - c_y - \omega^2}{s_1} q_1^0 e^{ \lambda t} - \frac{\lambda^2 - c_y - \omega^2}{s_1} p_1^0 e^{- \lambda t} - \frac{2 \omega \omega_p}{s_2} \left(p_2^0 \cos \omega_p t - q_2^0 \sin \omega_p t \right),\\
\bar q_2(t)&= -\frac{2 \lambda \omega}{s_1}  q_1^0 e^{ \lambda t} - \frac{2 \lambda \omega}{s_1} p_1^0 e^{- \lambda t} - \frac{\omega_p^2 + c_x + \omega^2}{s_2} \left(q_2^0 \cos \omega_p t+ p_2 ^0 \sin \omega_p t \right).\\
\end{split}
\end{equation}
Upon inspecting the general solution, we see that the solutions on the energy surface fall into different classes depending upon the limiting behavior of $\bar q_1(t)$ as $t$ tends to plus or minus infinity. As the $\bar q_1(t)$ expression is dominated by the $q_1^0$ term as $t\rightarrow + \infty$, $\bar q_1$ tends to minus infinity (staying on the left-hand side), is bounded (staying around the equilibrium point), or tends to plus infinity (staying on the right-hand side) for $q_1^0>0$, $q_1^0=0$ and $q_1^0<0$, respectively. 
The statement holds if $t \rightarrow - \infty $ and $-p_1^0$ replaces $q_1^0$. Varying the signs of $q_1^0$ and $p_1^0$, and following the procedures described in \cite{zhong2018tube,Conley1968}, one can also obtain the same nine classes of orbits grouped into the same four categories as in Section \ref{Ball rolling}.

\begin{figure}[!t]
	\begin{center}
		\includegraphics[width=5in]{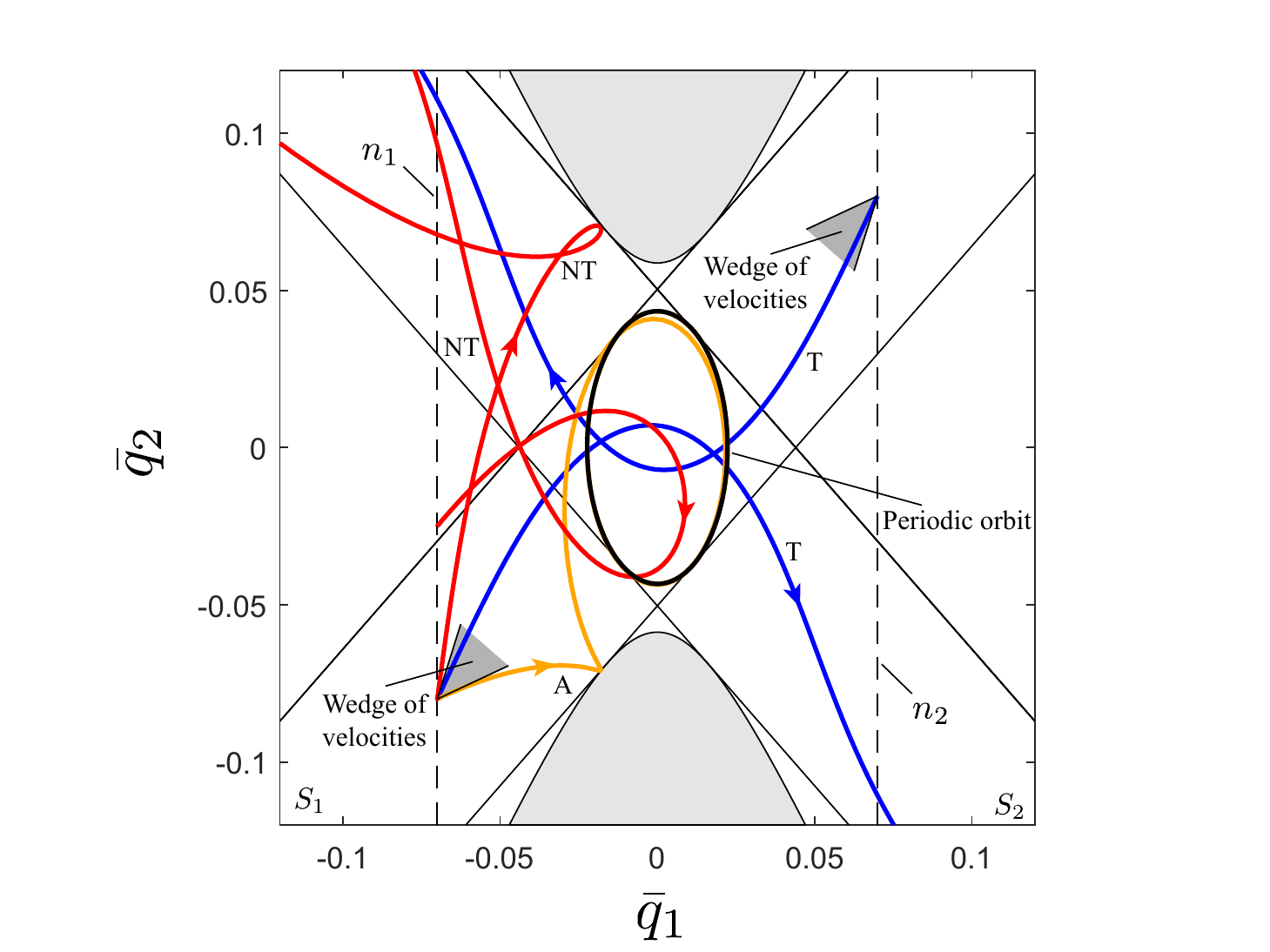}
	\end{center}
	\caption{{\footnotesize 
			The flow in the equilibrium region $\mathcal{R}$ projected onto position space $(\bar q_1, \bar q_2)$ in the conservative system with fixed positive energy, $\mathcal{H}_2=h>0$, for a ball rolling on a rotating surface. Shown are the periodic orbit acting as an ellipse; one asymptotic orbit gradually approaching the periodic orbit; two transit orbits; and two non-transit orbits, one starting inside the strips and the other outside the strips. Note that the dynamic behavior in the position space is identical to those in the circular restricted three-body problem \cite{jaffe2002statistical,KoLoMaRo2000}.
	}}
	\label{R: conserfative confguration flow}
\end{figure}

\begin{enumerate}
	\item If $q_1^0=p_1^0=0$, we obtain a periodic solution with the following projection onto the position space,
	\begin{equation}
	\frac{\bar q_1^2}{ \left(\frac{2 \omega \omega_p}{s_2} \sqrt{\frac{2 h}{\omega_p}} \right)^2} + \frac{\bar q_2^2}{\left( \frac{\omega_p^2 + c_x + \omega^2}{s_2} \sqrt{\frac{2 h}{\omega_p}} \right)^2}=1.
	\end{equation}
	Here, the initial energy is $h= \frac{1}{2} \omega_p \left[(q_2^0)^2 +(p_2^0)^2 \right]$. Identical to what has been proved by Conley \cite{Conley1968} for the restricted three-body problem, this periodic orbit, shown in Figure \ref{R: conserfative confguration flow}, projects onto the $\left(\bar q_1, \bar q_2 \right)$ plane as an ellipse. Note that the size of the ellipse goes to zero with $h$. It is different from the non-gyroscopic system where the periodic orbit projects to a straight segment in the position space.
	
	\item Orbits with $q_1^0 p_1^0=0$ are asymptotic orbits. They are asymptotic to the periodic orbits of category 1. The asymptotic orbit with $q_1^0=0$ projects into the strip $S_1$ in the $\left(\bar q_1 , \bar q_2\right)$ plane bounded by the lines,
	\begin{equation}
	\bar q_2=\frac{2 \lambda \omega}{\lambda^2 - c_y - \omega^2} \bar q_1 \pm \sqrt{\left(\frac{4 \lambda \omega_p \omega^2 }{s_2 \left(\lambda^2 - c_y - \omega^2 \right)} \right)^2 + \left( \frac{\omega_p^2 + c_x +\omega^2}{s_2} \right)^2} \sqrt{\frac{2h}{\omega_p}},
	\label{R: strip1}
	\end{equation}
	while orbits with $p_1^0 = 0$ project into the strip $S_2$ bounded by the lines,
	\begin{equation}
	\bar q_2=-\frac{2 \lambda \omega}{\lambda^2 - c_y - \omega^2} \bar q_1 \pm \sqrt{\left(\frac{4 \lambda \omega_p \omega^2 }{s_2 \left(\lambda^2 - c_y - \omega^2 \right)} \right)^2 + \left( \frac{\omega_p^2 + c_x +\omega^2}{s_2} \right)^2} \sqrt{\frac{2h}{\omega_p}}.
	\label{R: strip2}
	\end{equation}
	In fact, $S_1 $ is for stable asymptotic orbits, while $S_2$ is for unstable asymptotic orbits. Notice the width of the strips depend on $h$ and go to zero as $h\rightarrow 0$.
	
	\item Orbits with $q_1^0 p_1^0 >0$ are transit orbits because they cross the equilibrium region $\mathcal{R}$ from $- \infty$ (the left-hand side) to $+ \infty$ (the right-hand side) or vice versa.
	
	\item Orbits with $q_1^0 p_1^0 <0$ are non-transit orbits.
\end{enumerate}

\paragraph{The wedge of velocity}
To study the projection of the last two categories of  orbits in the restricted three-body problem, 
Conley \cite{Conley1968} proved a couple of propositions to determine whether at each position, $\left(\bar q_1, \bar q_2 \right)$, the wedge of velocity exists, in which $q_1^0 p_1^0 >0$. See the shaded wedges in Figure \ref{R: conserfative confguration flow}.  In the current problem, the same behavior is observed. In the next part, the derivation will be given by a more direct method than Conley's, developed in \cite{zhong2018tube} for the more general dissipative system. Note that the orbits with velocity on the boundary of a wedge  satisfy $q_1^0 p_1^0=0$, making them asymptotic orbits (which will be used in the derivation).

For initial conditions $\left(\bar q_1^0, \bar q_2^0, \bar p_1^0, \bar p_2^0 \right)$ in  the original phase space and $\left(q_1^0, q_2^0, p_1^0, p_2^0 \right)$ in the symplectic eigenspace, we can establish their relations by the symplectic matrix $C$ in \eqref{R: symp tranform}, i.e., $\left( \bar q_1^0, \bar q_2^0, \bar p_1^0, \bar p_2^0 \right)^T= C \left(q_1^0, q_2^0, p_1^0, p_2^0 \right)^T$. Note that we have $q_1^0=0$ and $p_1^0=0$ for stable and unstable asymptotic orbits, respectively.  We can then express $p_1^0$ (or $q_1^0$), $q_2^0$, $p_2^0$ and $\bar p_2^0$ in terms of $\bar q_1^0$, $\bar q_2^0$ and $\bar p_1^0$. After substituting $q_2^0$ and $p_2^0$ as a function of $\bar q_1^0$, $\bar q_2^0$ and $\bar p_1^0$ into the Hamiltonian normal form \eqref{phase-space Hamiltonian} we can rewrite \eqref{phase-space Hamiltonian} for asymptotic orbits as,
\begin{equation}
a_p \left( \bar p_1^0 \right)^2 + b_p \bar p_1^0+ c_p=0,
\label{R: quadratic wedge-strip}
\end{equation}
where $a_p$, $b_p$, and $c_p$ are found in Appendix \ref{app_ball} and depend on
$i=1,2$  for stable ($q_1^0=0$) and unstable ($p_1^0=0$) asymptotic orbits, respectively.
Thus, we can obtain the strips $S_i$ $(i=1,2)$ by taking the determinant, $\bar \Delta= b_p^2 - 4 a_p c_p$, of the quadratic equation \eqref{R: quadratic wedge-strip} to be zero (i.e., $\bar \Delta=0$) which are exactly the same expressions as that in \eqref{R: strip1} and \eqref{R: strip2}. 

For $\bar \Delta>0$, we obtain two real values for $\bar p_1^0$ as,
\begin{equation}
\bar p_1^0=\frac{-b_p \pm \sqrt{b_p^2 - 4 a_p c_p}}{2 a_p},
\end{equation}
and then the expression for $\bar p_2^0$ is obtained as,
\begin{equation}
\bar p_2^0=  \frac{\bar p_1^0 \lambda \left( 1+ c_x +\omega_p^2 \right)}{2 \left(1+ c_x \right)} + \frac{\left(1 + c_x -\omega_p^2 \right) \left[\bar q_2^0 \lambda + \left(1+ c_x \right) \bar q_1^0 \right]}{2 \left(1+ c_x \right)}.
\end{equation}
Therefore, the two initial velocities formed by the two asymptotic orbits can result in the wedge of velocity with wedge angle $\theta = \arctan \left(\bar p_2^0/ \bar p_1^0 \right)$.

Up to now, we have obtained the strips and wedge of velocity. In Figure \ref{R: conserfative confguration flow}, $S_1$ and $S_2$ are the two strips mentioned above. Outside of each strip $S_i \left(i=1,2 \right)$, the sign of $q_1^0$ and $p_1^0$ is independent of the direction of the velocity. These signs can be determined in each of the components of the equilibrium region $\mathcal{R}$ complementary to both strips. For example, in the left-most central components, $q_1^0$ is negative and $p_1^0$ is positive, while in the right-most central components $q_1^0$ is positive and $p_1^0 $ is negative. Therefore, $q_1^0 p_1^0 <0$ in both components and only non-transit orbits project onto these two components.

Inside the strips the situation is more complicated since the sign of $q_1^0 p_1^0$ depends on the direction of the velocity. For simplicity we have indicated this dependence only on the two vertical bounding line segments in Figure \ref{R: conserfative confguration flow}. For example, consider the intersection of strip $S_1$ with the left most vertical line. On this subsegment, 
there is at each point a wedge of velocity in which $q_1^0$ is positive. The sign of $p_1^0$ is always positive on this segment, so orbits with velocity interior to the wedge of velocity are transit orbits $\left(q_1^0 p_1^0 >0 \right)$. Of course, orbits with velocity on the boundary of the wedge are asymptotic $\left(q_1^0 p_1^0 =0 \right)$, while orbits with velocity outside of the wedge are non-transit $\left(q_1^0 p_1^0 <0 \right)$. In Figure \ref{R: conserfative confguration flow}, only one transit and one asymptotic orbit starting on this subsegment are illustrated. The situation on the remaining three subsegments is similar.

\subsubsection{Analysis in the dissipative system}
Recall that in the dissipative system of the rolling ball on a stationary surface the saddle projection and focus projection in the eigenspace of the conservative system (i.e., the symplectic eigenspace) are uncoupled. The transition is only determined by the location in the saddle projection and energy. However, when the surface is rotating, the situation is different. To compare the behavior in the different systems, we utilize the same change of variables as in \eqref{R: symp tranform}, i.e., $\bar z = C z$, and the equations of motion in the symplectic eigenspace are,
\begin{equation}
\dot z= \Lambda z + \Delta z,
\label{linear damping}
\end{equation}
where $\Lambda=C^{-1} M C$ from before, \eqref{Lambda_standard}, but the transformed damping matrix is now,
\begin{equation}
\Delta=C^{-1} DC = c_h K,  \label{K_matrix_transform}
\end{equation}
where $K$ is a $4 \times 4$ matrix with many non-zero  components, given in \eqref{K matrix} in Appendix \ref{app_ball}.

Notice that for the rolling ball on a stationary surface discussed in Section \ref{Dissipative system in eigenspace} and the dynamical buckling of a shallow arch \cite{zhong2018tube} in the dissipative system, the
canonical planes $\left(q_1,p_1\right)$  and $\left(q_2,p_2\right)$ have their dynamics uncoupled. 
Here, however, the dynamics on the $\left(q_1,p_1\right)$  and $\left(q_2,p_2\right)$ planes {\it are coupled} due to the combination of dissipative and gyroscopic forces. 
We see this coupling via several coupling terms which are no longer zero in \eqref{K matrix}, e.g., $K_{12}$, $K_{14}$, $K_{21}$ and $K_{23}$, etc. 
Because of the coupling between the $\left(q_1,p_1\right)$ and $\left(q_2,p_2\right)$ planes, it is difficult to obtain simple analytical solutions in the symplectic eigenspace variables. 
Thus, the semi-analytical method which substitutes all the parameters into the equations will be used to analyze the linear behavior near the saddle point.

One can obtain a fourth-order characteristic polynomial for the matrix $\Lambda  + \Delta$ from which to obtain eigenvalues. Here we denote the four eigenvalues as $\beta_1, -\beta_2, \beta_{3,4}=-\delta \pm i w_d$, where $\beta_1, \beta_2, \delta$ and $\omega_d$ are all positive real numbers. Note that the saddle $\times$ center type equilibrium point in the conservative system becomes a saddle $\times$ focus type equilibrium point in the dissipative system. The four corresponding generalized eigenvectors are denoted as $u_1$, $u_2$ and $u_3 \pm i u_4$, where $u_i$ are all real vectors. Thus, the general solutions to system \eqref{linear damping} can be expressed as,
\begin{equation}
z(t)=k_1^0 u_1 e ^{\beta_1 t} + k_2^0 u_2 e^{-\beta_2 t} + e^{-\delta t} \mathrm{Re} \left[ k_0 e^{-i \omega_d t} \left(u_3 - i u_4 \right) \right],
\end{equation}
where $k_1^0$ and $k_2^0$ are real and  $k_0=k_3^0 + i k_4^0$ is complex ($k_3^0$ and $k_4^0$ are real).

\paragraph{The flow in the equilibrium region}
Analogous to the discussion for the conservative system, we still choose the same equilibrium  region $\mathcal{R}$ determined by $\mathcal{H}_2= h$ and $|p_1 -q_1 | \leq c$ with positive $h$ and $c$. Due to the coupling between the saddle projection and focus projection, the behavior in the eigenspace is complicated. When $t \rightarrow +\infty$ and $t \rightarrow -\infty$, $z$ is dominated by the $k_1^0$ term and $k_2^0$ term, respectively. Thus, one can categorize the orbits into different groups  based solely on the signs of $k_1^0$ and $k_2^0$. However, the visualization of all the initial conditions for different types of orbits specified by a given energy is indirect. 
To do so, setting the initial conditions in the
symplectic
eigenspace as $z_0=(q_1^0, q_2^0, p_1^0, p_2^0)$, the following relation between the symplectic and dissipative eigenspace variables is obtained,
\begin{equation}
\begin{split}
\begin{pmatrix}
q_1^0\\ q_2^0 \\ p_1^0 \\ p_2^0
\end{pmatrix}
=
\begin{pmatrix}
\vdots & \vdots & \vdots & \vdots \\
u_1 & u_2 & u_3 & u_4 \\
\vdots & \vdots & \vdots & \vdots
\end{pmatrix}
\begin{pmatrix}
k_1^0 \\ k_2^0  \\ k_3^0  \\ k_4^0 
\end{pmatrix},
\label{R: damp initial}
\end{split}
\end{equation}
where the eigenvectors $u_i$ are written as column vectors.

As discussed for the conservative system, asymptotic orbits play an important role, acting as the separatrix of transit orbits and non-transit orbits. Moreover, the size of stable asymptotic orbits determines the amount of transit orbits. A straightforward method to obtain the stable asymptotic orbits, analogous to what was done for the conservative case, is as follows. For the stable asymptotic orbits, we have $k_1^0=0$. Then we can use \eqref{R: damp initial} to obtain $k_i^0$ $(i=2,3,4)$ and $p_2^0$ in terms of $q_1^0$, $q_2^0$ and $p_1^0$. Analogous to the situation for the conservative system in Section \ref{R: sec:separatrix}, we select the initial conditions on two sets $n_1$ and $n_2$ projecting to the line segments $p_1^0 = q_1^0 \pm c$. Substituting $p_2^0$ in terms of $q_1^0$, $q_2^0$ and $p_1^0$ and the relation $q_1^0 = p_1^0 \mp c$ into the Hamiltonian normal form \eqref{phase-space Hamiltonian}, we  rewrite it  in exactly the same form as in \eqref{R: quadratic wedge-strip}: $ a_p \left(p_1^0 \right)^2 + b_p p_1^0 + c_p=0$. Note that here $a_p$, $b_p$ and $c_p$ are functions of $q_{20}$ which are different to that in \eqref{R: quadratic wedge-strip}. To guarantee $p_1^0$ has real solutions, $\bar \Delta=b_p^2 - 4 a_p c_p>0$ should be true. Thus, we can obtain $q_{20}^{(l)}<q_2^0<q_{20}^{(u)}$, where $q_{20}^{(l)}$ and $q_{20}^{(u)}$ are the lower and upper bounds for $q_2^0$. For different $q_2^0 \in \left[q_{20}^{(l)}, q_{20}^{(u)} \right]$, we can obtain $p_1^0=\left(-b_p \pm \sqrt{b_p^2 -4 a_p c_p} \right)/(2 a_p)$ and thus obtain $q_1^0$ and $p_2^0$.

\paragraph{Null space method}
Another method to obtain the stable asymptotic orbits, here called the \textbf{null space method}, can also be utilized. The procedure is as follows: (1) using three generalized eigenvectors corresponding to the eigenvalues with negative real part (i.e., $u_2$, $u_3$, $u_4$), the null space of the stable eigenspace, $E^s=\mathrm{span} \{u_2, u_3, u_4 \}$, can be obtained, denoted as $u_n=\left(u_{n1}, u_{n2},  u_{n3}, u_{n4} \right)^T$, with the relation $u_n \cdot u_{i}=0$ $(i=2,3,4)$; (2) Since the initial conditions $z_0$ of forward asymptotic orbits (i.e., stable asymptotic orbits) should be normal to the null space, we have $u_n \cdot z_0=0$, which, along with the Hamiltonian function, will give the same quadratic equation, $ a_p \left(p_1^0 \right)^2 + b_p p_1^0 + c_p=0$; (3) following the same manipulation as described in the previous paragraph, we obtain the same results.

\begin{figure}[!t]
	\begin{center}
		\includegraphics[width=\textwidth]{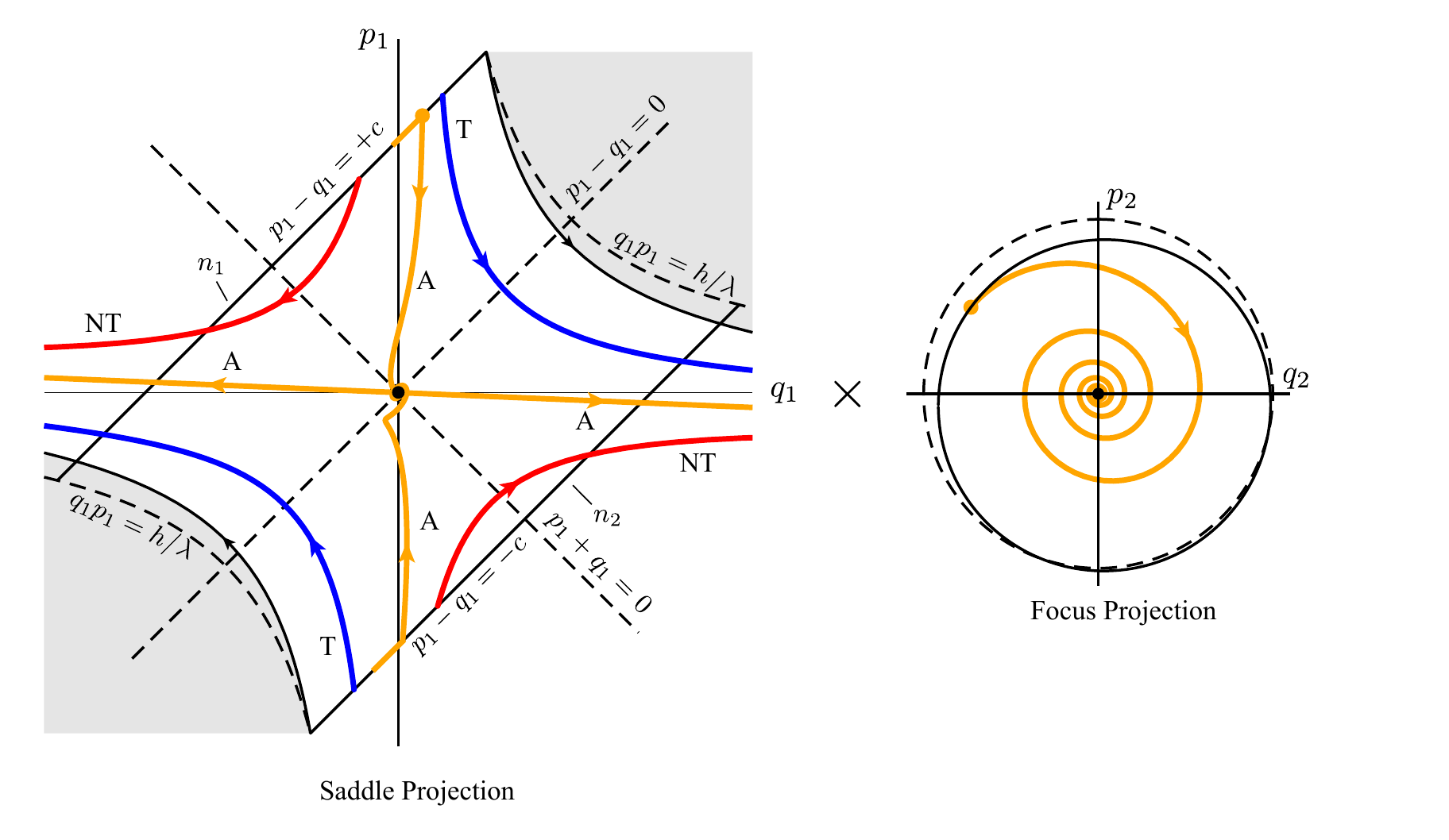}
	\end{center}
	\caption{{\footnotesize 
			The flow in the equilibrium region $\mathcal{R}$ projected onto $(q_1,p_1)$ plane and $(q_2,p_2)$ plane which are coupled has form saddle $\times$ focus. Shown are the saddle-type asymptotic orbits (labeled A), transit orbits (labeled T) and non-transit orbits (labeled NT). The dot at the origin of $(q_1,p_1)$ plane is the focus-type asymptotic orbits with projection only on $(q_2,p_2)$ plane which is a damped oscillator decaying to the origin. Due to the coupling between $(q_1,p_2)$ plane and $(q_2,p_2)$ plane, the initial conditions for the three-dimensional stable asymptotic orbit are dependent on the angle in focus projection. The one dimensional unstable asymptotic orbits are a straight line in the saddle projection.
	}}
	\label{R: dissipative eigenspace}
\end{figure}

\paragraph{Flow in the equilibrium region}
Different combinations of the signs of $k_1^0$ and $k_2^0$ gives nine classes of orbits which can be grouped into the same four categories as the dissipative system of the rolling ball on a stationary surface. All initial conditions on the bounding lines $n_1$ and $n_2$ for different types of orbits can be visualized based on the analysis listed below.
\begin{enumerate}
	\item Orbits with $k_1^0=k_2^0=0$ corresponds to a focus-type asymptotic orbit with motion in the $\left(q_2,p_2\right)$ plane (see black dot at the origin of the $\left(q_1,p_1 \right)$ plane in Figure \ref{R: dissipative eigenspace}). Due to the effect of energy dissipation, the periodic orbit does not exist.
	
	\item Orbits with $k_1^0 k_2^0=0 $ are saddle-type asymptotic orbits. For example, the bolded orange line on the bounding line $n_1$ in the saddle projection associated with the closed solid curve in the focus projection in Figure \ref{R: dissipative eigenspace} are all the initial conditions for the stable asymptotic orbits with initial conditions of initial energy $h$ on $n_1$. Because of the coupling between the saddle projection and focus projection, one point on the closed solid curve in the focus projection has a corresponding point on the bolded region in saddle projection which together give the initial condition for a specific asymptotic orbit of initial energy $h$. See the orange dots for the initial condition of the stable asymptotic orbit starting from $n_1$ and orange curve for the evolution. Of course, the bounding line $n_2$ has the behavior for the stable asymptotic orbits. Since the system just has one positive eigenvalue, the unstable asymptotic orbits just have one specific direction along each side of the saddle point. See the orange straight lines for the unstable asymptotic orbits. Four asymptotic orbits are shown in Figure \ref{R: dissipative eigenspace} labeled A.
	
	\item The segments determined by $k_1^0 k_2^0 <0 $ which cross $\mathcal{R}$ from the bounding line $n_1$ to the bounding line $n_2$ in the northern hemisphere, and vice versa in the southern hemisphere, correspond to the transit orbits with initial energy $h$ on $n_1$ and $n_2$, respectively. See the two example orbits labeled T of Figure \ref{R: dissipative eigenspace}.
	
	\item Finally the segments with $k_1^0 k_2^0 >0 $ which start from one hemisphere and bounce back are the non-transit orbits of initial energy $h$. See the two orbits labeled NT in Figure \ref{R: dissipative eigenspace}.
\end{enumerate}

\begin{figure}[!t]
	\begin{center}
		\includegraphics[width=\textwidth]{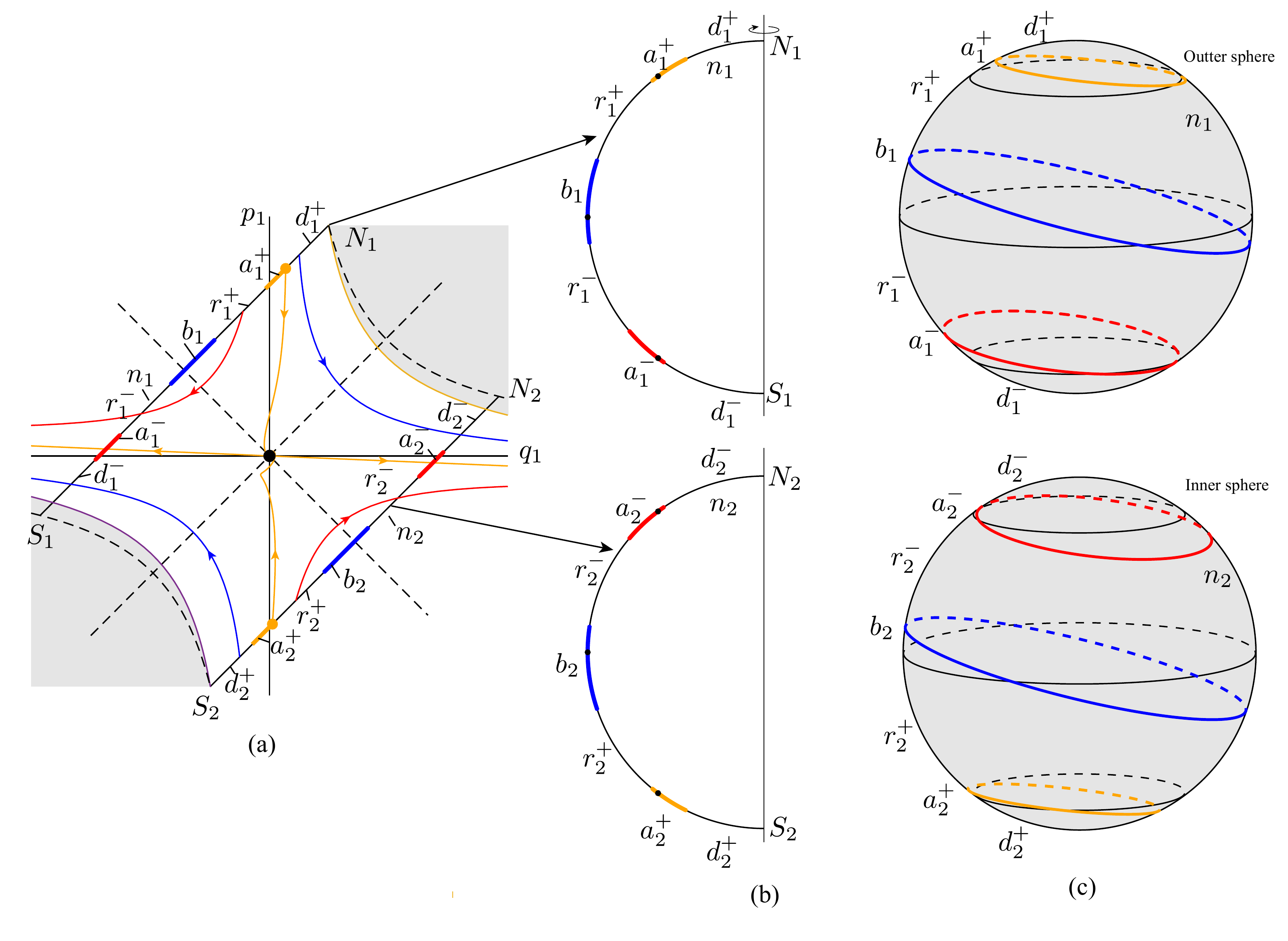}
	\end{center}
	\caption{{\footnotesize 
			The McGehee representation of the equilibrium region for dissipative system of rolling ball on a rotating surface. 
			(a) The projection of flow onto $(q_1,p_1)$ plane. 
			(b) The projection of the flow in the $\mathcal{R}$ region of the energy surface onto a cross-section. 
			(c) The McGehee representation of the flow on the boundaries of the $\mathcal{R}$ region, highlighting the features on the bounding spheres $n_1$ and $n_2$, the ``inner'' and ``outer'' spheres, respectively.
	}}
	\label{R: McGehee}
\end{figure}

\paragraph{McGehee representation}
The previous section gives the topological structure of initial conditions for different types of orbits in the dissipative system, but it still may not be intuitive. Thus, as we did in the rolling ball on a stationary surface, we introduce the McGehee representation to visualize the region $\mathcal{R}$ for easier interpretation. Since there are many curves on the two 2-spheres, $n_1$ and $n_2$, of initial energy $h$, we show the two spheres separately in Figure \ref{R: McGehee}(c).

As mentioned in the ball rolling on a stationary surface with damping, the McGehee representation gives the spheres with the same energy $h$ so that here the McGehee representation again just shows the initial conditions on each bounding sphere. The symbols in Section \ref{Conservative McGehee representaion} have the same meaning as used here. The previous four categories of orbits are interpreted as follows.
\begin{enumerate}

	\item There is a 1-sphere $S_h^1$ in the region $\mathcal{R}$, similar to that in the rolling ball on a stationary surface with dissipation, which is the equator of the 2-sphere given by $p_1-q_1=0$. The set $S_h^1$  gives the initial conditions for the focus-type asymptotic orbits with initial energy $h$. Readers are referred to the dot in Figure \ref{B: McGehee representation}(b) for interpretation. 
	
	\item There are two 1-spheres represented by the orange closed curves on each bounding sphere, denoted by $a^+_i$ and $a^-_i$ on sphere $n_i$. They give the initial conditions for stable asymptotic orbits. Compared to the ball rolling on a stationary surface, which has initial conditions for stable asymptotic orbits given by circles on the bounding spheres parallel to the corresponding equators, initial conditions for stable asymptotic orbits for the rotating surface are tilted. 
	This is due to {\it dissipation-induced coupling} of the saddle and focus projections of the symplectic eigenspace. 
	Note that the unstable asymptotic orbits are one-dimensional and  have different energy from the bounding sphere so that they cannot be given in the McGehee representation.
	
	\item Consider the two spherical caps on each bounding 2-sphere denoted by $d_1^{+}$, $d_1^{-}$ and  $d_2^{+}$, $d_2^{-}$. The transit orbits with initial conditions on spherical cap $d_1^+$, which is in $n_1^+$ and bounded by $a_1^+$, enter $\mathcal{R}$ and leave through $n_2$ at a different (lower) energy, due to dissipation.  On the other hand, the transit orbits with initial conditions on spherical cap $d_1^-$ in $n_1^-$ bounded by $a_1^-$ are leaving $\mathcal{R}$ having entered through $n_2$ at a different (higher) energy. An analogous situation holds on bounding sphere $n_2$.
	
	\item There is a 1-sphere of tangency points, denoted by $b$, with initial conditions on which the orbits do not enter $\mathcal{R}$ locally. To obtain the tangency points, first we need to the recognize the relation along each angle $\theta$ in the focus projection, i.e. $p_2^0 =q_2^0 \tan \theta$,  as well as the initial conditions on the bounding spheres $n_1$ and $n_2$, i.e. $p_1^0=q_1^0 \pm c$, and the tangency conditions, i.e. $\dot p_1^0=\pm \dot q_1^0$.  We then substitute such relations into the Hamiltonian normal form to yield a quadratic equation which will give two tangency points along that angle. Note that the 1-spheres here are not the equators of the bounding spheres, as they are in the non-rotating case, but are tilted by an angle compared with the conservative system, again, due to the coupling via the dissipation matrix $K$, from \eqref{K matrix}. The topological hemisphere above $b_1$ in $n_1$ is referred to as $n_1^+$ and below $b_1$ as $n_1^-$;  similarly for $n_2$, as illustrated in Figure \ref{R: McGehee}(c). Similar to before, the non-transit orbits with initial conditions of initial energy $h$  
	on spherical zone $r^+_i$, in $n^+_i$ bounded by $a^+_i$ and $b_i$, are entering $\mathcal{R}$ and  non-transit orbits with initial conditions 
	on spherical zone $r^-_i$,  in $n^-_i$ bounded by $a^+_i$ and $b_i$, are leaving $\mathcal{R}$.

\end{enumerate}

\paragraph{Trajectories in the equilibrium region}
Following the standard procedure to solve  
\eqref{linear_ball}, we  get the eigenvalues of the matrix $M + D$ (denoted as as $\bar \beta_1, - \bar \beta_2, \bar \beta_{3,4} = -\bar \delta \pm i \bar w_d$, where $ \bar \beta_1, \bar \beta_2, \bar \delta$ and $\bar \omega_d$ are positive real values) 
associated with the corresponding generalized eigenvectors (denoted as $\bar u_i \hspace{0.1in} (i=1,2,3,4)$). The general real solutions to \eqref{linear_ball} are, 
\begin{equation}
\bar z(t)=\bar k_1^0 \bar u_1 e ^{\bar \beta_1 t} + \bar k_2^0 \bar u_2 e^{-\bar \beta_2 t} + e^{-\bar \delta t} \mathrm{Re} \left[ \bar k_0 e^{-i \bar \omega_d t} \left(\bar u_3 - i \bar u_4 \right) \right],
\end{equation}
where $\bar k_1^0$ and $\bar k_2^0$ are real and  $\bar k_0=\bar k_3^0 + i \bar k_4^0$ is complex. By inspecting the limiting behavior of $\bar q_1$ as $t$ tends to plus or minus infinity, we can also obtain the following four categories of orbits:
\begin{enumerate}
	\item Orbits with $\bar k_1^0=\bar k_2^0=0$ are focus-type asymptotic orbits. When dissipation is considered in the system, the periodic orbit does not exist, but these initial conditions correspond to purely focus-like dynamics, with an amplitude decreasing proportional to $e^{-\bar \delta t}$.
	
	\item Orbits with $\bar k_1^0=0$ ( or $\bar k_2^0=0$) are stable (or unstable) saddle-type asymptotic to the saddle equilibrium point.
	
	\item Orbits with  $\bar k_1^0 \bar k_2^0>0$ are transit orbits.
	
	\item Orbits with $\bar k_1^0 \bar k_2^0<0$ are non-transit orbits.
	
\end{enumerate}

\paragraph{Wedge of velocity and ellipse of transition} 
As discussed previously, 
for this rotating system one also obtains an {\it ellipse of transition} which confines the existence of transit orbits. 
Inside the ellipse, the transit orbits exist, while outside the ellipse transit orbits do not exist. 
As before, a non-empty 
wedge of velocity, which divides the transit orbits from the non-transit orbits, 
can only exist inside the ellipse of transition. 

Taking $t=0$, one obtains the relation between the initial conditions $\bar z_0=\left(\bar q_1^0, \bar q_2^0, \bar p_1^0, \bar p_2^0 \right)$ and the coefficients $\bar k_i^0$ with a similar form as in \eqref{R: damp initial}. For stable asymptotic orbits, i.e., $\bar k_1^0=0$, we can determine the coefficients $\bar k_i^0 (i=2,3,4)$ and $\bar p_{20}$ in terms of initial conditions $\bar q_{10}, \bar q_{20}, \bar p_{10}$. With the substitution of $\bar q_1^0, \bar q_2^0,\bar p_1^0$ and $\bar p_2^0$ into \eqref{R:rescale Hamiltonian}, the quadratic Hamiltonian \eqref{R:rescale Hamiltonian} restricted by energy $h$ can be rewritten as a second order algebraic equation for $\bar p_1^0$ which has exactly the same form as \eqref{R: quadratic wedge-strip}, but with different $a_p, b_p$ and $c_p$ in terms of $\bar q_1^0$ and $\bar q_2^0$. On the one hand, for the critical condition, i.e., $\vartriangle=b_p^2 - 4 a_p c_p=0$, we can obtain an ellipse of transition which is different from the strips $S_1$ in the conservative system. The ellipse limits the location of transit orbit initial conditions.
 On the other hand, when the determinant satisfies $\vartriangle=b_p^2 - 4 a_p c_p>0$, $\bar p_1^0$ has two real solutions, $\bar p_1^0= \left(-b_p \pm \sqrt{b_p^2 - 4 a_p c_p}\right)/\left(2 a_p\right)$, associated with two real solutions for $\bar p_2^0$. Thus, the two pairs of results of $\left(\bar p_1^0, \bar p_2^0 \right)$ will determine two bounding directions of velocity, which form the wedge of velocity.

\begin{figure}[!t]
	\begin{center}
		\includegraphics[width=5in]{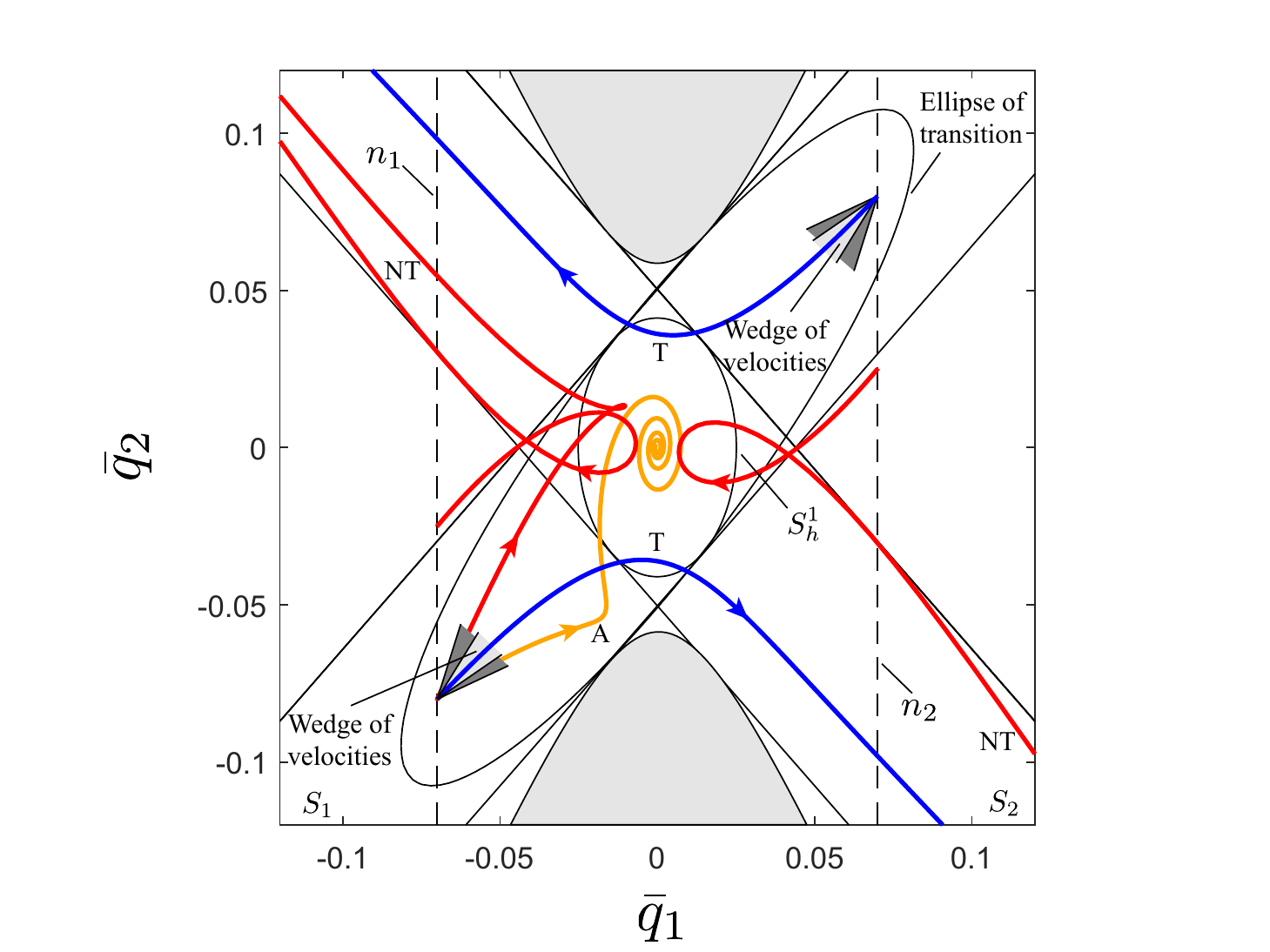}
	\end{center}
	\caption{{\footnotesize 
			The flow in the equilibrium region $\mathcal{R}$ of position space. Shown are the saddle-type asymptotic orbit; two transit orbits; three non-transit orbits. For the same given energy, the wedges of velocity for the dissipative system (the smaller light grey shaded wedges), restricted by the ellipse of transition, partially cover the wedges of velocity for the conservative system (the larger dark grey shaded wedges) restricted by a strip.
	}}
	\label{R: dissipative position space}
\end{figure}

Figure \ref{R: dissipative position space} shows the flow in the projection of the equilibrium region $\mathcal{R}$ to position space, taking gyroscopic and dissipative effects into consideration. Due to energy dissipation, the strips which are the boundaries of asymptotic orbits in the position space of the conservative system no longer exist. In particular, the strip for the stable asymptotic orbit is replaced by the ellipse of transition. The ellipse of transition, similar to the role in the rolling ball on a stationary surface, confines the existence of transit orbits. That is,  transit orbits of a given initial energy $h$ must have initial conditions inside the ellipse of transition, while only non-transit orbit initial conditions project onto the  area complementary of the ellipse. However, even if the initial condition of an orbit projects to a position inside the ellipse of transition, this alone does not guarantee the transition. This is a necessary but not a sufficient condition. The additional condition is that the velocity should be along certain directions. The wedge of velocity obtained above, which is non-empty only inside the ellipse, is exactly the condition providing the correct range of directions for the velocity of transit orbits. Orbits with initial conditions interior to the wedge can transit, while orbits with velocity outside the wedge cannot transit. The orbits with velocity on the boundary of the wedge are asymptotic to the equilibrium point.

The sizes of the wedge of velocity and ellipse of transition, which both represent the proportion of transit orbits compared to non-transit orbits, are dependent on the energy, $h$, and the amount of damping, $c_h$. An increase of energy gives more transit orbits, while an increase in damping reduces the proportion of transit orbits.  Furthermore, different positions inside the ellipse have different sizes of wedges of velocity. The closer the position is to the boundary of the ellipse of transition, the smaller the size of the wedge of velocity will be. From Figure \ref{R: dissipative position space}, we  find that the size of the wedge shrinks (light grey) compared with that of the conservative system (dark grey) which qualitatively indicates how  damping affects the wedge of velocity.

\subsubsection{Transition tube and transition ellipsoid}
We have discussed the flow in the position space for a rolling ball on a rotating surface near a saddle point.
 In this section, we will visualize the  structures governing transitions in the phase space, particularly on surfaces of constant initial energy $h$.

\begin{figure}[!t]
	\begin{center}
		\includegraphics[width=\textwidth]{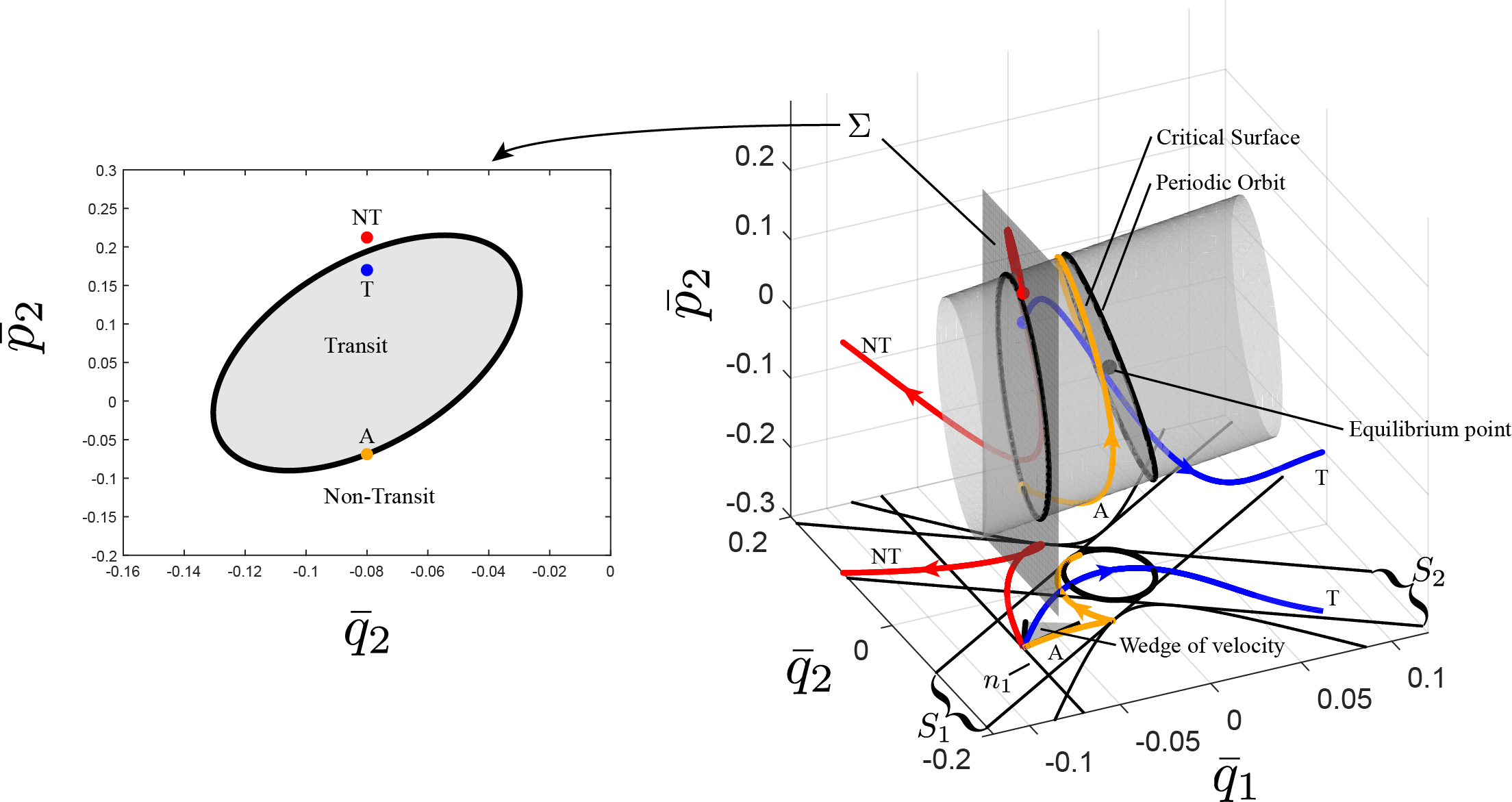}
	\end{center}
	\caption{{\footnotesize 
			Transition tube for the ball rolling on a rotating surface. The left figure gives the region for initial conditions of transit orbits on the Poincar\'e section $\Sigma$ with three initial conditions (the dots) for three types of orbits. The right shows the transition tube for a given energy. The critical surface playing the same role as the ball rolling on a stationary surface also exists here. Three types of orbits with initial conditions on the left figure are given.
	}}
	\label{R: transition tube}
\end{figure}

\begin{figure}[!t]
	\begin{center}
		\includegraphics[width=\textwidth]{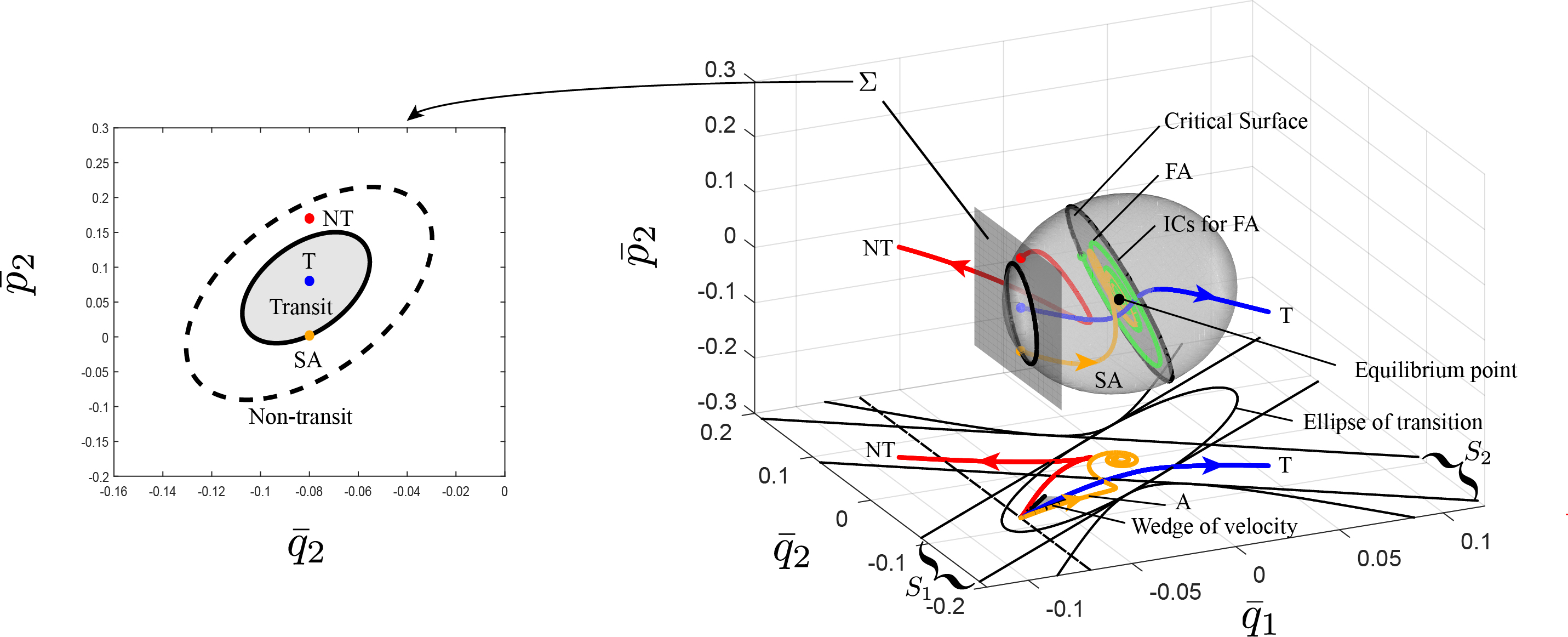}
	\end{center}
	\caption{{\footnotesize 
			Transition ellipsoid for the dissipative system in the rolling ball on a rotating surface. The left figure shows the Poincar\'e section $\Sigma$, where the dots are the initial conditions for the corresponding trajectories and the solid ellipse is the boundary of initial conditions for the transit orbits. For comparison, the dashed ellipse of tube boundary for the conservative system is also given. For The right figure shows the transition tube with three different types of orbits with initial conditions on the left figure.
	}}
	\label{R: transition ellipsoid}
\end{figure}

For the rolling ball on a stationary surface, we obtained the transition tube and transition ellipsoid that give all the initial conditions, starting at a given initial energy $h$, of transit orbits for the conservative system and dissipative system, respectively. In the current problem, we have similar phase space structures governing the transition which can be obtained by the semi-analytical method mentioned before. Figure \ref{R: transition tube} and Figure \ref{R: transition ellipsoid} show the transition tube and transition ellipsoid, respectively. As discussed in Section \ref{B: tube and ellipsoid}, for a specific energy all transit orbits in the conservative system and dissipative system must have initial conditions inside the transition tube and transition ellipsoid, respectively; all orbits with initial conditions outside the transition tube and transition ellipsoid are non-transit orbits. Furthermore, the critical surface divides the transition tube and transition ellipsoid into two parts. Orbits with initial conditions inside the left part will transit to the right  and orbits with initial conditions inside the right part will transit to the left. Orbits on the boundary are asymptotic to the periodic orbit (respectively, equilibrium point) in the  conservative (respectively, dissipative) system. Transit orbits can cross the critical surface, while non-transit orbits will bounce back before reaching the critical surface.

Figure \ref{R: transition tube} and Figure \ref{R: transition ellipsoid} also give different types of orbits with the initial conditions on the same Poincar\'e section in the corresponding system. This illustrates the discussion given above that transit orbits must have initial
conditions inside the transition tube or transition ellipsoid. In fact, the transit orbit (initial condition marked T in Figure \ref{R: transition tube}) in the conservative system and the non-transit orbit (initial condition marked NT in Figure \ref{R: transition ellipsoid}) in the dissipative system have the same initial condition. This demonstrates that initial conditions corresponding to a transit orbit in the conservative system may be non-transit orbits if damping is taken into account.

It is worth noting that the topological structures in phase space controlling the transition for both the non-rotating system and rotating system are almost the same. Nevertheless,  differences exist between these two systems. In the non-rotating system, the axes of the transition tube and transition ellipsoid are parallel to the position space axes, while in the rotating system,  the axes of the transition tube and transition ellipsoid is not parallel to the position space axes, but are tilted by an angle.

\subsection{Ship motion with unequal damping}
\label{Ship motion with unequal damping}
In Section \ref{Ship with equal damping}, we derived the equations of motion of a ship considering the coupled roll-pitch motion with damping.  
When the coefficients of damping happen to be equal, the equations of motion projected into the symplectic eigenspace have uncoupled saddle projection and focus projection dynamics. However, the situation becomes  more complicated if unequal damping is considered, since the saddle and focus projections in the symplectic eigenspace become coupled. This case is similar to the rolling ball on a rotating surface with dissipation. Since the analysis has been given in Section \ref{rotating surface}, here we simply give the equations of motion with unequal damping transformed into the symplectic eigenspace. 

The equations of motion linearized about the saddle point, \eqref{eq_ship_matrix}, under the same symplectic transformation as described in Section \ref{Ship with equal damping}, becomes,
\begin{equation}
\dot z=\Lambda z + \Delta z,
\end{equation}
where $\Lambda=C^{-1} M C$ is the standard matrix of the conservative part of the dynamics \eqref{Lambda_standard}, but the transformed damping matrix is,
\begin{equation}
\Delta=\Delta_{1} + \delta c_h \tilde{\Delta},
\end{equation}
where $\Delta_1$ is 
the standard uncoupled damping matrix \eqref{standard_damping_matrix}, with $c_h=c_{h_1}$, $\delta c_h = c_{h_2} - c_{h_1}$ is the difference between the two damping coefficients, and $\tilde{\Delta}$ is given in \eqref{tilde_Delta}.
The matrix  $\tilde{\Delta}$ contains  terms which couple the dynamics on the two canonical planes, $(q_1,p_1)$ and $(q_2,p_2)$, 
and this coupling vanishes only when $\delta c_h=0$, i.e., when the two damping coefficients are equal.  
See Appendix \ref{app_ship} for details.

\subsection{The restricted three-body problem with dissipation}

The restricted three-body problem is a classic problem which has attracted a lot of attention for the study of escape 
\cite{jaffe2002statistical,KoLoMaRo2000,KoLoMaRo2011,FuHe2000,ViSc2003,ross2003statistical,astakhov2003chaos,astakhov2004capture,WaBuWi2005,DeJuKoLeLoMaPaPrRoTh2005,DeJuLoMaPaPrRoTh2005,RoSc2007,GaMaDuCa2009,hasnain2012capturing,de2014escape,Onozaki2017,NaikLekienRoss2017}. 
Deeper understanding of the escape from a gravitational potential well due to multiple bodies can guide the design of trajectories for interplanetary space missions and also aid analysis of certain astronomical phenomena (e.g., galactic dynamics \cite{fukushige2000time,romero2007formation,athanassoula2009rings,athanassoula2010rings,athanassoula2012manifold,jung2016orbital,jung2016orbital2}).
When dissipation is taken into account, the situation becomes more complicated, but becomes applicable to a larger number of situations, such as drag due to interplanetary dust \cite{MuDe1999,DeGrDuJaKeKoWy2001}, and engineering applications, such as low thrust \cite{dellnitz2006target} and solar sails \cite{baoyin2006solar,mcinnes2013solar}. 

\begin{figure}
	\begin{center}
		\includegraphics[width=5in]{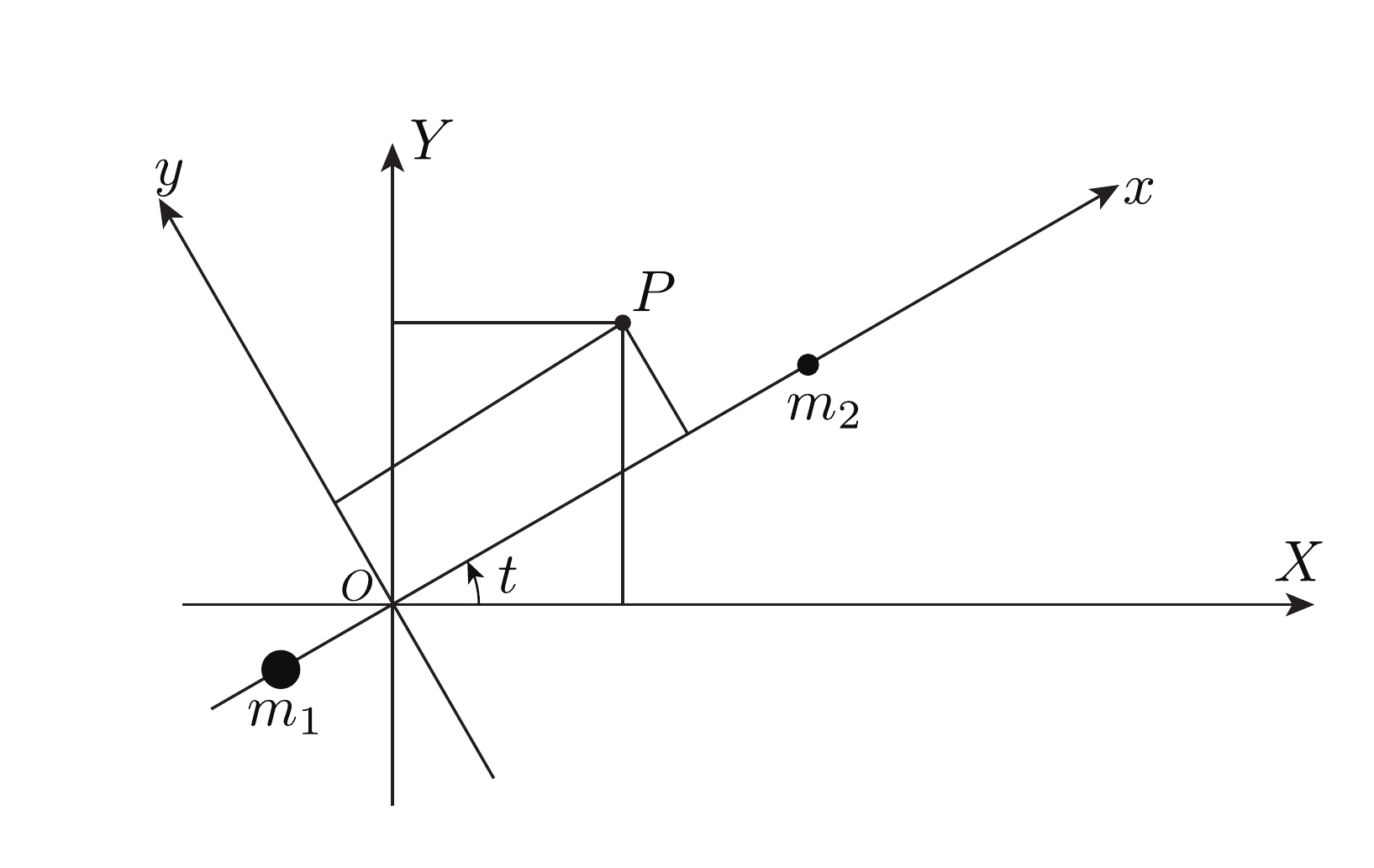}
	\end{center}
	\caption{\label{3bp_coordinate}{\footnotesize 
Inertial and rotating frames in restricted three-body problem. The rotating coordinate system of $x$ and $y$ axes moves counterclockwise with unit angular velocity relative to the inertial frame with $X$ and $Y$ axes.
	}}
\end{figure}

We formulate the planar circular restricted three-body problem (PCR3BP) using a standard approach \cite{Szebehely1967,MuDe1999,KoLoMaRo2011}. 
Without loss of generality, the total mass of the two main bodies are normalized to unity, denoted by $m_1=1- \mu$ and $m_2=\mu$, respectively, where $\mu$ is  the non-dimensional mass parameter. Masses $m_1$ and $m_2$ orbit in a plane counterclockwise in circles about their common center of mass with angular velocity also normalized to one. A third body (spacecraft or small object) denoted by $P$, whose mass is ignored as too small to influence $m_1$ or $m_2$ significantly, can move freely in the $m_1$-$m_2$ orbital plane under the effect of their combined gravitational field. Denote the position of $P$ by $(x,y)$ in a co-orbiting or {\it rotating frame} whose $x$-axis coincides with the line connecting the two main bodies, with the origin at the center of mass, as shown in Figure \ref{3bp_coordinate}. 

From the Lagrangian point of view, the non-dimensional equations of motion of $P$ are, 
\begin{equation}
\ddot x -  2 \dot y=-\mathcal{U}_{,x} + Q_x, \hspace{0.2in} 
\ddot y + 2 \dot x=-\mathcal{U}_{,y} + Q_y,
\end{equation}
where the (effective) potential energy, which includes both gravitational and centrifugal forces, is,
\begin{equation*}
\mathcal{U}(x,y)=-\tfrac{1}{2}(x^2 + y^2) - \frac{1-\mu}{r_1} - \frac{\mu}{r_2} - \tfrac{1}{2}\mu (1-\mu),
\end{equation*}
where $r_1 = \sqrt{(x + \mu)^2 + y^2}$ and $r_2=\sqrt{(x-1+\mu)^2 + y^2}$ are the distances of $P$ from $m_1$ and $m_2$, respectively. The generalized forces $Q_x$ and $Q_y$ are the components of damping along $x$ and $y$, respectively. After applying the Legendre transformation to the Lagrangian formulation, the Hamiltonian function is, 
\begin{equation}
\mathcal{H}=\tfrac{1}{2} \left[\left(p_x + y \right)^2 + \left(p_y - x \right)^2 \right] +\mathcal{U}(x,y),
\label{T: full Hamiltonian}
\end{equation}
which yields the following Hamilton's equations,
\begin{equation}
\begin{aligned}
\dot x &= p_x +y,\\
\dot y&=p_y -x,\\
\dot p_x&= 	p_y - x - \mathcal{U}_{,x} + Q_x,\\
\dot p_y&= - 	p_x - y - \mathcal{U}_{,y} + Q_y.
\label{T: Nonlinear Hamiltonian ODEs}
\end{aligned}
\end{equation}

\paragraph{Conservative system} 
For the conservative system, we have $Q_x=Q_y=0$. In this case, the system has five equilibrium points, three collinear equilibria on the $x$-axis, labeled $L_1$, $L_2$, $L_3$ and two equilateral points labeled $L_4$ and $L_5$. 
These equilibrium points are critical points of the (effective potential) function $\mathcal{U}$. 
In this study, we focus on the behavior of particle trajectories near the two {\it Lagrange 
points}, $L_1$ and $L_2$, on either side of the (smaller) secondary mass, $m_2$. Here $L$ is used to denote either point $L_1$ or $L_2$. To find the linearized equations around the collinear libration point $L$ with coordinates $(x_e,y_e,p_{xe},p_{ye})=(x_e,0,0,x_e)$, we do the following. After making a coordinate change with $(x_e,0,0,x_e)$  as the origin, the quadratic terms 
form a Hamiltonian function for the linearized equations, given by \cite{KoLoMaRo2011}, 
\begin{equation}
\mathcal{H}_2=\tfrac{1}{2} (p_x^2 + p_y^2) + y p_x - x p_y - \bar \mu  x^2 + \tfrac{1}{2} \bar \mu  y^2, 
\label{H2_pcr3bp}
\end{equation}
where the only parameter, $\bar \mu >1$, 
is defined by, 
\begin{equation}
\bar \mu = \mu |x_e -1 + \mu|^{-3} + (1-\mu) |x_e + \mu|^{-3}.
\end{equation}

A short computation gives the linearized equations in the canonical Hamiltonian form,
\begin{equation}
\begin{aligned}
\dot x&=p_x + y,\\
\dot y&=p_y -x,\\
\dot p_x&=2 \bar \mu  x +p_y,\\
\dot p_y&=-\bar \mu  y -p_x.
\end{aligned}\label{pcr3bp_linear_cons}
\end{equation}
Following the same procedure as other problems, one can find the change of variables via a symplectic matrix $C$ given by \eqref{C matrix PCR3BP},
which transforms the quadratic Hamiltonian \eqref{H2_pcr3bp} into the  simple form  \eqref{phase-space Hamiltonian Equations}.
See Appendix \ref{app_pcr3bp} and \cite{KoLoMaRo2011} for details.
The dynamic behavior near the equilibrium in both the symplectic eigenspace and configuration space are the same as that of the rolling ball on a rotating surface from Section \ref{rotating surface}.

\paragraph{Dissipative system} 
In this part, we consider the linearization around the point $L$ with damping included. As is mentioned in Section \ref{rotating surface}, there are two types of damping in a rotating system, internal damping and external damping. Internal damping, proportional to the relative velocity in the rotating frame, is also called simple nebular drag \cite{murray1994dynamical} in celestial mechanics. To obtain the equilibrium point, one seeks the set corresponding to $\dot x=\dot y=0$. Thus, one can conclude the equilibrium point $L$ does not shift compared with the conservative system. Since its dynamical behavior is similar to those discussed in the rolling ball on a rotating system with internal damping, readers can refer to Section \ref{rotating surface} for  details. 

\begin{figure}[!t]
	\begin{center}
		\includegraphics[width=\textwidth]{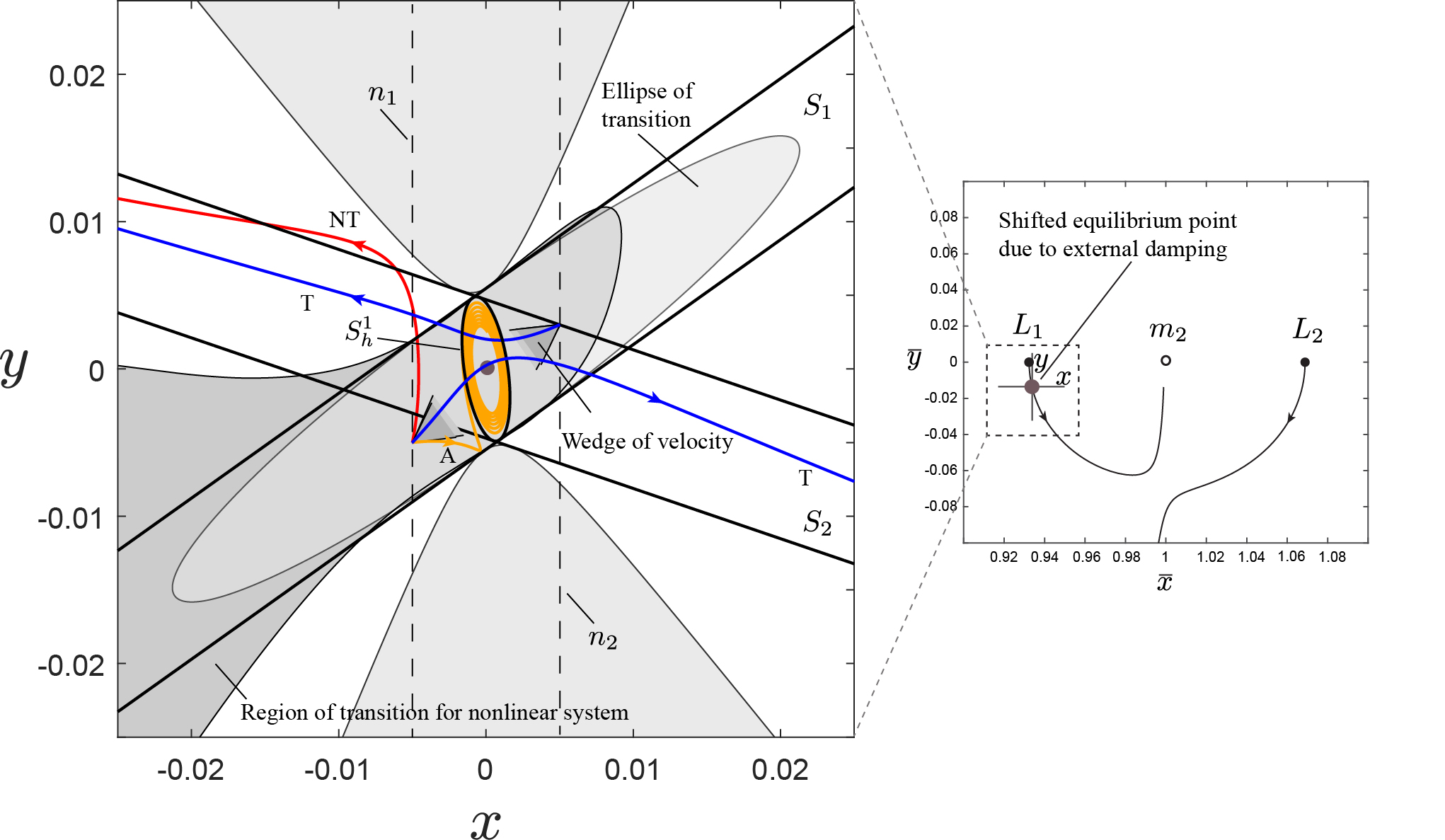}
	\end{center}
	\caption{{\footnotesize 
			In the right panel, we show the original rotating axes with bars, $(\bar x, \bar y)$, for 
			the PCR3BP with external damping, where here the $\bar x$-axis is along the $m_1$-$m_2$ line.  
			We also show the curve of equilibrium point positions  $L_1$ and $L_2$ as a function of the damping coefficient $c_h$, starting at $c_h=0$ for the black dots on the $\bar x$-axis, with $c_h$ increasing away from these points.
			In the left panel, we show 
			the flow in the equilibrium region $\mathcal{R}$ projected onto position space 
			$(x,y)$, which are the  displacement from the now-shifted $L_1$ equilibrium point (gray dot in both panels). 
			Shown are a saddle-type asymptotic orbit (labeled A); two transit orbits (labeled T) and a non-transit orbit (labeled NT). The larger light grey shaded ellipse and the dark grey shaded region are boundaries of initial conditions of all possible transit orbits for linearized system and nonlinear system with equal quadratic terms whose common area is the medium gray shaded region. The smaller ellipse close to the origin is periodic orbit and the two strips, labeled $S_1$ and $S_2$, are the boundary of asymptotic orbits in effective conservative system which does not exist physically. In this figure, the Sun-Jupiter system is used with mass parameter $\mu=9.537 \times 10^{-4}$; the damping coefficient is taken as $c_h=0.05$.
	}}
	\label{T: dissipative position space}
\end{figure}

On the other hand, external damping, proportional to the velocity with respect to the inertial frame, is also called the inertial drag force \cite{murray1994dynamical}. In this case, generally the equilibrium points will shift. Ref.\ \cite{murray1994dynamical} gives some approximate formulas to determine the location of the equilibrium point $L$ for systems with sufficiently small drag forces.  In general, for small drag, the equilibrium point $L$ keeps $x_e$ approximately constant, but $y_e$ moves away from zero, becoming increasingly negative with increasing $c_h$. 
For a large drag force, numerical tools have to be used to find the equilibrium points. 
See Figure \ref{T: dissipative position space}, the right panel, for a curve, a 1-parameter family, of how the equilibria corresponding to $L_1$ and $L_2$ shift for increasing $c_h$.
We note that there are other types of drag forces, like solar radiation pressure, Poynting-Robertson drag,  solar wind drag \cite{beauge1994capture,liou1995radiation}, and Stokes drag \cite{jain2015study}. For internal and external linear damping in the non-dimensionalized PCR3BP problem, the formulas are generally, 
\begin{equation}
\begin{split}
Q_x^{\rm int} &= -c_h\dot x, \\
Q_y^{\rm int} &= -c_h\dot y,
\end{split}
\hspace{0.1in} \text{ for internal damping}
\end{equation}
and,
\begin{equation}
\begin{split}
Q_x^{\rm ext} &= -c_h( \dot x  -  y ), \\
Q_y^{\rm ext} &= -c_h( \dot y +  x ),
\end{split}
\hspace{0.1in} \text{ for external damping}
\label{inter-exter damping2}
\end{equation}
%
where $c_h$ is the non-dimensional coefficient of damping.

For simplicity, only  external damping will be considered in the numerical computations that follow, as it leads to new phenomena. We focus on  the equilibrium point $L_1$.  We denote the shifted equilibrium point by
$(x_e,y_e,-y_e,x_e)$ , where $y_e$ is now non-zero, and shifted from $L_1$ in the conservative system. Linearizing the Hamilton's equations in \eqref{T: Nonlinear Hamiltonian ODEs} about the equilibrium point results in the following linearized equations in Hamiltonian form,
\begin{equation}
\begin{aligned}
\dot x&= p_x + y,\\
\dot y&=p_y - x,\\
\dot p_x&= b_1 x + b_2 y + p_y - c_h p_x,\\
\dot p_y&= b_2 x + b_3 y- p_x -  c_h p_y,
\label{T: Linearization in dissipative}
\end{aligned}
\end{equation}
where $(x,y)$ are now the small displacement from the shifted equilibrium $(x_e,y_e)$ with $(p_x,p_y)$ the conjugate momenta.  The matrix representation of \eqref{T: Linearization in dissipative} is in the form $M + D$, where $M$ is the  conservative part coming from a Hamiltonian function, and $D$ comes from damping.
This system can be written as resulting from Hamilton's equations with external damping and the following {\it effective quadratic Hamiltonian},
\begin{equation}
\mathcal{H}_2^{\mathrm{eff}}
=\tfrac{1}{2}(p_x^2 + p_y^2) + p_x y- p_y x\underbrace{- \tfrac{1}{2} \left( b_1 x^2 + 2b_2 x y + b_3 y^2 \right)}_{\text{quadratic effective potential, $\mathcal{U}_2$}},
\label{T: H2 in dissipative}
\end{equation}
where the coefficients in the potential are,
\begin{equation}
b_1=-\mathcal{U}_{xx}(x_e,y_e,-y_e,x_e)-1, \hspace{0.05in} b_2=-\mathcal{U}_{xy}(x_e,y_e,-y_e,x_e), \hspace{0.05in} b_3=-\mathcal{U}_{yy}(x_e,y_e,-y_e,x_e)-1.
\end{equation}
Since the coefficients $b_i$ are dependent on the new equilibrium position, $(x_e(c_h),y_e(c_h))$, which is dependent on the damping coefficient, $c_h$,  the coefficients are therefore functions of $c_h$, i.e.,  $b_i(c_h)$.

Note that if no damping ($c_h=0$), or only internal damping, is considered, we have $b_1=2 \bar \mu$, $b_2=0$, and $b_3=-\bar \mu$, since no damping and internal damping both do not shift the equilibrium points away from their zero-damping locations. 
When $c_h$ is non-zero, $b_1(c_h)$, $b_2(c_h)$ and $b_3(c_h)$ change from their values at $c_h=0$, in particular,  $b_2(c_h)\ne0$.  

For small damping,
the eigenvalues of $M$ consist of a real pair, $\pm \lambda$, and a purely imaginary pair, $\pm i \omega_p$, which means the effective conservative system (\eqref{T: Linearization in dissipative} with $c_h=0$) can be put into the standard Hamilton's equations form for an index-1 saddle \eqref{phase-space Hamiltonian Equations} with a Hamiltonian quadratic normal form \eqref{phase-space Hamiltonian}. 
The symplectic transformation matrix $C$ is given by \eqref{effective sympectic matrix}, which reduces to \eqref{C matrix PCR3BP} for the non-damping case ($c_h=0$).  See Appendix \ref{app_pcr3bp} for details.

Taking the derivative of \eqref{T: H2 in dissipative} with respect to time and applying \eqref{T: Linearization in dissipative}, we have,
\begin{equation}
\frac{d \mathcal{H}_2^{\mathrm{eff}}}{d t} = - c_h \left(p_x + y \right)^2 - c_h \left(p_y -x \right)^2 ,
\label{T: internal derivative}
\end{equation}
for the system with internal damping and,
\begin{equation}
\frac{d \mathcal{H}_2^{\mathrm{eff}}}{d t} = - c_h \left(p_x^2 + p_y^2 \right)  - c_h y p_x + c_h x p_y,
\label{T: external derivative}
\end{equation}
for the system with external damping. We see that the system with internal damping always has its energy decrease, while with external damping, the energy could increase or decrease, depending on the specific trajectory path. 

\paragraph{Trajectories in the equilibrium region of position space}
To obtain numerical results, the method described in Section \ref{rotating surface} is applied. 
Since external damping  shifts the locations of the equilibrium points, the dynamic behavior in the conservative system and dissipative system  cannot be compared by the same method as done for other systems where the equilibrium stayed in the same location. 
For the case of $c_h$ non-zero but small, we assume the effective conservative system relative to the equilibrium is a good approximation, that is, the canonical Hamilton's equations coming from the effective Hamiltonian \eqref{T: H2 in dissipative} using $b_1(c_h)$, $b_2(c_h)$ and $b_3(c_h)$.


Figure \ref{T: dissipative position space}, the left panel, shows the projection onto position space in the equilibrium region for the system with external damping. 
For simplicity we just select two bounding line segments on which all trajectories have initial conditions. In the effective conservative system, there are two strips bounding the asymptotic orbits and confining the existence of the wedge of velocity which determines the transit orbits (if any). 
The corresponding wedges of velocity are dark grey shaded, for the nonlinear system, which are partially covered by the light grey shaded wedges for the linear system. 
For the dissipative system with external damping, we obtain the ellipse of transition which is the boundary of the existence of the transit orbits in the linearized approximation. 

Note that the zero velocity curves are no longer a boundary of all trajectories with the specified Hamiltonian value $h$. These curves are just the boundary of {\it initial conditions} in the position space. 
As shown in \eqref{T: external derivative}, the rate of change of $\mathcal{H}_2^{\mathrm{eff}}$ for a system with external dissipation is not always decreasing, but depends on the trajectory path. 
Thus, the Hamiltonian of a trajectory with certain initial conditions on the permissible side of the zero velocity curves 
may increase within an interval of time and may cross the initial zero velocity curve.

To illustrate the results in the linearized effective Hamiltonian system with external damping, we give the results for the nonlinear system by using the bisection method introduced in \cite{zhong2018tube}. 
First, we choose a Poincar\'e section in the linearized system with a fixed $x$ value and select a point on the Poincar\'e section specified by the given fixed $\mathcal{H}_2^{\mathrm{eff}}$ in \eqref{T: H2 in dissipative} which is regarded as an initial condition for the linearized system. 
The initial condition for the linearized system added to the equilibrium point $(x_e,y_e,-y_e,x_e)$ is taken as the initial condition for the full nonlinear system in \eqref{T: Nonlinear Hamiltonian ODEs}. 
The trajectory of the nonlinear system with this initial condition should be a transit orbit. 
Then we select a direction on the Poincar\'e section and another initial condition with the same $\mathcal{H}_2^{\mathrm{eff}}$, which, added to the equilibrium point, gives another initial condition for the full nonlinear system \eqref{T: Nonlinear Hamiltonian ODEs}. 
The trajectory of the nonlinear system with this initial condition should be a non-transit orbit in the limit that the linearization is a good approximation. 

We carry out the bisection method along this direction on the Poincar\'e section in the linearized system until the distance of initial conditions between the transit orbit and non-transit orbit on the Poincar\'e section reaches a 
small tolerance value. 
What we get is the boundary for the nonlinear system with the same initial Hamiltonian value $h$ in the corresponding linearized system. 
In other words, the bisection method is used in the case where the full Hamiltonian in \eqref{T: full Hamiltonian} approximated to quadratic order has a constant value. 
Notice the trajectories in the nonlinear system with fixed magnitude of the quadratic terms of the Hamiltonian have a different `true' Hamiltonian value, since they may have different values for higher order terms in the expansion of the effective Hamiltonian. 

The difference between the present problem and the dynamic snap-through problem in \cite{zhong2018tube} should be pointed out, in that the Hamiltonian for the current nonlinear system has a constant value in the quadratic term {\it only}, while that in \cite{zhong2018tube} has a constant value of the `true' Hamiltonian to arbitrary order. 
The boundary of transition for the nonlinear system with the same quadratic Hamiltonian obtained by the bisection method is given by an area shaded with dark gray in  Figure \ref{T: dissipative position space}. The region  shared by both the nonlinear system and linearized system is shaded medium gray. 
From Figure \ref{T: dissipative position space}, we  conclude that for the region close to the equilibrium point, the boundaries of  the transition region for the nonlinear system and linearized system agree well. 
When one goes farther from the equilibrium point, the agreement becomes worse. 
Since the linearization should only predict the behavior in a small neigborhood of the equilibrium point, this level of (dis)agreement is considered acceptable.


\section{Conclusions and future work}
We have summarized the escape (or transition) geometry in several 
physical problems with two degrees of freedom when dissipative and/or gyroscopic forces are both present: the ball rolling on a stationary or rotating surface, the snap-through of a shallow arch, the roll and pitch dynamics of a ship near capsize, as well as the planar circular restricted three-body problem with drag. 
Since  escape 
occurs through a saddle point in all of these problems, 
we focused on the local behavior near the equilibrium neck region around the saddle. 
The problems are classified into two categories based on the coupling conditions between the saddle and focus canonical planes in the symplectic eigenspace. 

We define a transition region, $\mathcal{T}_h$, as the region of initial conditions of a given initial energy $h$ which transit from one side of a saddle to the other.  The boundary of the transition region, 
$\partial \mathcal{T}_h$, is a co-dimension 1 boundary on each surface of initial energy. For conservative systems $\partial \mathcal{T}_h$ is a tube (topologically, a cylinder), while for dissipative systems, $\partial \mathcal{T}_h$ is an ellipsoid (topologically, a 2-sphere).
These topological results carry over to the nonlinear setting via the stable manifold theorem \cite{Wiggins2003} and a theorem of Moser \cite{Moser1958,Moser1973}, for the dissipative and conservative cases, respectively.
Trajectories with initial conditions outside of $\partial \mathcal{T}_h$ 
do not escape from one side of the saddle to the other. 
The transition tube and transition ellipsoid are divided into two parts by a critical surface; trajectories with initial conditions on the left part (respectively, right part) can transit to the right part (respectively,  left part). 
The projection of the transition tube and transition ellipsoid  onto configuration space  are a strip and ellipse of transition, respectively.  
An initial configuration within the strip (respectively, ellipse) is a necessary condition for a transit orbit. 
The necessary velocity conditions at a specific configuration point within the strip (respectively, ellipse) are given by via a wedge of velocity at that point. 

In this paper, we  investigated only the local behavior around the saddle equilibrium revealing the phase space structure that govern the escape or transition, and did not consider the global behavior. 
A continuation of the study on escape dynamics can apply this theory to more complicated applications. 
However, based on the theorems given above, all the qualitative results of our discussion carry over to the full nonlinear equations, including the topology of $\partial \mathcal{T}_h$.
The bisection method presented in \cite{zhong2018tube} is a useful tool to find $\partial \mathcal{T}_h$ in the global setting. 
A more direct method is to determine the stable manifold of the saddle point, as foliated by energy, which provides another way to compute $\partial \mathcal{T}_h$.
In  future work, both methods will be carried out. 
Furthermore,  higher degree of freedom systems will be considered and the topological results are expected to generalize for the dynamics across index-1 saddles; that is,
the $({2N-2})$-dimensional boundary of transit orbits starting at same initial energy $h$ in $N$ degrees of freedom,  $\partial \mathcal{T}_h$, which are hyper-cylinders (topology $S^{2N-3} \times \mathbb{R}$) in the conservative setting become hyper-ellipsoids in the phase space (topology $S^{2N-2}$) with the addition of dissipation.

\small

\paragraph*{Acknowledgements}
This work was supported in part by the National Science Foundation under award 1537349.  We thank Yue Guan, Shibabrat Naik, Lawrie Virgin, and Yawen Xu for several stimulating conversations on these topics and suggestions of examples to consider.  We also thank an anonymous reviewer who gave several suggestions improving the quality of the paper.

\section*{Appendix}

\begin{appendices}

\section{Details for Example Systems}\label{appendix}

In this appendix, we provide further detailed calculations for the example systems discussed in the text.


\subsection{Snap-through buckling of a shallow arch}\label{app_arch}

Integration of \eqref{odes} can generate the effective kinetic and potential energies,
\begin{equation}
\begin{split}
\mathcal{K}(\dot X, \dot Y)=& \frac{1}{2} M_1 \dot X^2 + \frac{1}{2} M_2 \dot Y^2,\\
\mathcal{U}(X, Y)=& - K_1 \gamma_1 X - K_2 \gamma_2 Y + \frac{1}{2} K_1 X^2 + \frac{1}{2} K_2 Y^2 - \frac{1}{2} N_T\left( G_1 X^2 +  G_2 Y^2 \right) \\
& - \frac{EA}{2L}G_1^2 \left(\frac{1}{2}\gamma_1^2 X^2 - \frac{1}{4}X^4 \right) - \frac{EA}{2L}G_2^2 \left(\frac{1}{2} \gamma_2^2 Y^2 - \frac{1}{4}Y^4 \right) \\
& - \frac{EA}{2L} \frac{G_1 G_2}{2} \left(\gamma_2^2 X^2 + \gamma_1^2 Y^2 -X^2 Y^2 \right).\\
\end{split}
\end{equation}
With the Lagrangian function \eqref{Lagrangian} in hand, \eqref{odes} can also be obtained by Lagrange's equations \eqref{general Lagrange equations} when $q_1=X$ and $q_2=Y$. One can also write the equations of motions in a Hamiltonian form. To do so, one first needs to use the Legendre transformation in \eqref{general Legendre stransformation} to obtain the Hamiltonian function, $\mathcal{H}=\mathcal{K} + \mathcal{U}$, with kinetic energy rewritten by $\mathcal{K}= p_X^2/(2 M_1) + p_Y^2/(2 M_2)$. $p_x=M_1 \dot X$ and $p_Y=M_2 \dot Y$ are the generalized momenta obtained by $p_i=\partial \mathcal{L}/ \partial \dot q_i$. From \eqref{general Hamilton equations}, one can obtain the Hamilton's equations (with damping) for the current problem, 
\begin{equation}
\begin{aligned}
\dot X &=\frac{p_X}{M_1}, \hspace{0.5in} && \dot p_X =- \frac{\partial \mathcal{U}}{\partial X} - C_H p_X, \\
\dot Y &=\frac{p_y}{M_2}, \hspace{0.5in} && \dot p_Y =- \frac{\partial \mathcal{U}}{\partial Y} - C_H p_Y, 
\label{eq:eomHam}
\end{aligned}
\end{equation}
where,
\begin{equation}
\begin{split}
\frac{\partial \mathcal{U}}{\partial X}=& K_1 \left(X - \gamma_1 \right) - N_T G_1 X - \frac{EA}{2L}G_1^2 \left(\gamma_1^2 X -X^3 \right) - \frac{EA}{2L} G_1 G_2 \left(\gamma_2^2 X -X Y^2 \right),\\
\frac{\partial \mathcal{U}}{\partial Y}=& K_2 \left(Y - \gamma_2 \right) - N_T G_2 Y - \frac{EA}{2L}G_2^2 \left(\gamma_2^2 Y -Y^3 \right) - \frac{EA}{2L} G_1 G_2 \left(\gamma_1^2 Y -X^2 Y \right),
\end{split}
\end{equation}
and $C_H=C_1/M_1=C_2/ M_2$ is the single damping coefficient in the Hamiltonian system which can be easily found by comparing \eqref{odes} and \eqref{eq:eomHam}, and using the relations of $M_i$ and $C_i$ in  \eqref{Galerkin-coefficient}.

The parameters in \eqref{linearization} are,
\begin{equation}
\begin{split}
& A_{31}= -K_1 + N_T G_1 + \frac{E A G_1^2 \left(\gamma_1^2 - 3 X_e^2 \right)}{2L} + \frac{E A G_1 G_2 \left(\gamma_2^2 -Y_e^2 \right)}{2L},\\
& A_{32}= - \frac{E A G_1 G_2 X_e Y_e}{L},\\
& A_{42}= -K_2 + N_T G_2 + \frac{E A G_2^2 \left(\gamma_2^2 - 3 Y_e^2 \right)}{2L} + \frac{E A G_1 G_2 \left(\gamma_1^2 -X_e^2 \right)}{2L}.
\label{lin paras}
\end{split}
\end{equation}

\paragraph{Non-dimensional equations of motion}
In order to reduce the number of the parameters, some non-dimensional quantities are introduced,
\begin{equation}
\begin{split}
&\left( L_x, L_y \right)= L  \left(1,\sqrt{ \frac{M_1}{M_2}} \right), \omega_0= \frac{ \sqrt{- A_{32} }}{ \left( M_1 M_2\right)^ \frac{1}{4}}, \tau= \omega_0 t,  \left(\bar q_1 , \bar q_2 \right)= \left( \frac{x}{L_x}, \frac{y}{L_y} \right),\\
&\left(\bar p_1 , \bar p_2 \right)= \frac{1}{\omega_0} \left(  \frac{p_x}{ L_x M_1},\frac{p_y}{ L_y M_2}\right), \left( c_x  , c_y \right)= \frac{1}{ \omega_{0}^2} \left( \frac{A_{31}}{M_1}, \frac{A_{42}}{M_2} \right), c_h= \frac{C_H}{\omega_0}.
\label{dimless quan}
\end{split}
\end{equation}

Using the non-dimensional quantities in \eqref{dimless quan}, the non-dimensional form of the linearized equations of motion, \eqref{linearization}, can be written as,
\begin{equation}
\begin{split}
\dot {\bar q}_1 &= \bar p_1,\\
\dot {\bar q}_2 &= \bar p_2,\\
\dot {\bar p}_1 &= c_x \bar q_1 - \bar q_2 - c_h \bar p_1,\\
\dot {\bar p}_2 &=  -  \bar q_1 + c_y \bar q_2 - c_h \bar p_2
\label{nond eq}.
\end{split}
\end{equation}
Written in matrix form, with column vector $\bar z=(\bar q_1 , \bar q_2 , \bar p_1 , \bar p_2)^T$, 
we have,
\begin{equation}
\dot {\bar z} =M \bar z + D \bar z,
\label{arch_nondim_lin}
\end{equation}
where,
\begin{equation}
M = \begin{pmatrix}
0     	 & 0 & 1 & 0 \\
0	  & 0 & 0 & 1 \\
c_x	  & -1 & 0 & 0 \\
-1	  & c_y & 0 & 0 
\end{pmatrix},
\hspace{0.5in}
D = \begin{pmatrix}
0     	 & 0 & 0 & 0 \\
0	  & 0 & 0 & 0 \\
0	  & 0 & -c_h & 0 \\
0	  & 0 & 0 & -c_h 
\end{pmatrix}
\label{A_and_D_matrix}
\end{equation}
are the Hamiltonian part and damping part of the linear equations, respectively.

We remark that \eqref{arch_nondim_lin} can be transformed into the standard form of \eqref{B:EOM with damping in phase space} 
in the symplectic eigenspace via a symplectic transformation, $\bar z=C z$, where $C$ is now,
\begin{equation}
\begin{split}
C= \begin{pmatrix}
\frac{1}{s_1} & \frac{1}{s_2} & - \frac{1}{s_1} & 0\\
\frac{c_x - \lambda^2}{s_1} & \frac{ \omega_p^2 + c_x}{s_2} & \frac{\lambda^2 - c_x}{s_1}  & 0\\
\frac{\lambda}{s_1} & 0 & \frac{ \lambda}{s_1} & \frac{\omega_p}{s_2} \\
\frac{c_x  \lambda - \lambda^3}{s_1} & 0 & \frac{c_x \lambda - \lambda^3}{s_1} &  \frac{c_x \omega_p + \omega_p^3 }{s_2} 
\end{pmatrix},
\label{arch_sym matrix}
\end{split}
\end{equation}
where $s_1=\sqrt{d_{\lambda}}$ and $s_2=d_{\omega_p}$, and,
\begin{equation}
\begin{aligned}
& d_{\lambda}= \lambda [4-2(c_x-c_y)(\lambda^2 - c_x)], \hspace{0.2in} &&d_{\omega_p}=\tfrac{\omega_p}{2} [4+2(c_x - c_y)(\omega_p^2+c_x)],\\
& \lambda=\sqrt{(c_x + cy+\sqrt{(c_x - c_y)^2 + 4})/2}, \hspace{0.2in} &&\omega_p=\sqrt{-(c_x + cy\sqrt{(c_x - c_y)^2 + 4})/2}.
\end{aligned}
\end{equation}


\subsection{Ship motion}\label{app_ship}

Rescaling the equations of motion \eqref{S: dimensional ODE} by introducing the following parameters,
\begin{equation*}
X=\frac{\phi}{\phi_e}, \hspace{0.3in} Y=\frac{\theta}{2\theta_e}, \hspace{0.3in} \bar t=\omega_{\phi}t,  \hspace{0.3in} F_X=\frac{m_{\phi}}{\omega_{\phi}^2 \phi_e}, \hspace{0.3in} F_Y=\frac{m_{\theta}}{2 \omega_{\phi}^2 \theta_e},
\end{equation*}
one can rewrite  \eqref{S: dimensional ODE}, in the non-dimensional form,
\begin{equation}
\begin{split}
\ddot X &= -X +2 XY+F_X,\\
\ddot Y&=-R^2 Y + R^2 X^2/2 +F_Y,
\label{S:non-d-eqns}
\end{split}
\end{equation}
where $R=\omega_{\theta}/\omega_{\phi}$ denotes the ratio of pitch to roll natural frequencies, and $ \dot{(\ )}=\frac{d}{d \bar t}$. The corresponding Lagrangian is,
\begin{equation}
\mathcal{L} \left(X,Y, \dot X, \dot Y \right) = \frac{1}{2} \dot X^2 +\frac{1}{2} \left( \frac{2 \dot Y^2}{R^2} \right) -  \left(\frac{1}{2} X^2 + Y^2 -X^2 Y \right).
\label{S:Langrangian}
\end{equation}
Using the generalized momenta defined in \eqref{general Legendre stransformation}, the Hamiltonian can be given by,
\begin{equation}
\mathcal{H} \left(X,Y, p_X, p_Y \right)= \frac{1}{2} p_X^2 + \frac{1}{2} \left(\frac{ R^2}{2} \right) p_Y^2 + \left(\frac{1}{2} X^2 + Y^2 -X^2 Y \right).
\label{S:Nonlienar Hamiltonian}
\end{equation}
The Hamilton's equations defined in \eqref{general Hamilton equations} can be written as,
\begin{equation}
\begin{aligned}
\dot X&= p_X,\\
\dot Y&= \frac{R^2}{2} p_Y,\\
\dot p_X&=-X + 2 XY + F_X,\\
\dot p_Y&= X^2  - 2 Y +2 F_Y/R^2.
\label{S: nonlinear Hamiltonian ODE}
\end{aligned}
\end{equation}

For the torques we use viscous damping in roll and pitch, i.e. $\tau_{\phi} (t) = - c_1 \dot \phi$,$\tau_{\theta}(t)=- c_2 \dot \theta$, and obtain $F_X=- c_1\dot X /\left(I_{xx}  \omega_{\phi} \right)$ and $F_Y=- c_2\dot Y /\left(I_{yy}  \omega_{\phi} \right)$. In this setting, the system has two saddle-center equilibrium points at $(\pm 1, 1/2, 0, 0)$ and a center-center equilibrium point at $(0,0,0,0)$. The linearized equations about the saddle point $(1,1/2,0,0)$ can be written as,
\begin{equation}
\begin{pmatrix}
\dot x\\ \dot y\\ \dot p_x\\ \dot p_y
\end{pmatrix}
=
\begin{pmatrix}
0 & 0 & 1 & 0\\
0 & 0 & 0 & R^2/2\\
0 & 2 & -c_1/\left(I_{xx} \omega_{\phi}\right) & 0\\
2 & - 2 & 0 & -c_2/\left(I_{yy} \omega_{\phi}\right)
\end{pmatrix}
\begin{pmatrix}
x\\ y\\ p_x\\ p_y
\end{pmatrix},
\end{equation}
where $(x,y,p_x,p_y)=(X,Y,p_X,p_Y) - (1,1/2,0,0)$ is the displacement from the saddle point in phase space.

Introduce the following non-dimensional quantities,
\begin{equation}
\begin{aligned}
&\omega_0=2^{\tfrac{1}{4}} \sqrt{R}, \tau= \omega_0 \bar t, \bar q_1=x, \bar q_2=\frac{2y}{\omega_0^2}, \bar p_1=\frac{p_x}{\omega_0}, \bar p_2=\frac{\omega_0 p_y}{2}, \\
& c_y= - \frac{\omega_0^2}{2},c_{h_1}=\frac{c_1}{I_{xx} \omega_{\phi} \omega_0},c_{h_2}=\frac{c_2}{I_{yy} \omega_{\phi} \omega_0}.
\end{aligned}
\end{equation}

The equations can be written in the non-dimensional form,
\begin{equation}
\begin{aligned}
\bar q_1&= \bar p_1,\\
\bar q_2&= \bar p_2,\\
\bar p_1&= \bar q_2 - c_{h_1} \bar p_1,\\
\bar p_2&= \bar q_1 + c_y q_2 - c_{h_2} \bar p_2,
\label{ship: EOM in position}
\end{aligned}
\end{equation}
which can be written in matrix form, as in \eqref{eq_ship_matrix}.

\paragraph{Conservative system} 
For the conservative system, i.e. $c_{h_1}=c_{h_2}=0$, one can introduce a change of variables \eqref{change of variables} with the  symplectic matrix $C$ given by,
\begin{equation}
\begin{split}
C= \left( \begin{array}{cccc}
\frac{1}{s_1} & \frac{1}{s_2} & - \frac{1}{s_1} & 0\\
\frac{\lambda^2 }{s_1} & - \frac{ \omega_p^2 }{s_2} & - \frac{\lambda^2}{s_1}  & 0\\
\frac{\lambda}{s_1} & 0 & \frac{ \lambda}{s_1} & \frac{\omega_p}{s_2} \\
\frac{\lambda^3 }{s_1} & 0 & \frac{ \lambda^3 }{s_1} & - \frac{\omega_p^3 }{s_2}
\end{array} \right),
\label{ship sym matrix}
\end{split}
\end{equation}
where,
\begin{equation}
\begin{aligned}
& \lambda= \sqrt{\alpha_1}, \hspace{0.1in} \omega_p=\sqrt{-\alpha_2}, \hspace{0.1in} s_1 = \sqrt{d_{\lambda}}, \hspace{0.1in} s_2 = \sqrt{d_{\omega_p}},\\
& \alpha_1=(c_y + \sqrt{c_y^2 +4})/2, \hspace{0.1in} \alpha_2=(c_y - \sqrt{c_y^2 +4})/2,\\
& d_\lambda =2 \lambda (2+ c_y \lambda^2), \hspace{0.1in} d_{\omega_p}= \omega_p (2 - c_y \omega_p^2).
\end{aligned}
\label{Ship_coefficient}
\end{equation}

\paragraph{Different damping coefficients along the roll and pitch directions}
Considering  unequal damping along the roll and pitch directions and using the change of variables defined by the symplectic matrix in \eqref{ship sym matrix}, the linearized equations \eqref{ship: EOM in position} becomes, in the symplectic eigenspace,
\begin{equation}
\begin{split}
\dot q_1&= (K_1 + \lambda) q_1 + K_1 p_1 	+ K_2 p_2,\\
\dot q_2&=\omega p_2,\\
\dot p_1&= K_1 q_1 + (K_1 - \lambda )p_1 	+ K_2 p_2,\\
\dot p_2&=K_3 q_1- \omega_p q_2 + K_3 p_1 	+ K_4 p_2.
\end{split}
\end{equation}
Written in matrix form, we have,
\begin{equation}
\dot z=\Lambda z + \Delta z,
\end{equation}
where $\Lambda=C^{-1} M C$ from before and the transformed damping matrix is,
\begin{equation}
\Delta=C^{-1} D C= 
c_{h_1} \begin{pmatrix}
K_1 & 0 & K_1 & K_2\\
0& 0 & 0 & 0\\
K_1 & 0 & K_1 & K_2\\
K_3 & 0 & K_3 & K_4
\end{pmatrix},
\end{equation}
where,
\begin{equation*}
\begin{aligned}
K_1 =\frac{\alpha_2 -  \alpha_1 c_{h_2}/c_{h_1} }{2 \left(\alpha_1 - \alpha_2 \right)}, \hspace{0.1in}
K_2 =\frac{ \omega_p (c_{h_2}/c_{h_1}  -1)}{\sqrt{2}  (\alpha_1 - \alpha_2)} , \hspace{0.1in}
K_3 =\frac{c_{h_2}/c_{h_1}  -1}{\sqrt{2} \omega_p (\alpha_1 - \alpha_2)} ,  \hspace{0.1in}
K_4 = \frac{\alpha_2 c_{h_2}/c_{h_1} -  \alpha_1}{\alpha_1 - \alpha_2},
\end{aligned}
\end{equation*}
with $\alpha_i$ as in \eqref{Ship_coefficient}. The corresponding solution which is semi-analytical can be found in Section \ref{rotating surface}.  The matrix $\Delta$  can be re-written in terms of the difference between the two damping coefficients, $\delta c_h = c_{h_2} - c_{h_1}$, as,
\begin{equation}
\Delta=\Delta_{1} + \delta c_h \tilde{\Delta},
\end{equation}
where $\Delta_1$ is the same as the standard uncoupled damping matrix \eqref{standard_damping_matrix}, with $c_h=c_{h_1}$, and,
\begin{equation}
\tilde{\Delta}=
\frac{1}{\alpha_1 - \alpha_2}
\begin{pmatrix}
\tilde{K}_1 & 0 & \tilde{K}_1 & \tilde{K}_2\\
0& 0 & 0 & 0\\
\tilde{K}_1 & 0 & \tilde{K}_1 & \tilde{K}_2\\
\tilde{K}_3 & 0 & \tilde{K}_3 & \tilde{K}_4
\end{pmatrix}, \label{tilde_Delta}
\end{equation}
where,
\begin{equation*}
\begin{aligned}
\tilde{K}_1 =- \frac{ \alpha_1}{2  }, \hspace{0.1in}
\tilde{K}_2 =\frac{ \omega_p}{\sqrt{2}  } , \hspace{0.1in}
\tilde{K}_3 =\frac{1}{\sqrt{2} \omega_p } ,  \hspace{0.1in}
\tilde{K}_4 =  \alpha_2 , \hspace{0.1in}
\alpha_1 - \alpha_2 = \sqrt{c_y^2 +4}.
\end{aligned}
\end{equation*}
Notice that only $\tilde{\Delta}$ contains the terms $\tilde{K}_2$ and $\tilde{K}_3$ which couple the dynamics on the two canonical planes, $(q_1,p_1)$ and $(q_2,p_2)$, and this coupling vanishes when $\delta c_h=0$, i.e., when the two damping coefficients are equal.

\subsection{Ball rolling on a rotating surface}\label{app_ball}

Substituting the surface function $H(x,y)$ defined in \eqref{surface equation}, one can obtain the Lagrangian equations from \eqref{general Lagrange equations} as,
\begin{equation}
\begin{aligned}
& I \left(1 + k_1^2 x^2 \right) \ddot x + I k_1 k_2 x y \ddot y + I k_1^2 x \dot x^2- 2 I \omega \dot y + I k_1 k_2 x \dot y^2 - I \omega^2 x + g k_1 x \\
&+c_d \left[ \left(1+k_1^2 x^2 \right) \dot x + k_1 k_2 x y \dot y\right]=0,\\
& I k_1 k_2 x y \ddot x + I \left(1 + k_2^2 y^2 \right) \ddot y + 2 I \omega \dot x + I k_1 k_2 y \dot x^2 + I k_2^2 y \dot y^2 - I \omega^2 y + g k_2 y\\
&+c_d \left[ \left(1+k_2^2 y^2 \right) \dot y + k_1 k_2 x y \dot x\right]=0.
\end{aligned}
\end{equation}

Once the Lagrangian system is obtained, one can transform it to the Hamiltonian system by use of the Legendre transformation defined in 
\eqref{general Legendre stransformation} which gives the generalized momenta,
\begin{equation}
\begin{split}
p_x &= \frac{\partial \mathcal{L}}{\partial \dot x}= \dot x - y \omega + H_{,x}^2 \dot x + H_{,x} H_{,y} \dot y, \\
p_y &=\frac{\partial \mathcal{L}}{\partial y}= \dot y + x \omega + H_{,x} H_{,y} \dot x + H_{,y}^2 \dot y,       
\end{split}
\end{equation}
and the Hamiltonian function,
\begin{equation}
\begin{split}
\mathcal{H} &= \frac{1}{2 \left(1+ H_{,x}^2 + H_{,y}^2 \right)} \left[ p_x^2 \left(1+ H_{,y}^2\right) - 2 p_x p_y H_{,x} H_{,y} +  p_y^2 \left(1+ H_{,x}^2 \right)\right.\\
&  + 2 p_x \omega \left(y + x H_{,x} H_{,y} + y H_{,y}^2 \right) - 2 p_y \omega \left(x + y H_{,x} H_{,y} + x H_{,x}^2 \right) \\
& \left. - \left(y H_{,x} - x H_{,y} \right)^2 \omega^2 \right] + g H,
\end{split}
\end{equation}
where $p_x$ and $p_y$ are the momenta conjugate to $x$ and $y$, respectively, and the dependence of $H$ on $x$ and $y$ is understood.

The general form of Hamiltonian equations with damping are given by \eqref{general Hamilton equations}. For simplicity, the specific form of the Hamiltonian equations are not listed here. Following the same procedure as for the ball rolling on a stationary surface, we linearize the equations of motion around the equilibrium point at the origin which gives the linearized Hamiltonian equation as,
\begin{equation}
\begin{split}
\dot x &= \omega y + p_x/I,\\
\dot y &= - \omega x + p_y/I,\\
\dot p_x & = -g k_1 x + \omega p_y- c_d \left(\omega y + p_x/I \right),\\
\dot p_y &= -g k_2 y - \omega p_x - c_d \left(-\omega x + p_y/I \right),
\end{split}
\end{equation}
written in matrix form,
\begin{equation}
\begin{pmatrix}
\dot x\\ \dot y\\ \dot p_x \\ \dot p_y
\end{pmatrix}
=
\tilde{M}
\begin{pmatrix}
x \\ y \\ p_x \\p_y
\end{pmatrix}
+ 
\tilde{D}
\begin{pmatrix}
x \\ y \\ p_x \\p_y
\end{pmatrix},
\end{equation}
where,
\begin{equation}
\tilde{M}=
\begin{pmatrix}
0 & \omega & 1/I & 0\\
-\omega & 0 & 0 & 1/I\\
-g k_1 & 0 & 0 & \omega\\
0 & -g k_2 & -\omega & 0
\end{pmatrix}
,\hspace{0.2in}
\tilde{D}=c_d
\begin{pmatrix}
0 & 0 & 0 & 0 \\
0 & 0 & 0 & 0 \\
0 & -  \omega & -1/I  & 0\\
\omega & 0 & 0 & -1/I
\end{pmatrix}.
\end{equation}
The corresponding quadratic Hamiltonian is,
\begin{equation}
\mathcal{H}_2(x,y,p_x,p_y)=\frac{1}{2I}\left(p_x^2 + p_y^2 \right)+ \omega p_x y- \omega p_y x + \frac{g}{2}  \left(k_1 x^2 + k_2 y^2 \right).
\end{equation}

Using the same
re-scaled
parameters as in \eqref{B: nondimensional parameters}, the equations of motion can be rewritten in a 
re-scaled
form as,
\begin{equation}
\begin{split}
\dot{\bar{q}}_1 &= \omega \bar{q}_2 + \bar{p}_1,\\
\dot{\bar{q}}_2 &= - \omega \bar q_1 + \bar{p}_2,\\
\dot{\bar{p}}_1 &= c_x \bar{q}_1 + \omega \bar{p}_2- c_h \omega \bar q_2 - c_h \bar{p}_1,\\
\dot{\bar{p}}_2 & = c_y \bar{q}_2 - \omega \bar{p}_1 + c_h \omega \bar q_1 - c_h \bar{p}_2,
\label{R:linearized nondimen eqns}
\end{split}
\end{equation}
which can be written in matrix form \eqref{linear_ball}.

The characteristic polynomial  of the matrix $M$, the conservative part of the dynamics, from  \eqref{conservative and damping matrix}, which appears in the linear ODE, \eqref{R: conservative ODEs}, is,
\begin{equation}
p(\beta) = \beta^4 + \left(2 \omega^2 - c_x - c_y \right) \beta^2 + \omega^4 + \omega^2 c_x +  \omega^2 c_y + c_x c_y.
\end{equation}
Let $\alpha = \beta^2$, then the roots of $p(\alpha)=0$ are as follows,
\begin{equation}
\alpha_{1,2}= \tfrac{1}{2}\left(c_x + c_y -2\omega^2 \pm \sqrt{\left(c_x - c_y \right)^2 - 8 \left(c_x + c_y \right) \omega^2} \right).
\end{equation}
For the parameters listed in \eqref{surface equation}, one can conclude that $\alpha_1>0$ and $\alpha_2<0$. Here we define $\lambda=\sqrt{\alpha_1}$ and $\omega_p=\sqrt{- \alpha_2}$. 
Now, we want to find the eigenvectors of $M$ in \eqref{conservative and damping matrix} and use them to construct a symplectic linear change of variables which changes \eqref{R: conservative ODEs} into its simpler form \eqref{phase-space Hamiltonian Equations}. Denote the matrix $M- \beta I_4$ by $M_{\beta}$, then,
\begin{equation}
M_{\beta}=
\begin{pmatrix}
\bar M_{\beta} & I_2 \\
B & \bar M_{\beta}
\end{pmatrix}
,\hspace{0.1in}
\bar M_{\beta}=
\begin{pmatrix}
-\beta & \omega \\
- \omega & -\beta
\end{pmatrix}
,\hspace{0.1in}
B=\begin{pmatrix}
c_x & 0 \\
0 & c_y
\end{pmatrix},
\label{R: eigen equations}
\end{equation}
where $I_k$ is the $k \times k$ identity matrix.

Substituting the complex eigenvalues $\pm i \omega_p$ as $\beta$ into \eqref{R: eigen equations}, one obtains a pair of complex eigenvectors with the form $u_{\omega_p} \pm i v_{\omega_p}$. Separating the real and imaginary parts, it gives two generalized eigenvectors,
\begin{equation}
\begin{split}
u_{\omega_p} &= \left(0, \omega_p^2 + c_x + \omega^2, \omega \omega_p^2  - \omega c_x  -\omega^3, 0 \right),\\
v_{\omega_p} &= \left(-2 \omega \omega_p, 0, 0, \omega_p^3 + \omega_p c_x - \omega^2 \omega_p \right).
\end{split}
\end{equation}
Moreover, the remaining eigenvectors associated with eigenvalues $\pm \lambda$ can also be obtained similarly,
\begin{equation}
\begin{split}
u_{+ \lambda} &= \left(\lambda^2 - c_y -\omega^2, -2 \lambda \omega, \lambda^3 - \lambda c_y + \lambda \omega^2, - \omega \lambda^2 - \omega c_y -\omega^3 \right),\\
u_{- \lambda} &= \left(-\lambda^2 + c_y +\omega^2, -2 \lambda \omega, \lambda^3 - \lambda c_y + \lambda \omega^2,  \omega \lambda^2 + \omega c_y + \omega^3 \right).
\end{split}
\end{equation}

Initially, we consider the change of variables defined in \eqref{change of variables}. To find out whether the matrix $C$ is symplectic or not, we check $C^T J C = J$. After some algebra, we can find that, 
\begin{equation}
C^T J C =
\begin{pmatrix}
0 & \bar D\\- \bar D & 0
\end{pmatrix}
, \hspace{0.2in}
\bar D=\begin{pmatrix}
d_{\lambda} & 0\\
0 & d_{\omega_p}
\end{pmatrix},
\end{equation}
where,
\begin{equation}
\begin{split}
d_{\lambda} &=2 \lambda \left[(c_x - c_y -4 \omega^2) \lambda^2 - c_x c_y + c_y^2 - c_x \omega^2 -3 c_y \omega^2 - 4 \omega^4 \right],\\
d_{\omega_p} &=\omega_p \left[(c_x - c_y + 4 \omega^2) \omega_p^2 + c_x^2 - c_x c_y - 3 c_x \omega^2-  c_y \omega^2 - 4 \omega^4 \right].
\end{split}
\end{equation}
This implies that we need to apply some scaling on the columns of $C$ in order to have a symplectic change. Since it can be shown that $d_{\lambda}>0$ and $d_{\omega_p}>0$, the scaling is given by the factors $s_1=\sqrt{d_{\lambda}}$ and $s_2=\sqrt{d_{\omega_p}}$. Thus, the final change is given by the symplectic matrix,
\begin{equation}
C=\begin{pmatrix}
\tfrac{\lambda^2 - c_y -\omega^2}{s_1} & 0 & \tfrac{-\lambda^2 + c_y +\omega^2}{s_1} & \tfrac{-2 \omega \omega_p}{s_2}\\
\tfrac{-2 \lambda \omega}{s_1} & \tfrac{\omega_p^2 + c_x + \omega^2}{s_2} & \tfrac{-2 \lambda \omega}{s_1} & 0\\
\tfrac{\lambda^3 - \lambda c_y + \lambda \omega^2}{s_1} & \tfrac{ \omega \omega_p^2  - \omega c_x  -\omega^3}{s_2} & \tfrac{  \lambda^3 - \lambda c_y + \lambda \omega^2}{s_1}& 0\\
\tfrac{- \omega \lambda^2 - \omega c_y -\omega^3}{s_1} & 0 & \tfrac{ \omega \lambda^2 + \omega c_y + \omega^3}{s_1} &  \tfrac{\omega_p^3 + \omega_p c_x - \omega^2 \omega_p}{s_2}
\end{pmatrix}.
\label{R: symp tranform}
\end{equation}
By using the change of variables with the symplectic matrix in \eqref{R: symp tranform}, one obtains the Hamiltonian equations written in the simple standard form \eqref{phase-space Hamiltonian Equations} with the Hamiltonian function in a normal form \eqref{phase-space Hamiltonian} whose solutions are given in \eqref{phase-space solutions}. The corresponding results and discussion can be found in Section \ref{general phase space} which will not be repeated here.

The parameters in equation \eqref{R: quadratic wedge-strip} are,
\begin{equation}
\begin{split}
a_p&=\frac{s_2^2 \left[ \left(1+c_x \right)^2 + \lambda^2 \omega_p^2\right]}{8 \omega_p \left(1+c_x \right)^2 \left(\lambda^2 +\omega_p^2 \right)},\\
b_p&= \frac{s_2^2 \left[\left(1+c_x \right)^2 \left(\bar q_2^0 - \left(-1 \right)^i \bar q_1^0 \lambda \right) - \left(-1 \right)^i \lambda \omega_p^2 \left(\bar q_1^0 +\bar q_1^0 c_x + \left(-1 \right)^i \bar q_2^0 \lambda \right)\right] }{4 \omega_p \left(1+c_x \right)^2 \left(\lambda^2 +\omega_p^2 \right)},\\
c_p&=  \frac{s_2^2 \left[\left(1+c_x \right)^2 \left(\bar q_2^0 - \left(-1 \right)^i \bar q_1^0\lambda \right)^2 + \omega_p^2 \left(\bar q_1^0 +\bar q_1^0 c_x + \left(-1 \right)^i\bar q_2^0 \lambda \right)^2 \right] }{8 \omega_p \left(1+c_x \right)^2 \left(\lambda^2 +\omega_p^2 \right)} - h.
\end{split}
\end{equation}
Here $i=1,2$ are for stable ($q_1^0=0$) and unstable ($p_1^0=0$) asymptotic orbits, respectively.

The matrix $K$ from \eqref{K_matrix_transform} has components given as follows,
\begin{equation}
\begin{split}
& K_{11}=\frac{2}{S} - \frac{1}{2}, \hspace{0.2in} K_{12}=- \frac{1 + c_y}{S \lambda } \sqrt{\frac{2 (1+ c_x)}{\lambda \omega_p}}, \hspace{0.2in} K_{13}=\frac{c_y - c_x}{2 S},\\
& K_{14}=\sqrt{\frac{2 \omega_p (1 + c_x)}{S^2 \lambda}}, \hspace{0.2in} K_{21}=\frac{\lambda }{S} \sqrt{\frac{2 \omega_p \lambda}{1 + c_x}}, \hspace{0.2in} K_{22} = - \frac{1}{2} + \frac{c_x - c_y - 4}{2S},\\
& K_{23}=K_{21}, \hspace{0.2 in } K_{24}=0, \hspace{0.2in} K_{31}=K_{13}, \hspace{0.2in} K_{32}=K_{12}, \hspace{0.2in} K_{33}=K_{11}, \hspace{0.2in} K_{34} =- K_{14}, \hspace{0.2in} \\
& K_{41}= - \frac{1}{S} \sqrt{\frac{2 \lambda (1 + c_x) }{\omega_p}}, \hspace{0.2in} K_{42}=0, \hspace{0.2in} K_{43}=-K_{41}, \hspace{0.1in} K_{44}=-\frac{1}{2} - \frac{c_x - c_y + 4}{2 S},
\end{split}\label{K matrix}
\end{equation}
where,
\begin{equation*}
S=\sqrt{\left(c_x - c_y \right)^2 - 8 \left(c_x + c_y \right)}.
\end{equation*}

\subsection{The restricted three-body problem}\label{app_pcr3bp}

Following the same procedure as other problems, one can find the change of variables via a symplectic matrix $C$ given by,
\begin{equation}
C=
\begin{pmatrix}
\frac{2 \lambda}{s_1} & 0 & -\frac{2 \lambda}{s_1} & \frac{2 \omega_p}{s_2}\\
-\frac{\lambda^2 +1 + 2 \bar \mu}{s_1} & -\frac{ \omega_p^2  +1 + 2 \bar \mu}{s_2} & -\frac{\lambda^2 +1 + 2 \bar \mu}{s_1} & 0\\
\frac{\lambda^2 +1 + 2 \bar \mu}{s_1} &  \frac{-\omega_p^2  +1 + 2 \bar \mu}{s_2} &  \frac{\lambda^2 +1 + 2 \bar \mu}{s_1} & 0\\
\frac{\lambda^3 + (1-2 \bar \mu ) \lambda}{s_1} & 0  & -\frac{\lambda^3 + (1-2 \bar \mu ) \lambda}{s_1} &  \frac{-\omega_p^3 + (1-2 \bar \mu ) \omega_p}{s_2}
\end{pmatrix}.\label{C matrix PCR3BP}
\end{equation}
where,
$s_1=\sqrt{d_{\lambda}}$ and $s_2=\sqrt{d_{\omega_p}}$, and where,
\begin{equation}
\begin{aligned}
&d_{\lambda}=2 \lambda ((4+3 \bar \mu ) \lambda^2 + 4 + 5 \bar \mu  - 6 \bar \mu ^2), \hspace{0.2in }&& d_{\omega_p}=\omega_p ((4+3 \bar \mu ) \omega_p^2 -4 - \bar \mu  + 6 \bar \mu ^2),\\
& \lambda= \sqrt{\tfrac{1}{2}\left(\bar \mu  - 2 + \sqrt{9 \bar \mu ^2 - 8 \bar \mu}\right)}, \hspace{0.2in} &&\omega_p=\sqrt{-\tfrac{1}{2}\left(\bar \mu  - 2 - \sqrt{9 \bar \mu ^2 - 8 \bar \mu }\right)},
\end{aligned}
\end{equation}
which transforms the quadratic Hamiltonian \eqref{H2_pcr3bp} into the  simple form  \eqref{phase-space Hamiltonian Equations}.
See \cite{KoLoMaRo2011} for details.

\paragraph{Effective quadratic Hamiltonian.}
The quadratic effective potential, $\mathcal{U}_2$, in \eqref{T: H2 in dissipative}, can be written
as a quadratic form,
$\mathcal{U}_2(\mathbf{q};c_h)
=
-\tfrac{1}{2}
\mathbf{q}^T B(c_h) \mathbf{q}
$,
showing the phase space and parameter dependence of $\mathcal{U}_2$,
where $\mathbf{q}=(x,y)^T$ and,
\begin{equation}
B(c_h)=
\begin{pmatrix}
b_1 & b_2\\
b_2 & b_3
\end{pmatrix},
\end{equation}

We  obtain the eigenvalues $\beta$ of the Hamiltonian part of the linearization, $M$, 
\begin{equation}
\begin{aligned}
\beta_\pm^2=\alpha_\pm &= \tfrac{1}{2}\Big[(b_1+b_3-2) \pm \sqrt{(b_1+b_3-2)^2 -4(1+b_1+b_3+b_1 b_3-b_2^2)}\Big] \\
&=\tfrac{1}{2}\Big[({\rm tr}(B)-2) \pm \sqrt{({\rm tr}(B)-2)^2 -4(1+{\rm tr}(B)+{\rm det}(B))}\Big].
\end{aligned}
\end{equation}
Thus, for,
\begin{equation}
({\rm tr}(B)-2)^2 >  4(1+{\rm tr}(B)+{\rm det}(B)),
\label{condition for non-quartic eigenvalues}
\end{equation}
which holds for small damping,
the eigenvalues consist of a real pair, $\pm \lambda$, and a purely imaginary pair, $\pm i \omega_p$, which means  the effective conservative system (\eqref{T: Linearization in dissipative} with $c_h=0$) can be put into the standard Hamilton's equations form for an index-1 saddle \eqref{phase-space Hamiltonian Equations} with a Hamiltonian quadratic normal form \eqref{phase-space Hamiltonian}. The symplectic transformation matrix is given by,
\begin{equation}
C=
\begin{pmatrix}
\frac{b_2 + 2 \lambda}{s_1} 				& \frac{b_2}{s_2} 				 & \frac{b_2 - 2 \lambda}{s_1} 					& \frac{2 \omega_p}{s_2} \\
-\frac{\lambda^2 + 1 + b_1			}{s_1}& -\frac{ \omega_p^2 + 1 + b_1}{s_2} &-\frac{\lambda^2 			  +1 + b_1}{s_1} 	& 0\\
\frac{\lambda^2 + 1 + b_1 + b_2 \lambda }{s_1}&  \frac{-\omega_p^2  +1 + b_1}{s_2} & \frac{\lambda^2 -b_2 \lambda +1 + b_1}{s_1} 	& \frac{b_2 \omega_p}{s_2}\\
\frac{\lambda^3 +  (1-b_1 )\lambda + b_2}{s_1} & \frac{b_2}{s_2} 			 	 &-\frac{\lambda^3 + (1- b_1)\lambda -b_2}{s_1} 	& \frac{-\omega_p^3 +  (1 -b_1)\omega_p}{s_2}
\end{pmatrix},
\label{effective sympectic matrix}
\end{equation}
where $s_1=\sqrt{d_{\lambda}}$ and $s_2=\sqrt{d_{\omega_p}}$, and,
\begin{equation}
\begin{aligned}
&d_{\lambda} ~~=     2 \lambda \left[(4 + b_1 - b_3) \lambda^2  + 4 + 3 b_1 + b_3 -2 b_2^2  - b_1^2 + b_1 b_3 \right], \\
&d_{\omega_p}= \omega_p \left[(4 + b_1 - b_3) \omega_p^2 - 4 - 3 b_1  - b_3 +2 b_2^2 + b_1^2 - b_1 b_3 \right],\\
& \lambda	 ~~= \sqrt{~~\tfrac{1}{2}\left(b_1+b_3  - 2 + \sqrt{(b_1+b_3-2)^2 -4(1+b_1+b_3+b_1 b_3-b_2^2)}\right)},\\ 
& \omega_p=\sqrt{   -\tfrac{1}{2}\left(b_1+b_3  - 2 -  \sqrt{(b_1+b_3-2)^2 -4(1+b_1+b_3+b_1 b_3-b_2^2)}\right)}
\end{aligned}
\end{equation}
We remark that \eqref{effective sympectic matrix} reduces to \eqref{C matrix PCR3BP} for the non-damping case ($c_h=0$).
Note that if \eqref{condition for non-quartic eigenvalues} is not satisfied, then we have a quartic of complex eigenvalues and the conservative equilibrium is a complex saddle, rather than an index-1 saddle as considered throughout this paper.  The complex saddle case is not considered here.

\end{appendices}



\end{document}